\documentclass[9pt]{extarticle}       % onecolumn (second format)
\usepackage{graphicx}
\usepackage{tikz}
\usepackage{multirow}
\usepackage{amssymb,amsfonts,amsmath}
\usepackage{mathrsfs}
\usepackage{array}
\usepackage{color}

\DeclareSymbolFont{msbm}{U}{msb}{m}{n}
\DeclareMathSymbol{\R}{\mathalpha}{msbm}{'122}
\renewcommand{\S}{\mathbb{S}}

\begin{document}

\title{Are tumor cell lineages solely shaped by mechanical forces?}

%\author{Mathieu Leroy-Ler\^{e}tre \and Giacomo Dimarco \and Martine Cazales\textsuperscript \and Marie-Laure Boizeau\textsuperscript \and Bernard Ducommun \and
%Val\'{e}rie Lobjois \and Pierre Degond %\textsuperscript,
%}

\author{Mathieu Leroy-Ler\^{e}tre  \\
Institut de Math\'ematiques de Toulouse, Universit\'{e} de Toulouse, France.\\
ITAV-USR3505, Universit\'{e} de Toulouse, CNRS, UPS, France.  \\
\and 
Giacomo Dimarco \\
University of Ferrara, Department of Mathematics and Computer Science, Italy. \\
\and
Martine Cazales \\
ITAV-USR3505, Universit\'{e} de Toulouse, CNRS, UPS, France. \\
\and
Marie-Laure Boizeau \\
ITAV-USR3505, Universit\'{e} de Toulouse, CNRS, UPS, France.\\
\and
Bernard Ducommun\\ 
ITAV-USR3505, Universit\'{e} de Toulouse, CNRS, UPS, France.\\
CHU Toulouse, France.\\
\and
Val\'{e}rie Lobjois\\ 
ITAV-USR3505, Universit\'{e} de Toulouse, CNRS, UPS, France.
\and Pierre Degond\\
Department of Mathematics, Imperial college London, United Kingdom.
}

% \institute{M. Leroy-Ler\^{e}tre \at
% Institut de Math\'ematiques de Toulouse, Universit\'{e} de Toulouse, CNRS, UPS, France.\\
%  ITAV-USR3505, Universit\'{e} de Toulouse, CNRS, UPS, France.\\
%               \email{Mathieu.leroy-leretre@math.cnrs.fr}           
%            \and
%            G. Dimarco \at
% University of Ferrara, Department of Mathematics and Computer Science, Ferrara, Italy.
% \email{giacomo.dimarco@unife.it}
% \and
% M. Cazales \at  ITAV-USR3505, Universit\'{e} de Toulouse, CNRS, UPS, France.\\
% \and
% M.L. Boizeau \at  ITAV-USR3505, Universit\'{e} de Toulouse, CNRS, UPS, France.\\
% \and
% B. Ducommun \at
%  ITAV-USR3505, Universit\'{e} de Toulouse, CNRS, UPS, France.\\
% CHU Toulouse, France.\\
% \email{bernard.ducommun@itav.fr}
% \and
% V. Lobjois \at
% ITAV-USR3505, Universit\'{e} de Toulouse, CNRS, UPS, France.\\
% \email{valerie.lobjois@itav.fr}
% \and
% P. Degond \at
% Department of Mathematics, Imperial college London, United Kingdom.\\
% \email{p.degond@imperial.ac.uk}
% }
% 
% \date{Received: date / Accepted: date}

\maketitle

\begin{abstract}
This paper investigates cells proliferation dynamics in small tumor cell aggregates using an individual based model (IBM). The simulation model is designed to study the morphology of the cell population and of the cell lineages as well as the impact of the orientation of the division plane on this morphology. Our IBM model is based on the hypothesis that cells are incompressible objects that grow in size and divide once a threshold size is reached, and that newly born cell adhere to the existing cell cluster. We performed comparisons between the simulation model and experimental data by using several statistical indicators. The results suggest that the emergence of particular morphologies can be explained by simple mechanical interactions.
%\keywords{Cell proliferation \and Individual based model \and energy minimization \and lineage morphology}
\end{abstract}

\section{Introduction}
\label{intro}
Cancer cells proliferate at a high rate and can be considered as a dynamic population of agents that grow and divide without constraints \cite{HW00} (at least in the early phase of avascular growth). In the present work, we aim to investigate this preliminary stage of a tumor growth. As such, this study could also help in understanding the development of 3D microtumours. Here, we examine a population initially composed of several tens of cells that proliferate to reach several hundreds of cells within a few days. In this situation, the experimental model consists of cancer cells grown in a culture medium containing all the necessary nutrients for their growth and division. The proliferation of the population is restricted to a single layer, which permits the use of simple two dimensional simulation models for the comparisons. In the proposed simulation model we do not include a detailed description of the cell cycle and mitosis events. Instead we focus on the role of the orientation of the cell division planes in the morphology of the tumor cells cluster. We aim to investigate how the orientation of cell divisions influences the structure of the cell lineages. In particular, we would like to determine whether cell lineages break up and whether they have different morphologies according to their initial position in the cell population. The impact of the orientation of the division plane on the organization of the population has been suggested from recent studies. In particular, the influence of the geometry of the cell, the influence of neighboring cells and the role of external mechanical forces on the determination of the orientation of the division plane have been studied in \cite{Gib11,MBC11,TB06,Des13,Fin11}. Recent results also suggest that the orientation of the division plane plays a role in the differentiation of stem cells \cite{LF05,WSS07}. Moreover, recent discoveries tend to establish a link between cancer and disorientation of the division plane \cite{PT11}: dysfunction in cells can lead to disorientation and, conversely, disorientation can promote the development of cancer. The above observations lead us to focus on two main questions related to the organization of a growing tumor cell population: what is the impact of the multiplication of cells on the global organization of the entire cell population? Does the orientation of the division plane influence the evolution of the cell population and the organization of its lineages? The comparison of a mathematical model with biological experiments performed in this work shows that indeed a relation exists between division and organization and that lineages are strongly influenced by the initial position of the parental cell inside the population. 

There exists a large number of mathematical models in the literature describing cell proliferation and tumor growth. Some recent review papers can be found in \cite{AM04,Byr10,Maini06,Maini07}. Among the different ways of representing cells, one usually distinguishes between discrete models and continuous models. In a discrete model \cite{KRA12,RM10,BM07,Bon02,Byr09,Ubezio98,Ubezio01}, each element is treated as a separate entity.
This makes the comparison with experiments easy but the main drawback of this viewpoint is the huge computational cost when dealing with a large number of agents. On the other hand, continuous models \cite{Maini_pnas,Preziosi03,RCM07,AM06,Bre10,SC01}, which typically deal with an average density of cells, are more efficient when the system contains a large number of particles. However, it is difficult to establish direct links between the model parameters and the physical measures \cite{Mac10}. Comparisons between the two approaches can be found in \cite{SM06,BD09}, while hybrid models which employ both approaches at the same time can be found for instance in \cite{Colin06,ZAD07,Pat01}. 

Many models of the literature tend to be exhaustive in the description of the biological, physical and chemical phenomena. This leads to the introduction of many empirical parameters and makes the interpretation of the results difficult. Our approach is opposite: it relies on a simple mathematical model which focuses on few  determinants and attempts to explore some specific questions through comparisons with experiments. This approach permits to explore the influence of any single modeling choice more efficiently and to formulate hypotheses about which mechanisms are associated with given observations. We follow a bottom-up strategy which starts from simple rules and gradually adds complexity into the model until a good fit with the experiments is reached. Rather than quantitative agreement, we look for similar trends between the model and the experiments when some key parameters are varied. Here, we show that the sole growth and division mechanisms are not sufficient to explain the observed lineage morphologies and that additional phenomena must be taken into account in the model in order to reproduce the experimental results.
The biological situation we wish to investigate is a small population composed of 20 up to about 500 cells in which proliferation and movement are restricted to a two-dimensional plane. For this specific situation, the best modeling choice is an Individual Based Models (IBM) (see \cite{DH05,GH09,MPL01,Pal08,Rej07} for example). This permits to make the mathematical model reproduce the experiment and to consider objects moving only in a two-dimensional plane. In addition, since we seek to study the impact of the growth of cells and the influence of the orientation of division on the organization of the population, we need to be able to track individual entities. Indeed, a continuous model will not give access to such information. On the other hand, a discrete model on a grid \cite{ACR07,GG93,Shi09} could introduce artificial bias on the organization of the cells and it will not allow us to explore orientation issues in depth, since the number of possible orientations is limited by the underlying grid \cite{RCM07,Mac10,DD05}. 

The mathematical setting chosen is finally the following: cells aggregate spontaneously, grow and divide. After each growth or division event, a mechanical equilibrium between aggregation and cell-cell non overlapping is supposed instantaneously reached and gives the instantaneous configuration of the population. Thus motion arises from stresses between neighboring cells \cite{Tambe11,Laporta16,Levine01,Levine12}. This approach is different from more classical models based on introducing a repulsion potential between the cells \cite{BD09,DH05}. Indeed, the temporal scale associated with cell (quasi)-incompressibility is much faster than that involved in the growth of the tumor as a whole. Modelling cell-cell non-overlapping via a repulsion potential requires making these two scales closer than they are in reality (to ensure numerical stability), thereby introducing a bias in the numerical solution. In a current study \cite{Marina}, we intend to document precisely the difference between these two approaches. Since cell-cell non-overlapping is associated with a faster scale than growth an approach based on realizing a mechanical equilibrium at every time step permit to bypass the numerical stability issue. During the time evolution of the system different lineages are tagged and compared with experimental data as done for instance in \cite{Sepulveda13}. Comparisons between the mathematical model and the biological experiments show that a relation exists between geometric determinants of cell, division, and the organization of the cell population and that lineage shapes are strongly influenced by the initial position of the parental cell inside the population.

The paper is organized as follows. In Section \ref{sec:2} we discuss the mathematical model and the numerical method adopted. We also discuss the statistical indicators used to
measure the results of the numerical simulations, the experimental protocol and the image processing. In Section \ref{sec:3} we resume the principal results of the simulations and a first series of comparisons between the model and the experiments are presented and discussed. In Section \ref{sec:4} improvements introduced to the model and new comparisons with the data are analyzed. In Section \ref{sec:5} conclusions are drawn and future investigations are discussed.

\section{Mathematical model and experimental protocol}
\label{sec:2}
\subsection{Mathematical model: general description}
\label{sec:2.1}
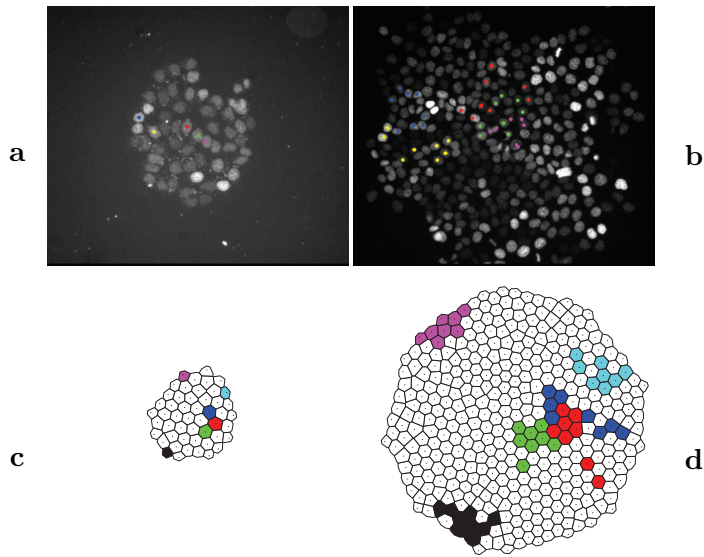
\begin{figure*}[ht]
\centering
\includegraphics[scale=0.68]{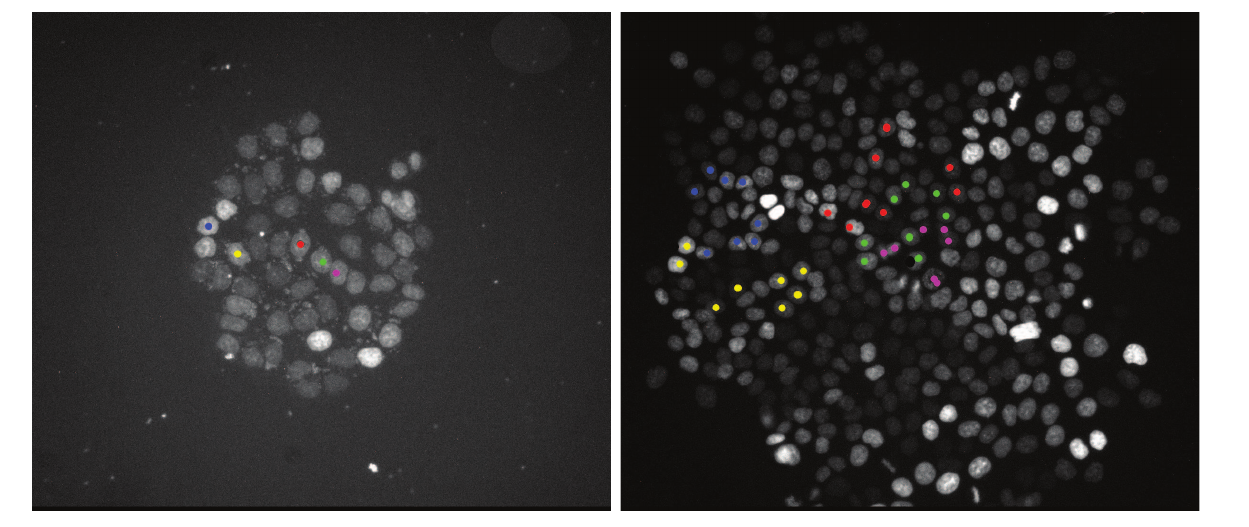}\\
\includegraphics[scale=0.9,trim={-0cm 0 0 0}]{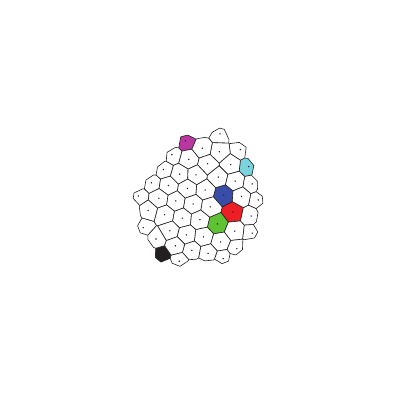}
\includegraphics[scale=0.9,trim={-0.5cm 0 0 0}]{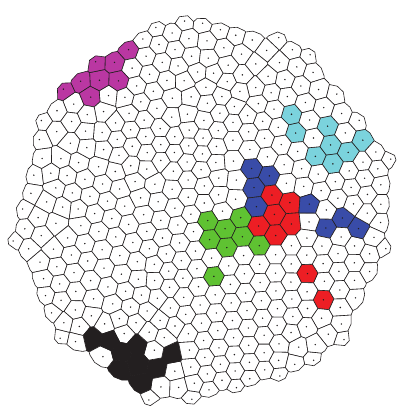}
\caption{Monitoring of cell lineages in a proliferating cell population. Top images (a) and (b) are experimental data obtained through video-microscopy monitoring of a HCT116 colon cancer cells population at time 0 (a) and 72h (b). Lineages are identified with colorful stickers that have been added after image segmentation and time-lapse analysis. Bottom images (c) and (d) illustrate numerical results: an example of the results of a numerical simulation, (c) initial state and (d) final state. Cells are represented by using the Vorono\"{\i} diagram which permits an easier definition of the concept of neighboring cells and periphery of the population and of the single lineage. The same color indicates cells of the same lineage. \label{fig1}}
 \begin{tikzpicture}[overlay]
    \node at (-4.5,9.55){
      {\Large {\bf a}}
    };
    \node at (-4.5,5.5){
      {\Large {\bf c}}
    };
    \node at (4.5,9.55){
      {\Large {\bf b}}
    };
    \node at (4.5,5.5){
      {\Large {\bf d}}
    };
  \end{tikzpicture}
\end{figure*}

Since the roles of division and growth in the lineage organization are at the center of this study, we must be able to track individual cells during time. In order to do that, we use an agent based mathematical model in which each cell is represented by a discrete entity. In this model, a cell/agent is defined by its center, its radius and its orientation. Only the plasma membrane is described. The details of the intracellular phenomena are omitted. The cell shape is chosen to be a two-dimensional disk which is incompressible and continuously growing in time. The notion of preferred orientation, linked to the alignment of chromosomes during mitosis, necessary to define the division plane, is not inherently modeled by the shape of the agent. Instead it is an internal parameter owned by each agent which may vary with time depending on the chosen division strategy and detailed next. The cell evolution is determined only by the growth and the division laws, respectively describing the interphase and the mitotic steps of a cell cycle. To reproduce the experimental setting, cells are free to move and have access to all the required nutrients. Cell cycle phases of growth and division alternate continuously; cell cycle checkpoints, $G1$ phase variability between cells and temporary cell cycle exits are excluded. The interphase is constituted solely of the cell growth phase: $G1$-, $S$- and $G2$-phases are combined into a single no-division phase during which growth is assumed to be linear in time. The mitosis trigger occurs as soon as the cell reaches a critical size, which is the only control condition. Conservation of the volume is imposed during the division process. Agents interact by minimizing at each time the global mechanical energy of the system subject to a non-interpenetration constraint modeling the fact that living cells cannot intermingle. This dynamic causes global as well as individual cell movement, which is thus the product of the combined actions of growth and division on the one hand, and the non-overlapping constraint on the other hand. Growth is modeled as a continuous phenomenon except at the time of division, which occurs when a cell reaches approximately twice the volume of a newborn daughter cell. A uniform probability distribution is added to the growth increment over a time-step to introduce some
randomness in the cell division starting time. When this process starts the mother cell deforms itself in a dumbbell shaped geometry to give birth to two identical daughter cells. Deformation occurs with total volume kept constant. The duration of mitosis is short compared to the interphase (around one over thirty units of time). Thus, when a division occurs, the other cells stop growing, i.e. we consider mitosis as an instantaneous phenomenon. During division an equilibrium between the mechanical adhesion forces and the non-overlapping constraint determines the state of the system at each instant of time. In order to explore the influence of the orientation of the division plane onto the lineage organization, we consider three different possibilities: divisions occur in \textit{1)} a random direction, \textit{2)} in the direction of the line joining the origin and the mother cell (radial direction) \textit{3)} in the direction orthogonal to the radial direction (tangential direction). Furthermore, in addition to the different division planes, we consider two different strategies. \textit{1)} Free orientation strategy: orientation is chosen at the beginning of the division but is free to change during the deformation into a dumbbell-like shape and finally into two daughter cells. This change of orientation is only due to the interactions with neighboring agents through the energy minimization procedure which defines the new configurations. \textit{2)} Constrained orientation strategy: Orientation is chosen at the beginning of the division and remains fixed up to the end of the division process. At each time step, a minimum of this mechanical energy subject to the non-overlapping constraint is computed. At the beginning of the next time step, cells radii grow and divisions may arise. This induces a disruption of the mechanical equilibrium and thus a new minimum of the energy is computed. Fig. \ref{fig1} (c) and (d) show a typical result of the simulation of the described model. A Vorono\"{i} representation, detailed in \ref{sec:2.4}, is used to define the concept of neighboring cells. Figs.~\ref{fig1} (a) and (b) illustrate the initial and the final phases of the proliferation and monitoring of the lineages. The cells tracking procedure is described in section \ref{sec:3}.

\subsection{Detail on the model}
\label{sec:2.2}
\paragraph{Rules for cells positioning.}
As already stated, each cell is described by a 2D incompressible disk  with a center positioned at $$X_i(t)=(x_i(t),y_i(t)),$$ a radius $R_i(t)>0$ and an orientation $\omega_i(t)\in \S^1$ (the set of two-dimensional vectors with unit length) depending on time $t$. In this setting, we use $\xi(t)$ to denote the vector whose elements are the positions of the cells, i.e. $\xi(t)=(X_1(t),X_2(t),..,X_{N(t)}(t))$, while $\rho(t)$ is the vector whose elements are the radii of the cells, i.e. $\rho(t)=(R_1(t),R_2(t),..,R_{N(t)}(t))$. The number of cells at time $t$ is denoted by $N(t)$. Each cell belongs to a lineage $\ell_i$ which defines the developmental history of a given initial mother cell and which does not evolve with time. What evolves in time is the number of cells $N_{\ell_i}(t)$, belonging to a given lineage $\ell_i$, due to mitosis. 

\begin{figure}
%\vskip-0.5cm
\hspace*{1.cm}
\includegraphics[scale=0.7,trim={0 0 0 0},clip]{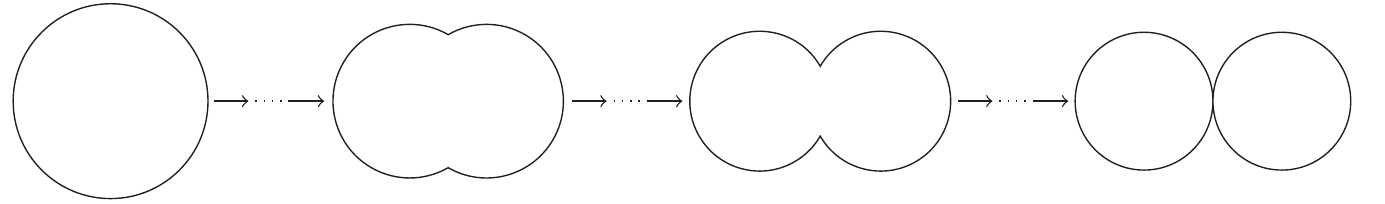}
\vskip0cm
\caption{Different steps of the division process. From one mother cell on the left, at initial time $t=\tau_0$, up to two daughter cells on the right at the end of the process $t=\tau_f$.\label{figd}}
 \begin{tikzpicture}[overlay]
    \node at (2.3,1.6){
      {\tiny $\tau_0$, {\bf Mother cell}}
    };
    \node at (4.4,1.6){
      {\tiny $\tau_1$}
    };
        \node at (6.7,1.6){
      {\tiny $\tau_{f-1}$}
    };

    \node at (9.4,1.6){
      {\tiny $\tau_{f}$, {\bf Daughter cells}}
    };
  \end{tikzpicture}
\end{figure}
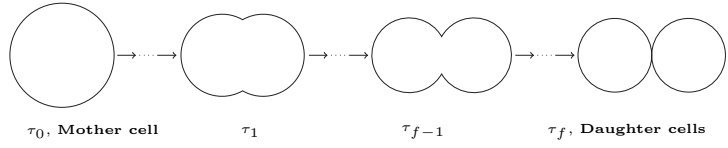

\begin{figure}
\vskip-0.5cm
\hspace*{1.5cm}
\includegraphics[scale=0.8,trim={-2cm 0cm 0cm 0cm}]{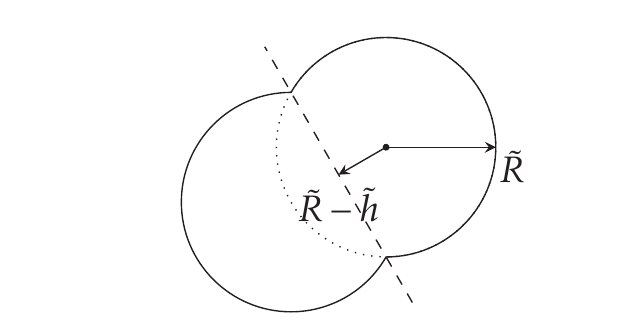}
%\vskip-4.5cm
\caption{Configuration of two daughter cells during the division process. The division plane is represented by a dashed line. The radius $\tilde R$ indicates the generic radius of two new born cells during the mitosis which is parametrized by $\tilde h$. \label{fig2}}
\end{figure}

The impenetrability condition between two cells $i$ and $j$ is expressed by an inequality constraint $\phi_{ij}$ with a suitable function $\phi_{ij}$ which expresses the fact that two cells should not overlap. Thus an admissible configuration $\mathcal A(t)$ for the system is a set of positions $\xi(t)$ such that $\phi_{ij}(\xi(t),\rho(t))\le 0$ for all possible indices $i$ and $j$: 
\begin{equation}\hskip-0.1cm\label{eq:set}
\begin{split}
&\mathcal A(t)=\{ \xi(t)\in(\R^2)^{N(t)} \ | \ \forall i,j\in[1,N(t)], \\
& i\ne j, \phi_{ij}(\xi(t),\rho(t))\le 0\}.  
\end{split}
\end{equation}
The global adhesion potential is expressed by a function 
\begin{equation}
W:(X_1(t), X_2(t),..,X_{N(t)}(t))\rightarrow \sum_{i=1}^{N(t)} V(X_i(t)) 
\end{equation}
where $X_i(t)\rightarrow V(X_i(t))$ is a convex function on $\R^2$. The instantaneous configuration at time $t$ is then given by a minimum $\xi^*(t)$ of the potential $W$ under the constraint that $\xi^*(t)$ belongs to the set of admissible configurations $\mathcal A(t)$, i.e. 
\begin{equation}\label{eq:min}
\xi^*(t)=\textrm{argmin}_{\xi(t)\in \mathcal A(t)} W(\xi(t)). 
\end{equation}
In this setting, the non-overlapping condition is defined by $\phi_{ij}(\xi(t),\rho(t))=(R_i(t)+R_j(t))^2-|X_i(t)-X_j(t)|^2$, where $|X_i(t)-X_j(t)|^2=(x_i(t)-x_j(t))^2+(y_i(t)-y_j(t))^2$ is the Euclidean distance on $\R^2$ between cell located at $X_i(t)$ and cell located at $X_j(t)$. The potential function models the trend of the cells to regroup themselves isotropically around a given position chosen to be the origin of the coordinate system. The potential function we consider is quadratic $V_Q(X_i(t))=x_i^2(t)+y_i^{2}(t)$. 
\paragraph{Growth law.}
We introduce the size of a new born cell $R_{min}$, the size of a cell just before mitosis $R_{max}$ and $T_G$ the mean duration of the growth phase. Even though the model is two dimensional, we consider cells as tridimensional structures whose volumes grow linearly in time. Thus the growth law of the i-th cell is given by 
\begin{equation}\label{eq:growth}                                                                                                                                                                                                                                                                                                                                                                                                                                                                                                                                                                                                                                                                                                                                                                                                                                                                                                                                                                                                                                                                                                                                                                                                                     R_i^3(t)=R^3_{min}+(1+\gamma)\frac{R^3_{max}-R^3_{min}}{T_g}t
\end{equation}                                                                                                                                                                                                                                                                                                                                                                                                                                                                                                                                                                                                                                                                                                             where $\gamma$ is a random variable sampled from an uniform distribution with support on $[-\alpha,\alpha]$. Once a cell reaches a radius $R(t)\geq R_{max}$ it starts to divide into two daughter cells. Eq. (\ref{eq:growth}) is discretized in small time steps $\Delta t$. After a time step cell growth leads to the violation of non overlapping constraints. Thus a new energy minimum must be computed through (\ref{eq:min}) resulting in a repositioning of the cells. Then a new growth step is performed followed by a repositioning step. The cycle of growth and repositioning is repeated until one cell starts to divide. 
\paragraph{Division rules.}
The initial orientation $\omega_{i_0}$ of the division plane of the cell $C_i$ is random, radial or tangential. The radial and tangential directions are computed relative to the origin supposed to be the center of the tumor. The division process starts when a cell $C_i$ reaches a size $R_{i_0}(t)\geq R_{max}$ at time $t$. The process is considered as discrete in time and at each time step the disk which describes the mother cell stretches apart in a peanut like shape until the final separation in two daughter cells as shown in Fig. \ref{figd}. 
During this process the volume is kept constant equal to the volume of the mother cell. At each discrete instant of time, $\tau_k, \ k\in [1,f]$ (where $f$ is the total number of intermediate steps in the division process) a new equilibrium of the whole system is computed by solving (\ref{eq:min}) with a modified set of admissible configurations $\mathcal A(\tau_k)$ (at step $\tau_k$) as described below. When the mitosis comes to an end, each daughter cell has reached half the size in volume of the mother cell. Moreover, they have the same size and shape and their position is symmetric with respect to the division plane. The orientation of the division plane is described by the unit vector $\omega_i(\tau_k), \ \forall i$. Since the division process is much faster than the cell cycle, we make the hypothesis that two cells cannot divide at the same time and that during division the other cells do not grow.\\  
Let $\omega_{i_0}(t)$ the orientation of the division plane of the mother cell when the division starts, $(x_{i_0}(t),y_{i_0}(t))$ its coordinates and $\ell_{i_0}$ its lineage. Then, initially, the two daughter cells occupy the same location in space as the mother cell, i.e. $(x^+(\tau_0),y^+(\tau_0))=(x^-(\tau_0),y^-(\tau_0))=(x_{i_0}(t),y_{i_0}(t))$; they share the same orientation $\omega^+(\tau_0)=\omega^-(\tau_0)=\omega_{i_0}(t)$ and they belong to the same lineage $\ell^+(\tau_0)=\ell^-(\tau_0)=\ell_{i_0}$, where the upper indices $+$ and $-$ refer to the two daughter cells and the $i_0$ index to the mother cell. During division the lineage of the two daughters remains unchanged while the two radii $R^+(\tau)$ and $R^-(\tau)$ are functions of the time during division $\tau$ (which is rather a degree of completion of the division process), and are such that the initial volume of the mother cell is preserved in time. During the division process the real time variable $t$ is kept constant. In particular, at the end of the process the two radii are such that $R^+(\tau_{f})=R^-(\tau_{f})=R_{i_0}(t)/\sqrt[3]2$. The transformation is parametrized by a function $\tilde h(\tau_k)=R_{i_0}-k\frac{R_{i_0}}{f}$, see Fig. \ref{fig2}, where for each step $\tau_k$, the $R^\pm(\tau_k)$ are obtained by solving the following equation $$R^\pm(\tau_k)^3-\frac{3\tilde h(\tau_k)^2}{4}R^\pm(\tau_k)+\frac{h(\tau_k)^3}{4}-\frac{R_{i_0}^3(t)}{2}=0,$$ which expresses the conservation of volume during the division process since the volume of a daughter cell is $$\mathcal{V}^\pm(\tau_k)=\frac{\pi}{3}(R^\pm(\tau_k)^3)-3\tilde h(\tau_k)^2 R^\pm(\tau_k)+\tilde h(\tau_k)^3)$$ at time $\tau_k$ while $\mathcal{V}_{i_0}(t)=2\mathcal{V}^\pm(\tau_k)$ with $\mathcal{V}_{i_0}(t)$ the volume of the mother cell at time $t$ before the division starts. This value then defines the new positions through
\begin{equation*}
\begin{cases}
&x^\pm(\tau_{k+1}) = \frac{x^+(\tau_k)+x^-(\tau_k)}{2}\pm(R^\pm(\tau_k)-\tilde h(\tau_k))\cos(\omega(\tau_k)) \\
&y^\pm(\tau_{k+1}) = \frac{y^+(\tau_k)+y^-(\tau_k)}{2}\pm(R^\pm(\tau_k)-\tilde h(\tau_k))\sin(\omega(\tau_k)),\\
\end{cases}
\end{equation*}
since the two new born cells are placed along the normal vector direction to the plane of division at the distance $R^\pm(\tau_k)-\tilde h(\tau_k)$ from this plane. Once the new positions are computed, the non overlapping constraint is likely to be violated. A new minimal energy configuration $\xi^*(\tau_{k+1})$ must be computed at step $\tau_{k+1}$ solving (\ref{eq:min}). Here the definition of the set of admissible configuration is different from (\ref{eq:set}) and incorporate equality constraints $\tilde\phi_{ij}$ associated with the maintenance of the peanut shape when the pair $(i,j)$ corresponds to two daughter cells, i.e. $(i,j)=(i_+,i_-)$. In addition in the case of fixed orientation strategy another constraint is added to the system which imposes $\omega^\pm(\tau_{k+1})=\omega^\pm(\tau_{k}), \ \forall k$, which means that the dividing cells do not change their orientation during the re-positioning. By contrast, in the free orientation case, $\omega^\pm(\tau_{k+1})\ne\omega^\pm(\tau_{k})$, i.e. no constraint is imposed on the new orientation at step $\tau_k$. This new constraints are defined in \ref{sec:2.3}.

\subsection{Numerical solution of the model}
\label{sec:2.3}
\begin{figure}
%\vskip-0.8cm
\hspace*{2cm}
\includegraphics[scale=0.62]{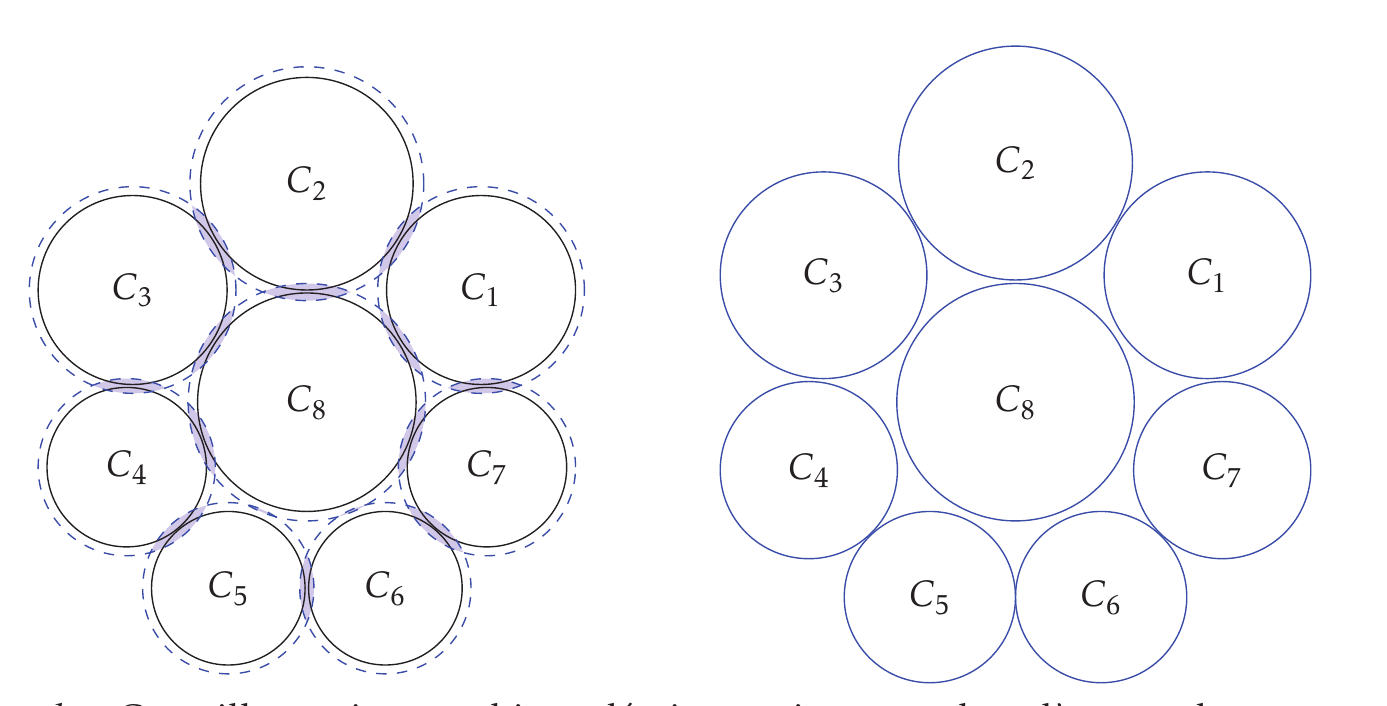}
%\vskip-3cm
\caption{Illustration of the sequence of growth and positioning steps for eight cells of different size. Left: a typical equilibrium configuration (black line) and the situation after the growth stage (blue dashed line). Right: a typical equilibrium configuration after the repositioning procedure.\label{fig3}}
\end{figure}

\begin{figure}
%\vskip-0.6cm
\hspace*{1.8cm}
\includegraphics[scale=0.62]{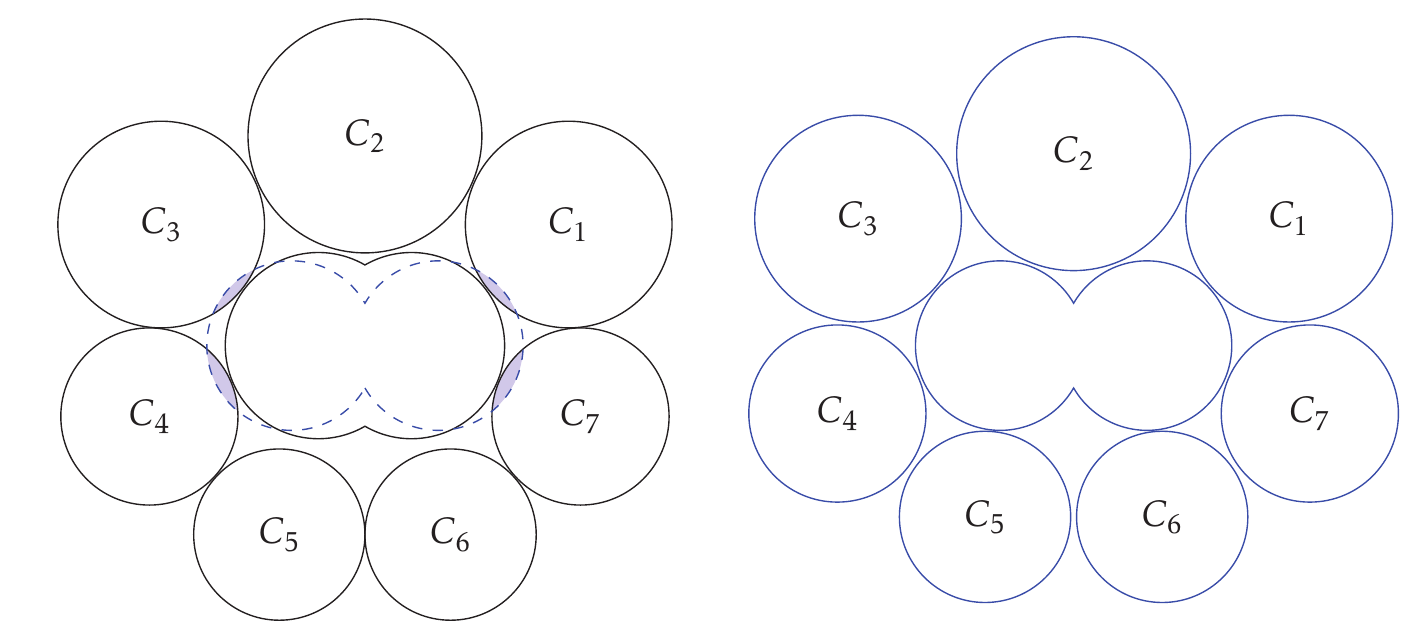}
%\vskip-3cm
\caption{Illustration of one substep of the division process.
We consider eight cells of different sizes with one (the central one) having begun its mitosis. 
Left image: the black line shows a typical configuration at the end of step $\tau_{k-1}$:
the cells are then all positioned without overlapping. The blue dashed line shows the $\tau_{k}$
step of the mitosis. This results in some overlap with some neighboring cells.
Right image: a typical configuration of the cells after the repositioning step. 
\label{fig4}}
\end{figure}

We now detail the numerical method used to solve our model.
%\subsection{General structure of the algorithm}
The general structure of the algorithm is the following
\begin{itemize}
 \item[a)] Initialization
 \item[b)] At each time step
 \begin{itemize}
 \item[i)] Growth step.
 \item[ii)] Test on size of the cell, cell by cell. If the threshold size is reached a division occurs.
 \item[iii)] For each mitosis up to the final division
 \begin{itemize}
  \item[1)] Partial division.
  \item[2)] Positioning step. 
  \item[3)] If necessary, depending on the chosen orientation strategy, orientation update.
 \end{itemize}
 \item[iv)] Positioning step.
 \end{itemize}
 \item[c)] Statistical quantifiers computation.
\end{itemize}
The computation of the statistical quantifiers is detailed in the next Section. 
\paragraph{Positioning step.}
We discuss now step $\textmd{(iv)}$. In order to find a solution to the minimization problem $\xi^*(t)=\textrm{argmin}_{\xi(t)\in \mathcal A(t)} W_Q(\xi(t))$, where $W_Q$ is the global adhesion potential relative to the quadratic choice of the potential function $V_Q$, we construct a method based on the Uzawa algorithm \cite{Uzawa_1958}. Given $N(t)$ cells, the number of constraint functions $\phi_{ij}(\xi(t),\rho(t))$ due to the non overlapping condition is $M=N(t)(N(t)-1)/2$. Then, the algorithm consists in finding a saddle point of the Lagrangian function $\mathcal L_Q (\xi(t),\lambda(t)): (\R^2)^{N(t)}\times \R^M\rightarrow \R$ defined by 
\begin{equation}
\begin{split}
&\mathcal L_Q (\xi(t),\lambda(t))=W_Q(\xi(t))\\
&+\sum_{1\leq i\leq j\leq N(t)}\lambda_{ij}(t)\phi_{ij}(\xi(t),\rho(t)), \ \forall (\xi(t),\lambda(t)),  
\end{split}
\end{equation}
where the $\lambda_{ij}$ are called the Lagrange multipliers. The algorithm constructs a sequence of approximate values $(\xi(t)^{(p)},\lambda(t)^{(p)})_p$ such that $\xi(t)^{(p)}\rightarrow \xi(t)^*$, when ${p\rightarrow\infty}$. Starting from an initial guess $(\xi(t)^{(0)},\lambda(t)^{(0)})$, the method reads as
\begin{equation*}
\begin{cases}
&\xi^{(p+1)} =X^{(p)}-\beta\nabla_x\mathcal L_Q\left(\xi^{(p)},\lambda^{(p)}\right), \\
&\phi^{(p+1)}_{ij} =\phi_{ij}\left(\xi^{(p+1)}\right), \ \forall \ i,j\in[1,N], \ i<j, \\
&\lambda^{(p+1)}_{ij} = \max\left(0,\lambda^{(p)}_{ij}+\mu \phi^{(p)}_{ij}\right), \forall \ i,j\in[1,N], \ i<j,\\
\end{cases}
\end{equation*}  
where $\beta$ and $\mu$ are numerical parameters and where the dependence on $t$ has been omitted for simplicity and will also be omitted in the sequel of this paragraph if not strictly necessary for comprehension. After some computations, the first equation of the above system can be rewritten for $k\in[1,N]$ as
$$X^{(p+1)}_k =(1-2\beta)X_k^{(p)}+2\beta\sum_{j=1}^{N}\lambda^{(p)}_{kj}\left(X_k^{(p)}-X_j^{(p)}\right),$$ which clarifies the role of the numerical parameter $\beta$ in the scheme, it is related to the displacement of the cells during the search of an equilibrium position. Two stopping criteria, which need to be satisfied at the same time, are used in order to advance to the next step. They are based on measuring the following quantities $$\varepsilon^{(p+1)}_\phi=\max_{\substack{1\leq k\leq N \\ 1\leq l\leq k-1}}(\phi_{lk}^{(p+1)}),$$ $$\varepsilon^{(p+1)}_W=\left|\frac{W_Q^{(p+1)}(\xi)-W_Q^{(p)}(\xi)}{W_Q^{(p)}(\xi)}\right|.$$ 
Then new equilibrium state is considered to be valid if $\varepsilon^{(p+1)}_\phi< \textmd{tol}_\phi$ and $\varepsilon^{(p+1)}_W<\textmd{tol}_W$ where $\textmd{tol}_\phi$ and $\textmd{tol}_W$ are two tolerances the values of which are given below. These criteria permit to control the largest overlapping permitted between the cells and to exit the algorithm when two consecutive values of the total mechanical energy of the system are very close to each other, indicating that a saddle point is likely to have been reached. Finally, the parameter $\mu$ is related to the speed at which the constraints are updated.\\ 
In order to reach a solution to the minimization problem as fast as possible, an adaptive $\beta$ has been chosen which depends on the number of cells considered. In practice, $\beta=3 \ 10^{-4}$ for $1\leq N\leq 100$, $\beta=3 \ 10^{-5}$ for $100\leq N\leq 300$ and $\beta=6 \ 10^{-6}$ for $300\leq N\leq 500$, while $\mu$ is kept fixed to $\mu=100$. This reflects the observation that the Lagrange multipliers values grow with the number of cells $N$. Consequently the value of $\beta$ should diminish when $N$ grows in order to avoid too large displacements of the cells which may lead to saddle points very far from the initial configuration and thus unrealistic. However, it may happen that when constraints are strongly violated, these choices for $\beta$ are not sufficient to prevent ejection of cells from the aggregate. This is measured by computing the distance traveled by a cell between two consecutive steps $(p)$ and $(p+1)$ of the minimization algorithm. If this distance goes beyond a fixed tolerance $\textmd{tol}_X$, this is repaired by repeating the positioning algorithm with a choice of $\beta$ which avoids too large displacements. Details on the value used for $\textmd{tol}_X$ are given below.  
\paragraph{Growth step.}
Step $\textmd{(i)}$ consists of the simple implementation of the growth law discussed in the previous Section. Given the parameters $R_{min}$, $R_{max}$, $\gamma$, $T_G$ and the time step $\Delta t$, we just sample a random number $u$ between $[-\alpha,\alpha]$ and we compute $R_i(t)=\left(R^3_{min}+(1+\gamma)\frac{R^3_{max}-R^3_{min}}{T_g}\Delta t\right)^{1/3}$. After the growth, in general, an overlapping between cells is produced which is resolved by the repositioning step described in Section 3.1. Fig. (\ref{fig3}) reports a typical repositioning computation. The left picture shows the situation before and after the growth, while the right picture show the equilibrium configuration after repositioning.
\paragraph{Division step.}
We assume that the cell $C_{i_0}$ is ready to start the division, i.e. $R_{i_0}(t)\geq R_{\max}$. For each simulation, we fix the number of steps of the division process $k=[1,f]$, the initial direction $\omega_{i_0}(t)$ of the division plane and the division strategy. As soon as the cell begins its division, the cell $C_{i_0}$ is replaced by two new cells. The algorithm can be summarized as follows. For each $\tau_k,\ k\in[1,f]$, compute $\tilde h(\tau_k)=R_{i_0}(t)-k\frac{R_{i_0}(t)}{f}$, the radii $R^\pm(\tau_k)$, the new positions $(x^\pm(\tau_{k+1}),y^\pm(\tau_{k+1}))$, where the variable $t$ is fixed during all the division process. This series of actions causes the cells to partially overlap with their neighbors. This is corrected by a new application of the positioning algorithm. Fig. \ref{fig4} shows a typical situation which arises during the division process: partial division, overlap and repositioning.\\ 
The additional constraints imposed by the mitosis are, for both free and fixed orientation strategies, the change in the non overlapping constraint between the cells is changed into an equality constraint between the two daughter cells (indexed by $i^+,i^-$) which, for the iteration $(p)$ of the minimization algorithm relative to the generic division step $\tau_k$, reads as follows: $$\phi_{i^+i^-}^{(p)}=4(R_{i_0}-\tilde h)^2-(x^{(p)}_{i^+}-x^{(p)}_{i^-})^2-(y^{(p)}_{i^+}-y^{(p)}_{i^-})^2,$$
while the corresponding Lagrange multiplier is updated accordingly to $\lambda_{i^+i^-}^{(p)}=\lambda_{i^+i^-}^{(p-1)}+\mu\phi_{i^+i^-}^{(p)}$ and where the dependences on $\tau_k$ and $t$ have been omitted for simplicity. In the constrained strategy case, two additional constraints should be added to the positioning algorithm to take into account that $\omega^\pm(\tau_k)$ remains constant equal to $\omega_{i_0}(t)$ for all $k\in[1,f]$. They read
$$\phi_{1}^{(p)}=(x^{(p)}_{i^+}-x^{(p)}_{i^-})\sin(\omega^\pm)-(y^{(p)}_{i^+}-y^{(p)}_{i^-})\cos(\omega^\pm)$$
$$\phi_{2}^{(p)}=-(x^{(p)}_{i^+}-x^{(p)}_{i^-})\sin(\omega^\pm)+(y^{(p)}_{i^+}-y^{(p)}_{i^-})\cos(\omega^\pm),$$
while the new positions of the two daughters cells take into account these constraints through the corresponding Lagrange multipliers $\lambda_1^{(p)}$ and $\lambda_2^{(p)}$ as follows:
\begin{equation*}
\begin{cases}
&\tilde x_{i^\pm}^{(p+1)} =x_{i^\pm}^{(p+1)}\mp\beta(\lambda_1\mp \lambda_2)\sin(\omega^{\pm}), \\
&\tilde y_{i^\pm}^{(p+1)} =y_{i^\pm}^{(p+1)}\pm\beta(\lambda_1\pm \lambda_2)\cos(\omega^{\pm}), \\
\end{cases}
\end{equation*}  
where $x_{i^\pm}^{(p+1)}$ and $y_{i^\pm}^{(p+1)}$ are the positions computed during the iteration $(p+1)$ of the positioning algorithm without divisions and where once again dependence on $\tau_k$ and $t$ are omitted.

\paragraph{Initialization and numerical parameters.}
The initialization is done by inserting $N_0$ cells in the computational domain with radius $R_{\min}$, at a random location, each cell defining a different lineage $ell_i, \ i=[1,N_0]$. Then, a positioning step finds a first saddle point of the Lagrangian function $ \mathcal L_Q (\xi(t=0),\lambda(t=0))$ which furnishes the initial positions of the cells in the tumor. \\ We list now all numerical values given to the parameters. We distinguish the model parameters listed in Table 1 from the numerical parameters listed in Table 2. In particular, considering Table 1, the choice of $N_0$ represents the effective number of initial cells used in average in the experiments. In the same spirit, $T_G$ and $T_{\max}$ are also chosen to be as close as possible to experimental values. The HCT116 line used in the experiments has a cycle of an average duration of 24 hours and they are typically tracked up to three generation times. Then cell cycle can vary from 18 hours to 30 hours which justifies the choice of $\gamma$. The value $R_{\min}$ sets a reference value. This choice fixes consequently $R_{\max}$. Finally, the value $f$ (number of steps during the division process) is chosen to avoid too large cell overlapping during the division process. Concerning the numerical parameters, the time step is chosen to guarantee small enough cell size increments which avoids too large cell overlapping before repositioning. The values assigned to $\beta$ have been already discussed in Section 3.2 together with $\mu$. The value chosen for $\textrm{tol}_W$ is directly related to the values of $\beta$. The tolerance $\textrm{tol}_\phi$ is chosen to permit only very small overlaps (of the order of $5 \times 10^{-2}$ for values of the radius $R$ between $R_{\min}$ and $R_{\max}$). The value of $\textrm{tol}_X$ detects too large displacements of cells (of the order of twice the radius $R_{\min}$).
\begin{table}
\centering
\begin{tabular}{lll}
\hline
 Parameter&  Value&  Meaning  \\
 \hline
 $N_0$& 50 & Number of initial cells     \\
 \hline
 $T_G$&  24&  Mean duration of the cell cycle    \\
\hline
$ T_{\max}$&  72&  Duration of the simulations\\   
 \hline
  $\alpha$&  0.25&  Support of the uniform distribution    \\
\hline
$ R_{\min}$& 1 & Minimal radius of a cell     \\
\hline
$ R_{\max}$&  $\sqrt[3]{2}$&  Larger value assumed by a cell    \\
\hline
$ f$&  8&  Number of discrete mitosis steps     \\
\hline
\end{tabular}
\caption{Model parameters
\label{table1}}
 \end{table}
 
\begin{table}
\centering
\begin{tabular}{lll}
\hline
 Parameter&  Value&  Meaning  \\
 \hline
 $\Delta t$& 0.25 & Time step     \\
 \hline
 $\beta$&  $\mathcal{O}(\frac{1}{N^2})$& Cell displacement rate   \\
\hline
$ \mu$&  100& Lagrange multipliers change rate \\   
 \hline
  $\textmd{tol}_W$&  0.25& Energy minimum tolerance    \\
\hline
$\textmd{tol}_\phi$& $\mathcal{O}(\frac{\rho}{10})$ & Overlapping tolerance     \\
\hline
$ \textmd{tol}_\xi$&  2&  Cell displacement tolerance   \\
\hline
\end{tabular}
\caption{Numerical parameters
\label{table2}}
 \end{table}

\subsection{Statistical indicators}
\label{sec:2.4}
\begin{figure}
%\vskip-.6cm
\hspace*{1.5cm}
\includegraphics[scale=0.62,trim={-1cm 0 0 0}]{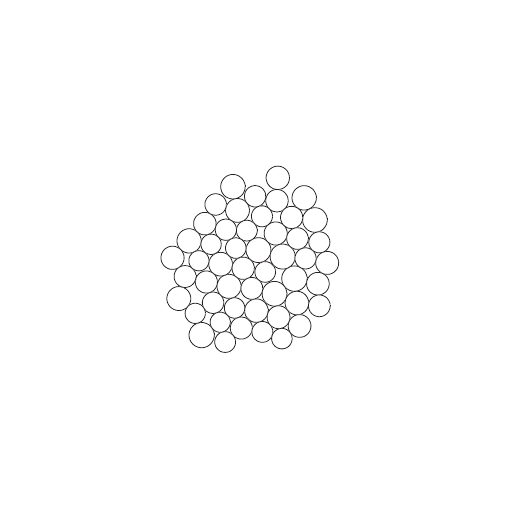}
\includegraphics[scale=0.62,trim={-1cm 0 0 0}]{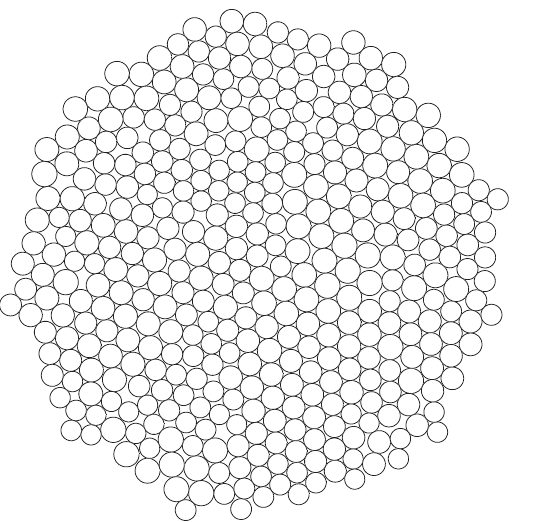}
%\vskip-3cm
\caption{Typical result of a simulation without the Vorono\"{\i} diagram. Left: initial step. Right: final step.
\label{fig5}}
\end{figure}

\begin{figure}
%\vskip-0.4cm
\hspace*{1.5cm}
\includegraphics[scale=0.62,trim={-1cm 0 0 0}]{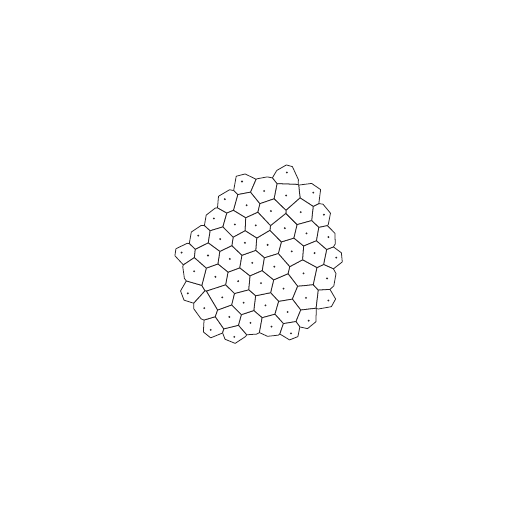}
\includegraphics[scale=0.62,trim={-1cm 0 0 0}]{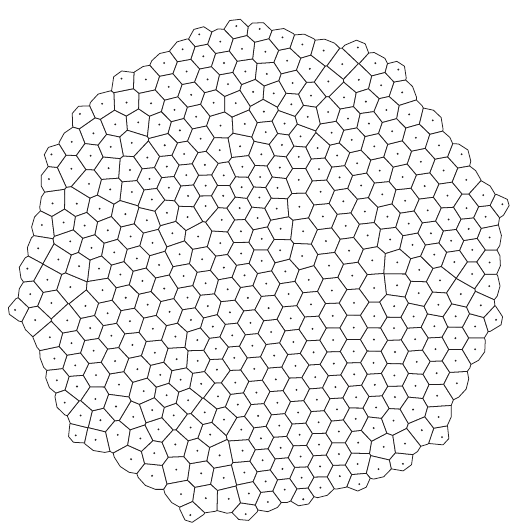}
%\vskip-3cm
\caption{Representation of the solution by the Vorono\"{\i} diagram. Left: initial step. Right: final step.
The Figures report the same simulation result as Fig. (\ref{fig5}). Now, the definitions of area, perimeter and neighborhood of cells or groups of cells are easier.
\label{fig6}}
\end{figure}

\begin{figure}
%\vskip-0.4cm
\hspace*{2.cm}
\includegraphics[scale=0.72,trim={-3cm 0 0 0}]{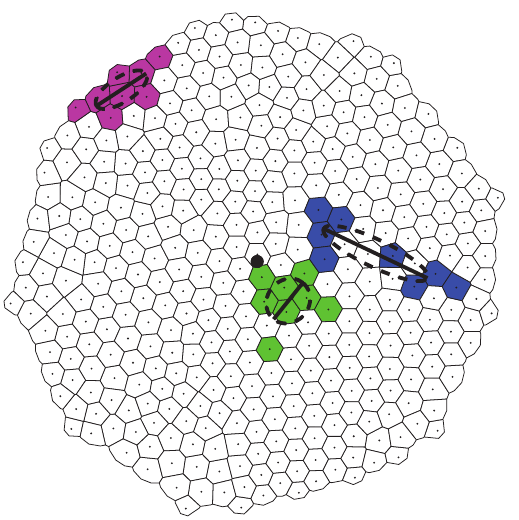}
%\vskip-3cm
\caption{Ellipses of inertia representing the inertia matrices for three different lineages (dotted lines). Major axes of inertia for the three different lineages (continuous line). 
\label{fig7}}
\end{figure}

To get insight into both the experimental and the numerical results, we develop several indicators to measure the characteristics and the morphologies of the single lineages or of the whole population. 
\paragraph{The Vorono\"{\i} diagram.}
In Fig. \ref{fig5} is reported a typical result of a simulation. The left picture shows the initialization after the first positioning phase, with $N_0=50$ cells, while the right picture shows the solution at $T_{\max}$ which corresponds to a situation with about $N=400$ cells. This representation of the solution has several limitations due to the difficulty in defining the notions of cell neighborhood, perimeter and area. In order to overcome this problem we use a modified Vorno\"{\i} diagram representation. This approach is frequently used in the context of growing cell populations, see for instance \cite{Osborne} where this method is used to determine the interaction forces between two cells.
We recall some basics about the Vorono\"{\i} diagram in the present setting. A Vorono\"{\i} site is defined as a point $p_i$ belonging to a predefined subset $S = \{p_i | 1 \leq i \leq n\}$ of $\R^2$. A Vorono\"{\i} region relative to the site $p_i\in S$ is a subset $\mathcal{R}_i$ of $\R^2$ such that $\mathcal{R}_i=\{x\in\R^2|\forall j\ne i, \textmd{d}(x,p_i)\leq \textmd{d}(x,p_j)\}$, where $ \textmd{d}(\cdot,\cdot)$ denotes the Euclidean distance on $\R^2$. A Vorono\"{\i} diagram for $S$ is the set of regions $\mathcal{R}_i$ for $p_i \in S$, i.e. $\cup_{i=1}^{n}[\mathcal{R}_i]$. Thus, given a set of sites $S$, the Vorono\"{\i} diagram partitions the plane by which site is the closest. Two sites $p_i$ and $p_j$ are considered as neighbors if $\mathcal{R}_i$ and $\mathcal{R}_j$ share a common edge. The intersection of three regions, if not empty, is called a Vorono\"{\i} vertex or node. \\
In order to use this representation in our model, we define the Vorono\"{\i} sites as being the cell centers, i.e. $S=\{X_i | 1 \leq i \leq N\}$. But this leads to a problem at the boundary of the cellular aggregate, since the tumor does not occupy the entire plane. The Vorono\"{\i} regions corresponding to these outer sites are consequently unbounded. To overcome this problem, we add fictitious sites on the borders of the population. These additional sites are located on the polygon whose boundaries are the edges of the convex hull of the tumor slightly enlarged by dilation. More precisely, given a small $\delta>0$ and $\mathcal{P}$ the polygon obtained by enlarging the boundaries of the convex hull by an increment $\delta$, we place $n$ additional sites equally spaced along the segment of $\mathcal{P}$, where $n$ corresponds to the number of outer sites. The result of this procedure leads to Fig. \ref{fig6} where the Vorono\"{\i} diagram has been traced for the same situation as in Fig. \ref{fig5}. In this Figure the Vorono\"{\i} regions related to the fictitious sites are not represented. We will use the same Vorono\"{\i} approach for the experimental data to have similar definitions of neighborhoods, areas and perimeters in the experimental and numerical results.
\paragraph{Diagnostic definitions.}
One of the main questions we address is the influence of the orientation of the division plane during mitosis on the morphology of the lineages. We consider all descendants of a given ancestor cell present at the initial time and store this information in the lineage $\ell_i, \ i=[1,N_0]$. Thus, if cell $C_i$ is present at the initial time then $\ell_i = i$, otherwise when a cell is created its $\ell_i$ becomes the lineage of its mother cell. To define diagnostics, we use several concepts of graph theory. We briefly recall them. A graph $G$ is an ordered pair comprising a set of points or vertices and a set of edges or links where an edge is related with two vertices and two vertices of the graph may or may not be connected by a link. A chain in $G$ is a finite sequence of vertices connected by links with possible repetitions. A cycle is a closed chain. We are now ready to introduce some definitions:\\
\textbf{Neighboring cells.} Two cells $C_i$ and $C_j$ are said to be neighbors if their Vorono\"{\i} regions $R_i$ and $R_j$ share a common edge.\\
\textbf{Graph of N cells.} The set of all cells is a graph where the vertices are the cells centers.\\
\textbf{Connected set of cells.} The cells $C_{i_1}, C_{i_1},..,C_{i_p}$ are said to be connected when the sub-graph induced by these cells is a connected graph.\\
\textbf{Connected component of a lineage.} A connected component of a lineage is a connected component of the subgraph generated by the cells of the same lineage. \\
\textbf{Cell cycle.} A cell cycle is a finite sequence of cells corresponding to a cycle in the graph. In other words, for all $k\in[1,p+1]$, $C_{i_k}$ and $C_{i_k+1}$ are neighbors and $C_{i_{p+1}} = C_{i_1}$.\\
\textbf{Cell polygon associated with a cell cycle.} Let $p\geq 3$ and a cell cycle $C_{i_1}, C_{i_1},..,C_{i_{p+1}}$. The line passing from the centers of the cells $X_{i_1}, X_{i_2},.., X_{i_{p+1}}$ forms a polygon which is called the cell polygon associated to $C_{i_1}, C_{i_1},..,C_{i_{p+1}}$.\\ 
\textbf{Boundaries of a set of connected cells.} Given a set of connected cells, the boundary is defined as a cell cycle whose associated cell polygon contains all cells of the connected set. This polygon is called the boundary polygon.\\ 
\textbf{Perimeter and area of ​​a set of connected cells.} Let us consider a set of connected cells with $p\geq 1$ and its boundary denoted by $C_{i_1},C_{i_1},..,C_{i_{p+1}}$. The perimeter of this array of cells is defined as the perimeter of the cell polygon associated to $C_{i_1}, C_{i_1},..,C_{i_{p+1}}$. It is given by $\mathcal{P}=\sum_{k=1}^p\textmd{d}(C_{i_k},C_{i_{k+1}})$. The area of ​​this set is defined as the area of ​​the cell polygon. It is given by $\mathcal{A}=\frac{1}{2}\left|\sum_{k=1}^p(x_{i_{k}}y_{i_{k+1}}-x_{i_{k+1}}y_{i_{k}})\right|$.\\
\textbf{Convex hull of a set of cells.} The convex hull a set of cells is defined as the convex hull of the point cloud consisting of all the cell centers.\\
We are now ready to define the statistical indicators used for studying the morphology of the cellular aggregate.\\
\textbf{(1) Sphericity of the population:} ratio between the area and the perimeter squared of the entire cell population:
$R_1=\frac{4\pi\mathcal{A}}{\mathcal{P}^2}$ where $R_1\in[0,1]$. $R$ is one when the boundary of the population is a perfect circle.\\
\textbf{(2) Convexity of the population:} ratio between the area $\mathcal{A}$ of all cells and $\mathcal{A}_{\textmd{conv}}$, the area of their convex hull: $R_2=\frac{\mathcal{A}}{\mathcal{A}_{\textmd{conv}}}, \ R_2\in[0,1]$. Its value is one when the boundary of the population coincides with the boundary of its convex hull. \\
\textbf{(3) Sphericity of a lineage:} for each connected component (of at least two cells) of a lineage, ratio between the area and the perimeter squared of this connected component: $R_3=\frac{4\pi\mathcal{A}}{\mathcal{P}^2}$ where now $\mathcal{P}$ and $\mathcal{A}$ are respectively the perimeter and area of the boundary polygon of the considered connected component.\\
\textbf{(4) Lineage fragmentation:} number of connected components in which a lineage is split: $R_4$. It permits to understand whether the cells of the same lineage tend to remain grouped or to scatter.\\
\textbf{(5) Size of the fragments of a lineage:} Ssze of a connected component of a lineage. It counts the number of cells in a connected component of a given lineage ($R_5$).\\
\textbf{(6) Lineage orientation.} Orientation direction for a given lineage $R_6$. The main direction of orientation is computed by using the inertia matrix of the cells composing the lineage.\\
\textbf{Inertia Matrix.} Let $p\geq 2$ and for $k\in[1,p]$,  $X_{i_{k}}=(x_{i_{k}}, y_{i_{k}})$. Denoting $X_G = (x_G, y_G)$ the barycenter of $X_{i_{k}}:$ $X_G=\sum_{k=1}^p\frac{X_{i_{k}}}{p}$, the inertia matrix of the cloud is defined by $$\mathcal{E}=\frac{1}{p}\sum_{k=1}^p(X_{i_{k}}-X_G)(X_{i_{k}}-X_G)^T.$$
Assuming that the $X_{i_{k}}$ are not aligned, this matrix is ​​symmetric and positive definite with strictly positive eigenvalues. Thanks to these values we can measure the angle $\theta_2$ formed by the semi-major axis of the inertia ellipse and a reference direction, and the angle $\theta_0$ formed by the line joining the origin to the barycenter of the lineage this a reference direction. The angle $\theta_2$ is the angle which measures the direction of the eigenvector $v_2$ associated to the larger eigenvalue $\lambda_2$ . We then measure the quantity $R_6=\theta_2-\theta_0-\frac{\pi}{2}$ where the shifting of $\frac{\pi}{2}$ is done in order to have an angle always between $-\frac{\pi}{2}$ and $\frac{\pi}{2}$. In Fig. \ref{fig7} is reported an example where the main direction of the ellipse (continuous line) representing the inertial matrix together with the ellipse (dotted line) are shown for three different lineages. \\
% \begin{figure}
% \vskip-0.4cm
% \hspace*{1cm}
% %\includegraphics[scale=0.52,trim={-1cm 0 0 0}]{Voronoi2.pdf}
% \includegraphics[scale=0.82,trim={-0cm 0 0 0}]{Lineage1.pdf}
% \includegraphics[scale=0.82,trim={-0.5cm 0 0 0}]{Lineage2.pdf}
% %\vskip-3cm
% \caption{Example of lineage tracking. Left: the initial step. Right: the final step. The solution is represented by means of the Vorono\"{\i} diagram.
% \label{fig6bis}}
% \end{figure}
\subsection{Experimental protocol}
\label{sec:2.5}
In order to generate reference experimental data to which mathematical modeling can be confronted, we set up the following protocol. The experiments are performed on cells of the colon adenocarcinoma HCT116 cell line, modified by lentiviral transduction to express a histone $H_2B$ fused with the mCherry fluorescent protein. This allows visualizing by fluorescence microscopy the nuclei of the cells. The cells are seeded in culture chambers (Lab-Tek, Dutscher) at a density of 7500 cells / cm$^2$ in an OPTIMEM medium supplemented with $3\%$ of fetal calf serum (FCS) and penicillin / streptomycin. After 48 hours, the chosen cell density provides in the bottom of the chamber a culture of islets composed of about 50 cells, isolated from one another. This allows following the individual cell evolution in real time by an inverted fluorescence microscope. Before making microscopy acquisitions, isolated groups of cells are selected and the bottom of the culture chamber manually processed, to prevent neighboring cells to join the main group. Acquisitions are performed on an Axiovert microscope (Zeiss) fitted with a CoolSnap HQ camera (Roper Scientific) and piloted by the MetaView software. For every acquisition, several individual groups are followed in parallel by videomicroscopy (1 image every 10 minutes). The images are processed using Metamorph and Image J before being analyzed. 

\begin{figure}%[h!]
%\vskip-.2cm
\hspace*{1cm}
\includegraphics[width=90mm,trim={0cm 0 0 0}]{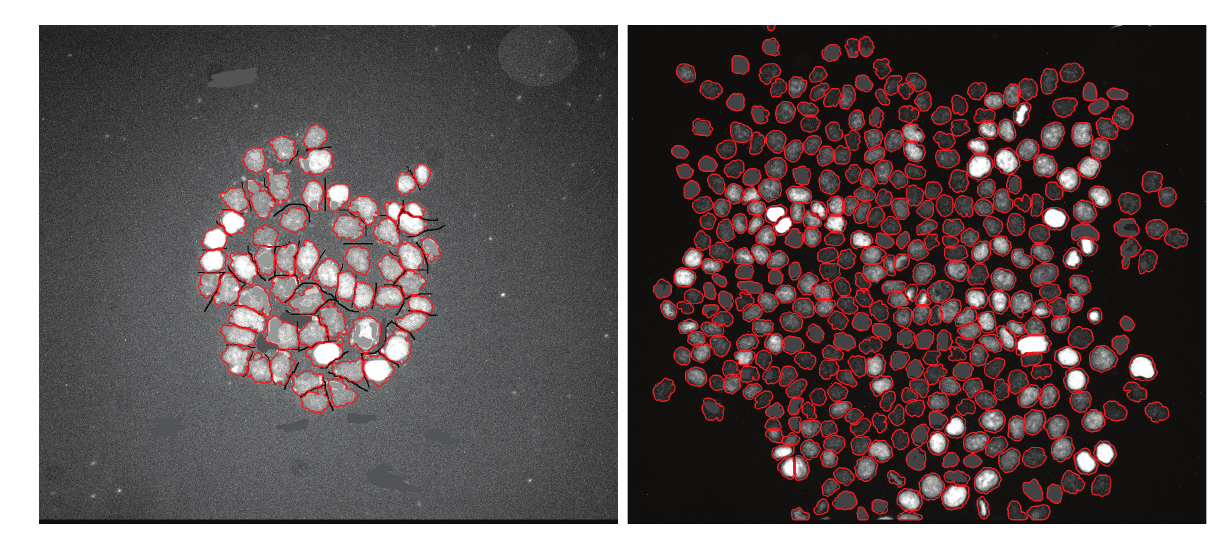}
%\vskip-3cm
\caption{Example of segmentation. Left: initial tracking time. Right: final tracking time. The figure reports the nuclei of the cells. The black lines and the gray zones correspond to corrections done after the automatic segmentation procedure. Same experiment as in Fig. \ref{fig1} (a) and (b).
\label{fig19}}
\end{figure}

The data post-processing consists of the following steps repeated for each single cell culture. Cells are selected at the initial time and the filiation tracked manually through direct labeling. Fig. \ref{fig1} (a) and (b) shows an example of this procedure. Segmentation is first performed automatically thanks to the level of fluorescence, then checked and, if necessary, results are corrected by post-processing the data manually. This is possible thanks to the fact that the analysis is made on a relatively small number of cells. A segmentation result is shown in Fig. \ref{fig19}.
\begin{figure}%[h!]
%\vskip-.4cm
\hspace*{1cm}
\includegraphics[width=90mm,trim={0cm 0 0 0}]{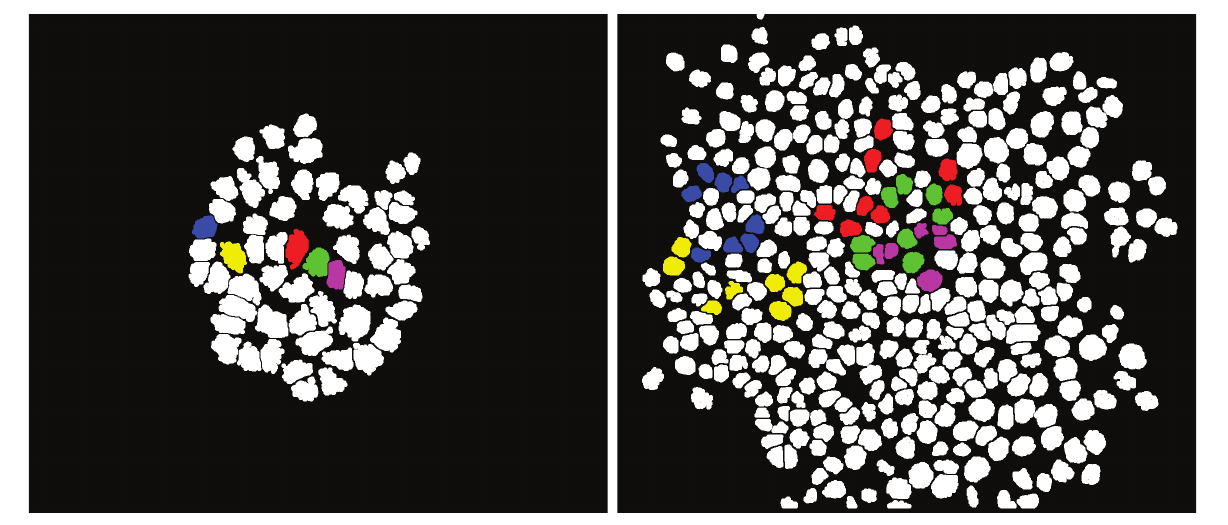}
%\vskip-3cm
\caption{Representation of the experimental data after segmentation. Left: initial tracking time. Right: final tracking time. The same experiment as for Figs. \ref{fig1} (a) and (b) and \ref{fig19} is considered. The pixels corresponding to nuclei are in white while the others are in black. The nuclei corresponding to a lineage which will be tracked are colored. 
\label{fig20}}
\end{figure}
A post-treatment is performed after the segmentation procedure. It consists of identifying each single lineage. The result is shown in Fig. \ref{fig20} where the same experiment as in Figs. \ref{fig1} (a) and (b) and \ref{fig19} is considered.\\
We describe now how the experimental data are processed. The areas detected during segmentation correspond to the nuclei of the cells. For each of these areas, the coordinates of the center of mass and the equivalent circular diameter, i.e. the diameter of the disc that has the same surface as the area of interest, is computed. This permits to have a first representation of the data by discs (Fig. \ref{fig21}), similar to that used in the simulations. However, this gives large gaps between some disks and overlapping between others and it does not correspond to the observed experiments. Indeed, this is only an artifact of the representation because cells are stuck together and do not overlap. We then choose to represent the population by the same adapted Vorono\"{\i} diagram as that used for the simulations. This result is reported in Fig. \ref{fig23}. %Such representation of the experimental data is frequently used in literature (see \cite{SM_05} and \cite{Honda} for examples).
\begin{figure}%[h!]
%\vskip-.4cm
\hspace*{1cm}
\includegraphics[width=90mm,trim={0cm 0 0 0}]{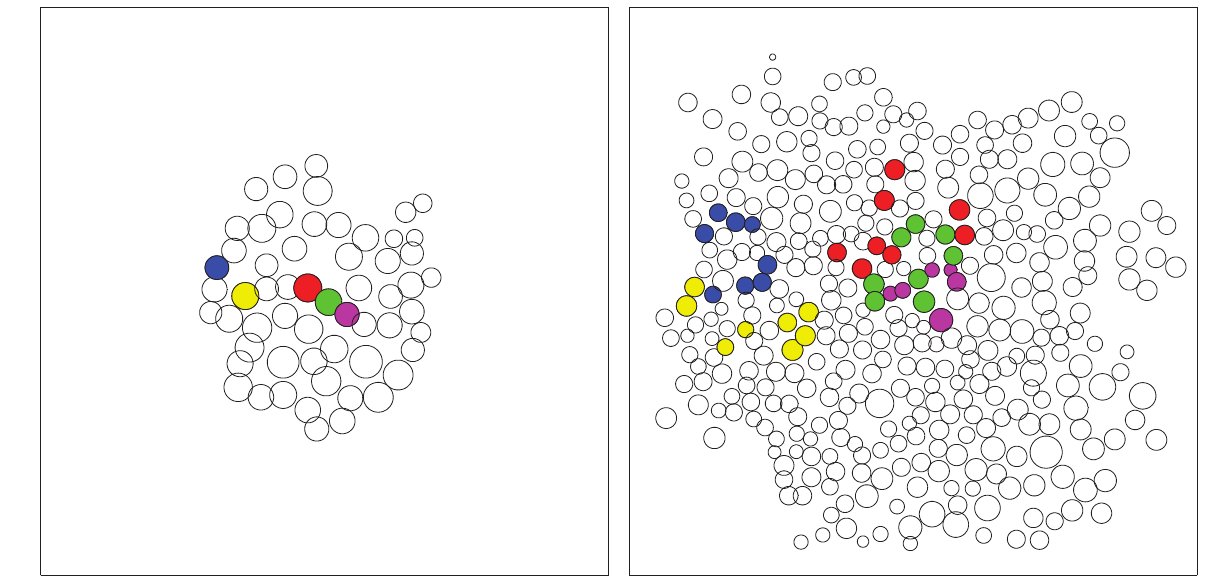}
%\vskip-3cm
\caption{Representation of cells by discs. Left: initial tracking time. Right: final tracking time. The colored discs correspond to the cells whose lineage is tracked. The same experiment as for Fig. \ref{fig1} (a) and (b) and \ref{fig19} is reported.
\label{fig21}}
\end{figure}
\begin{figure}%[h!]
%\vskip-.4cm
\hspace*{1cm}
\includegraphics[width=90mm,trim={0cm 0 0 0}]{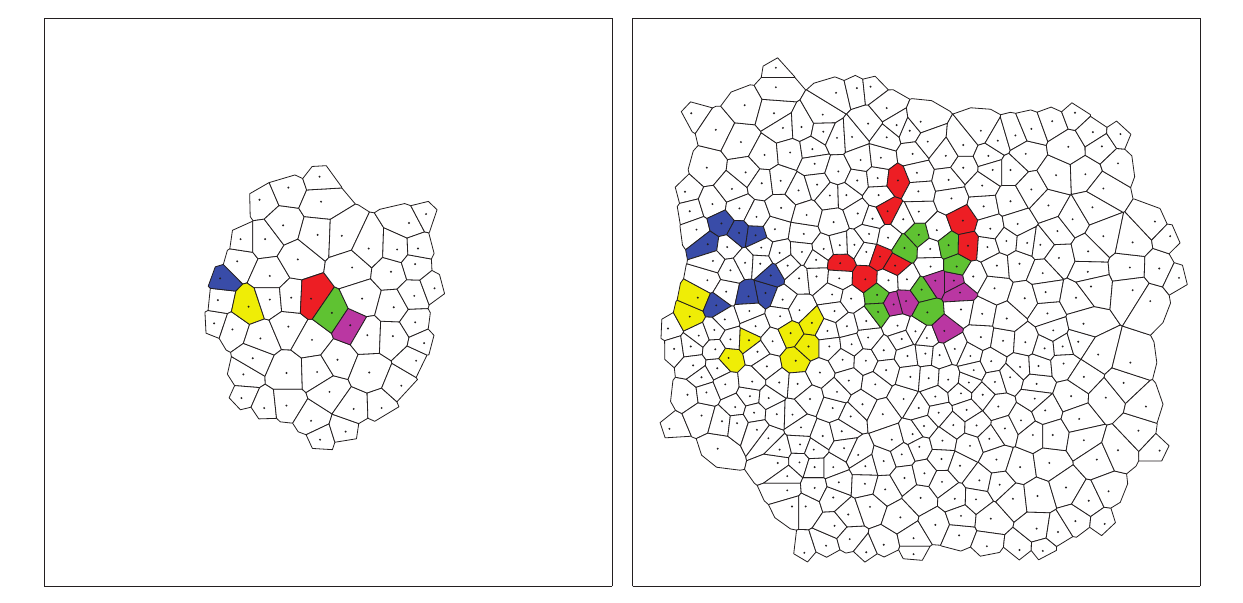}
%\vskip-3cm
\caption{Representation of cells by the Vorono\"{\i} diagram. Left: initial tracking time. Right: final tracking time. The colored Vorono\"{\i} regions correspond to the cells whose lineage is tracked. The same experiment as in Fig. \ref{fig1} (a) and (b) and \ref{fig19} is reported.
\label{fig23}}
\end{figure}\\
The lineages used for the comparisons with the simulations are either at the periphery of the aggregate or in the central region. A lineage is considered as peripheral if, at initial time, the progenitor cell is situated at the periphery of the aggregate, while it is considered as central if, at initial time, there are at least two cells between the progenitor cell and the boundary of the tumor. The data at our disposal are based on fourteen cellular cultures. This corresponds directly to the size of the samples which has been used in the first and the second diagnostic. In these fourteen experiments we have followed twenty-two central and fifty-four peripheral lineages which corresponds to the size of samples used in diagnostics $4$ and $6$. The twenty-two central lineages split up into fifty-three connected components, while fifty-four peripheral lineages divided into one hundred and fifteen connected components which correspond to the size of samples for diagnostics $3$ and $5$.

% \subsection{Quantitative analysis}
% \label{sec:3.1}
% We propose different indicators to compare the results of the simulations and the outcomes of the experiments. These indicators measure both the geometrical characteristics of the entire population and those of a single lineage. These measures are repeated a large number of simulations or experiments and the corresponding results are averaged. The indicators are the following: \textit{1)} Circularity of the population $R_1$: ratio between the area and the perimeter squared of the entire cell population. \textit{2)} Convexity of the population $R_2$: ratio between the area of all cells and the area of their convex hull. \textit{3)} Circularity of a lineage $R_3$: for each connected component (at least two cells) of a given lineage, ratio between the area and the perimeter squared of the connected sets. \textit{4)} Fragmentation of a lineage $R_4$: for each lineage, number of connected components in which a lineage is divided. \textit{5)} Size of the fragments of a lineage $R_5$: for each connected component of each lineage, size of the component. \textit{6)} Orientation of a lineage $R_6$: for each lineage, main direction of orientation through the computation of the eigenvectors of the inertia matrix (See SI for definition and details).

\section{Simulation results and comparisons with the experimental data}
\label{sec:3}

\subsection{Identification of the different morphologies} 
\label{sec:3.1}
Since we have chosen three possible orientation planes and two division strategies, for each of the six possible situations we perform $100$ numerical simulations and we compute the values of the statistical indicators $R_1$ up to $R_6$. Concerning the study of the lineages, we extract two lineages, one in a central position and one on the boundaries of the cell population for which the different diagnostics are computed. This choice of the different position of the lineages is done in order to highlight the strong disparity between the behaviors of internal and external cells. This disparity which is one of the major results of this work is also confirmed by in vitro experiments as detailed next and not known before. The difference between the central and peripheral lineages is an information which is extrapolated from diagnostics $R_3$ up to $R_6$. In the appendix we report the detailed results of the simulations. 
In this part, we summarize the principal results. \textit{1)} \textit{Presence of larger numbers of lineage fragments at the periphery, typically of smaller sizes.} This can be explained by the fact that, as cells are farther from the center, their contribution
to the total energy of the system grows. Thus, when they are pushed from the interior of the cellular aggregate because of the growth, they preferably move in tangential direction not to excessively increase the energy of the system. This result in an intercalation of new cells coming from the interior between originally neighboring cells. \textit{2) Preferred tangential orientation of the lineages at the periphery.} An interpretation of this phenomenon is that far from the center, the potential energy obliges cells to place themselves on an equipotential energy curve, i.e. along a single shell, all other equilibrium states being unstable. \textit{3) In the central zone, except for the constrained radial division strategy, there are no privileged directions for the lineages.} This suggests that in the central area, the division rule, when radial, plays a more important role than the positioning rule, this feature being opposite for the other division rules. \textit{4) In general all strategies which leave freedom to the division plane orientation to change during division have a very small impact on the overall population shape as well as on the shape of a single lineage.} On the contrary, the constrained strategy has much more influence on the shape of the entire population and of single lineages. \textit{5) Concerning the global population, only the constrained strategy permits to detect differences between the radial and tangential directions of the division plane.} The boundaries of the growing cell population are smoother for this second choice. \textit{6) With the constrained radial orientation strategy for the division plane, the lineages in the center of the population are radially oriented with few fragments of large sizes.}

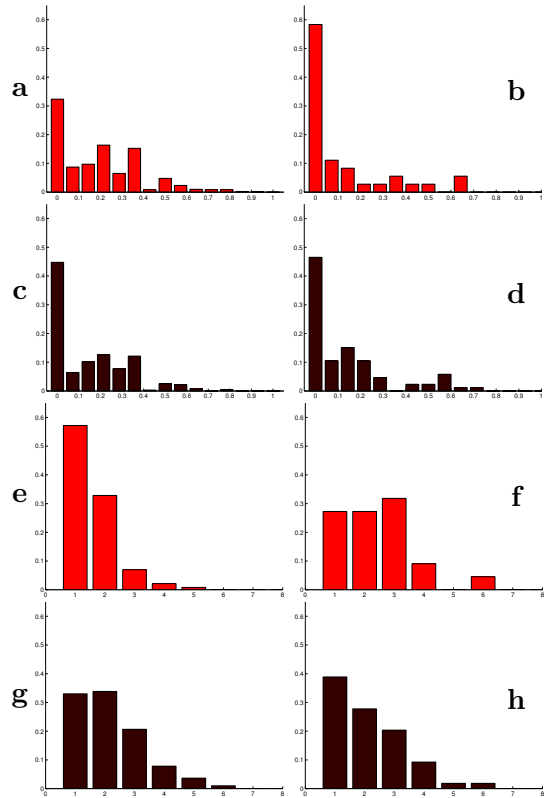
\begin{figure}
\centering
\includegraphics[scale=0.2]{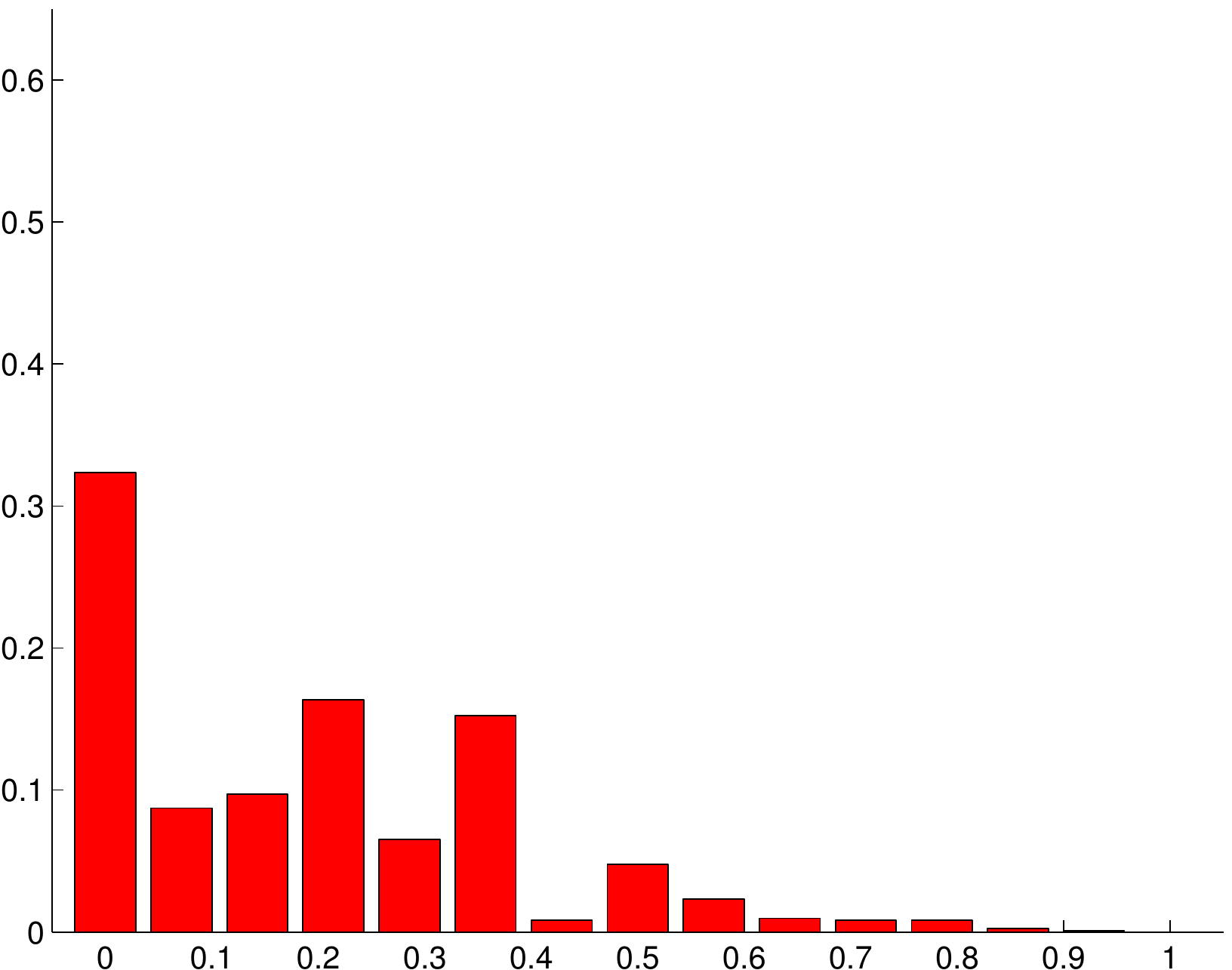}
\includegraphics[scale=0.2]{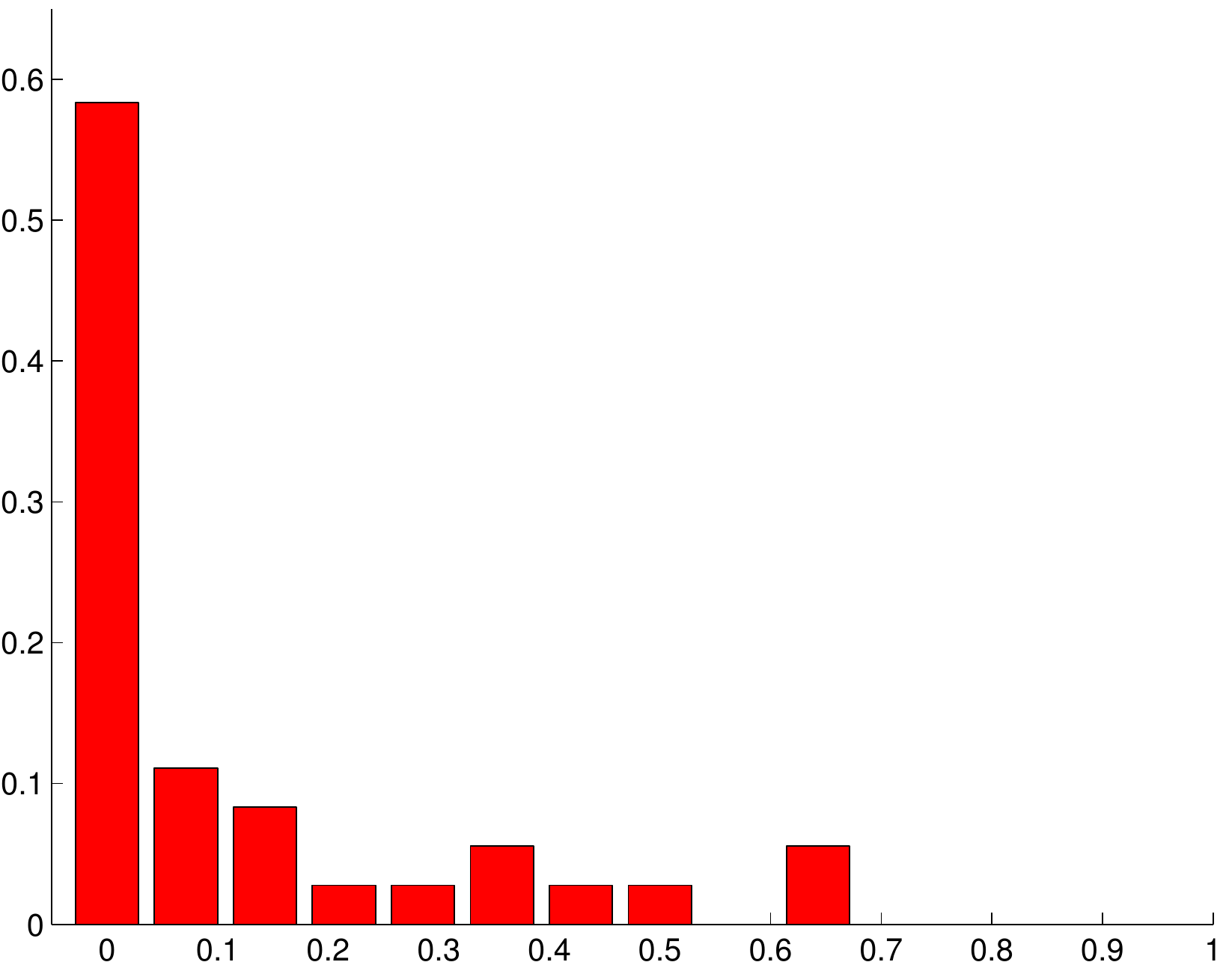}\\
\includegraphics[scale=0.2]{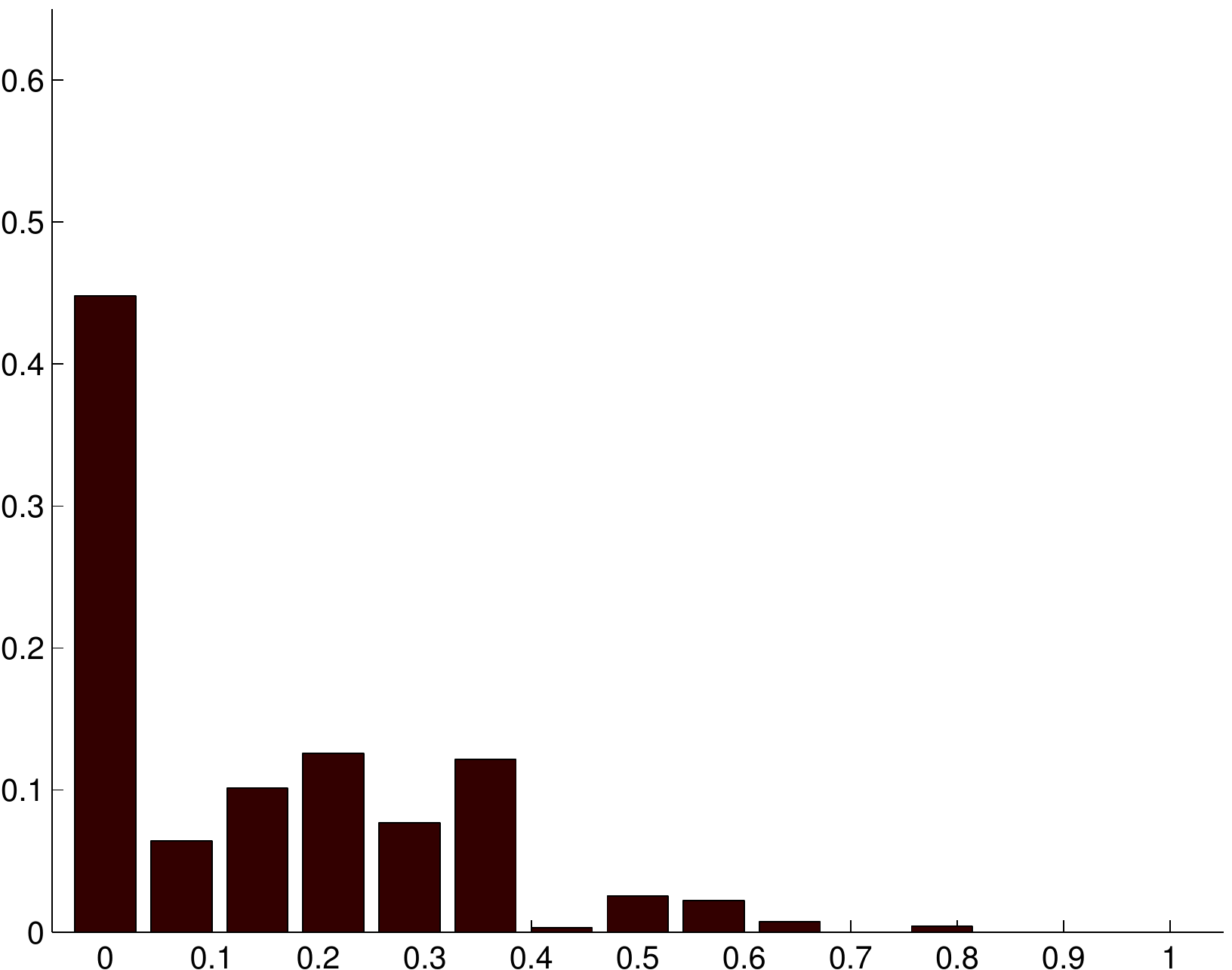}
\includegraphics[scale=0.2]{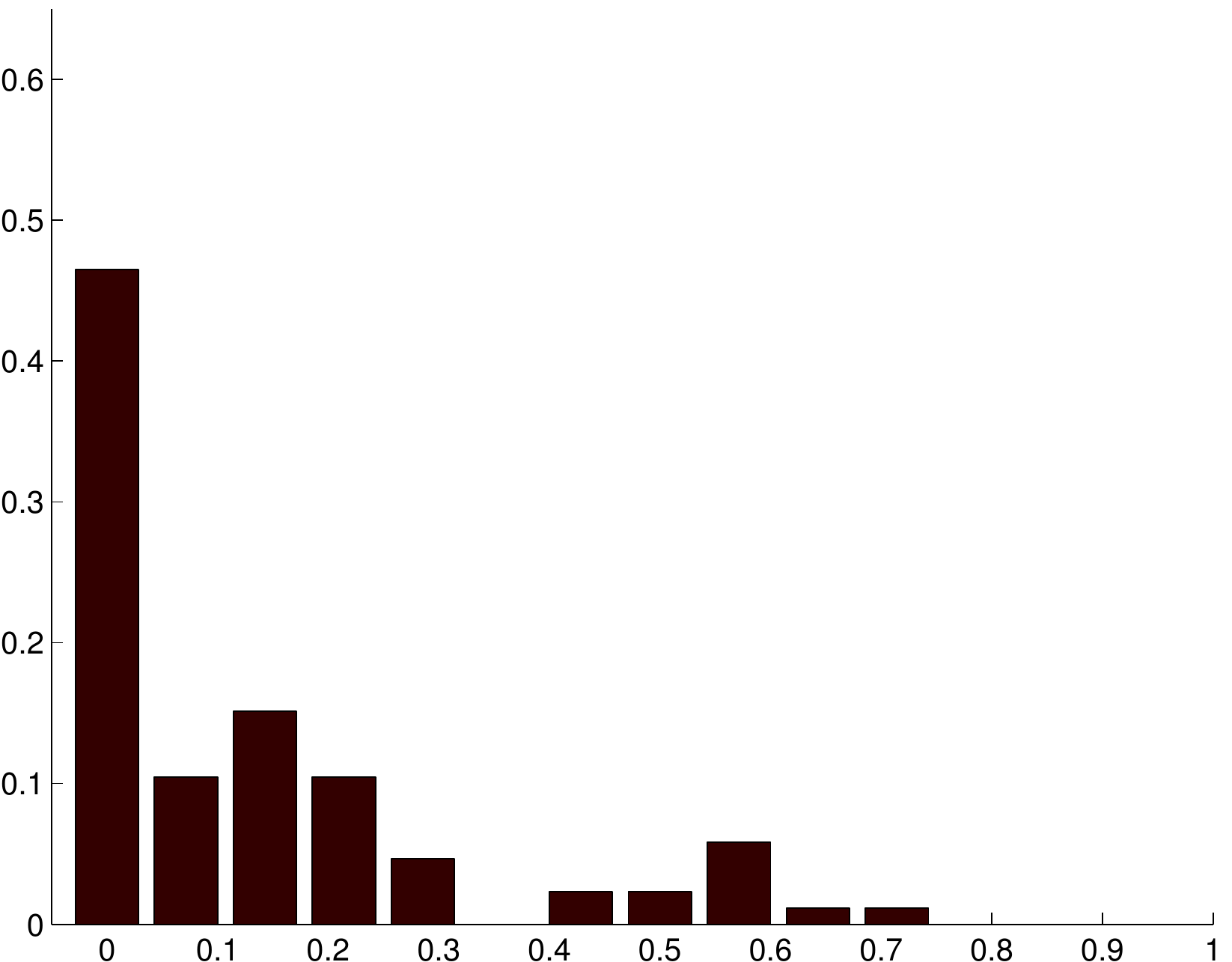}\\
\includegraphics[scale=0.2]{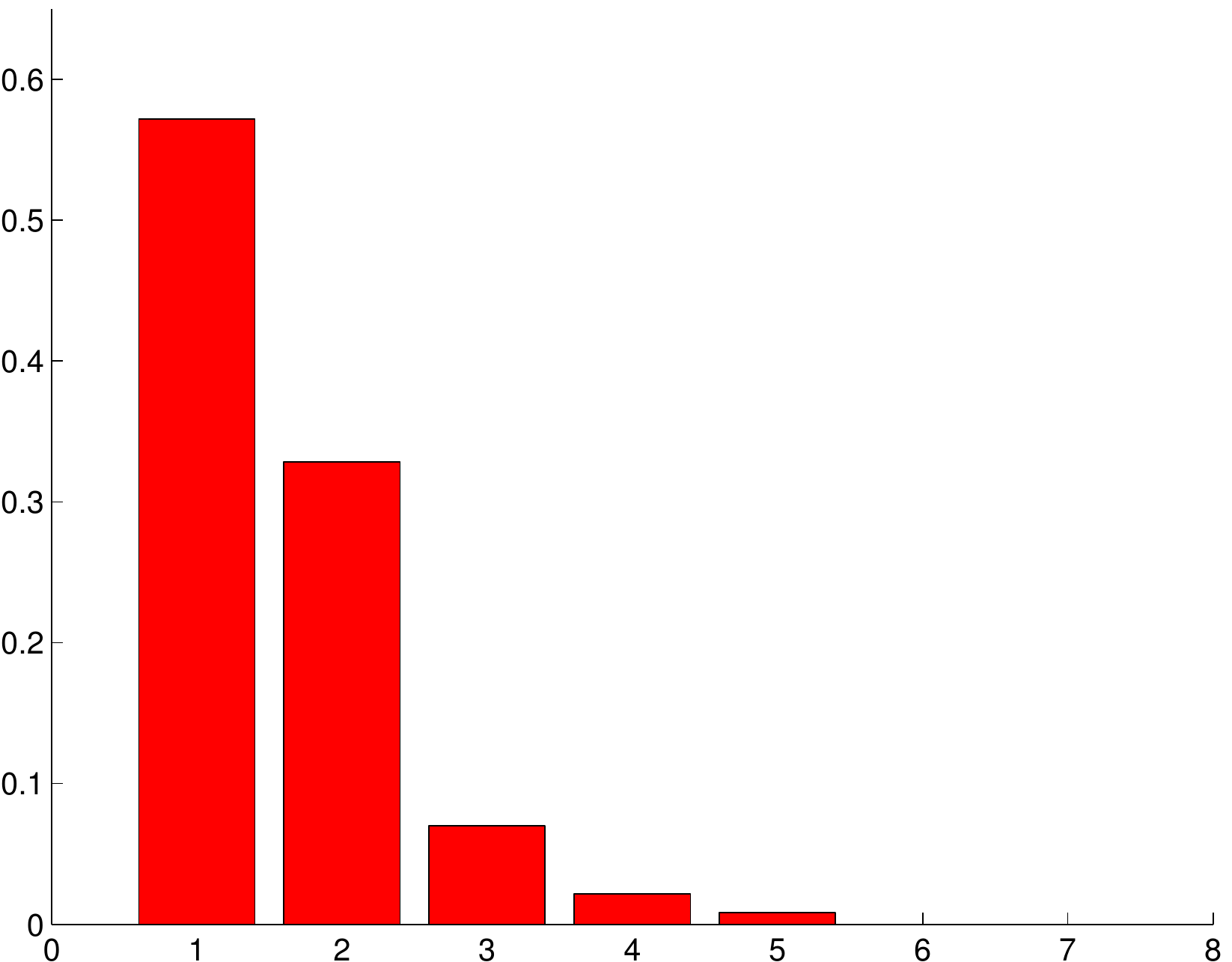}
\includegraphics[scale=0.2]{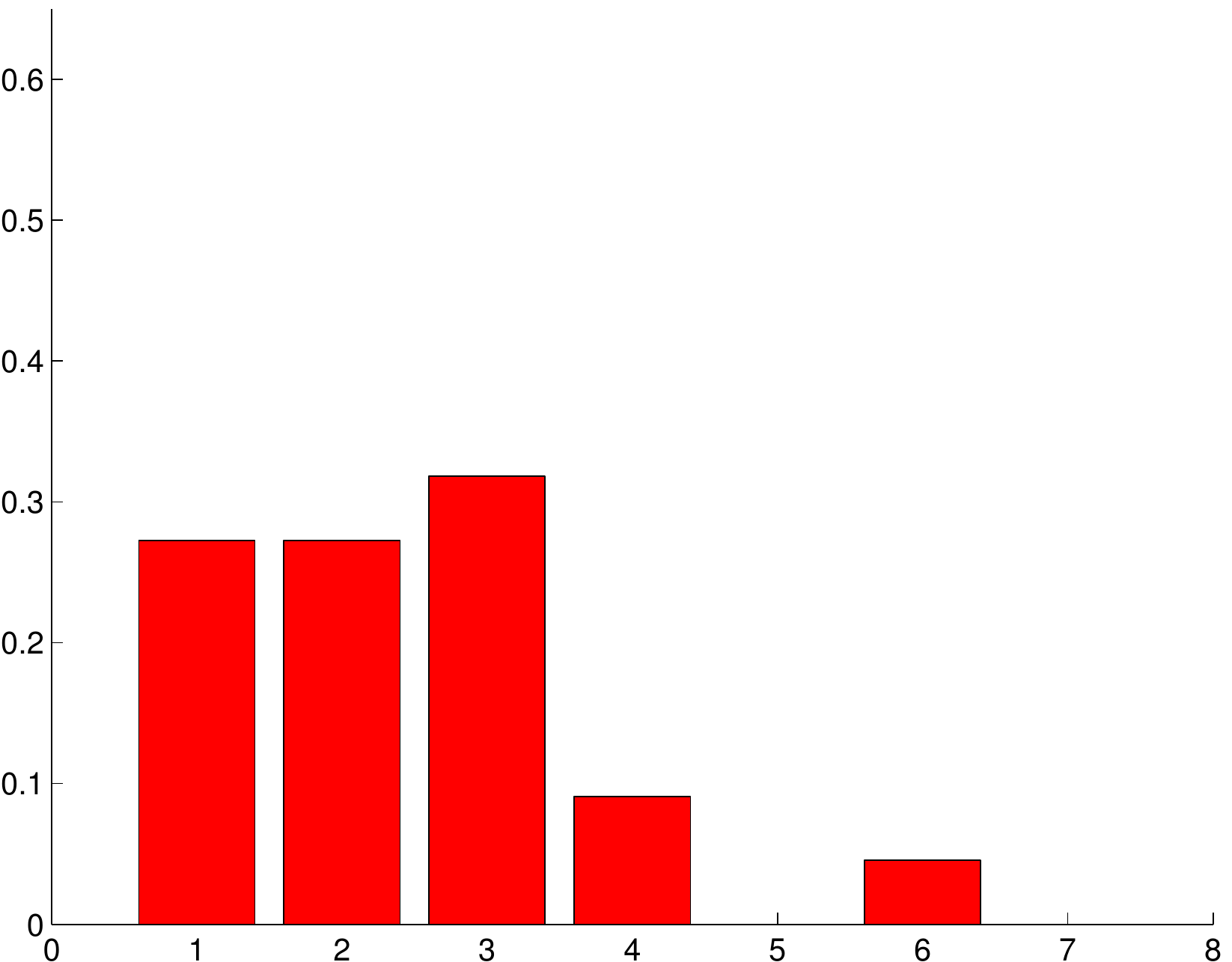}\\
\includegraphics[scale=0.2]{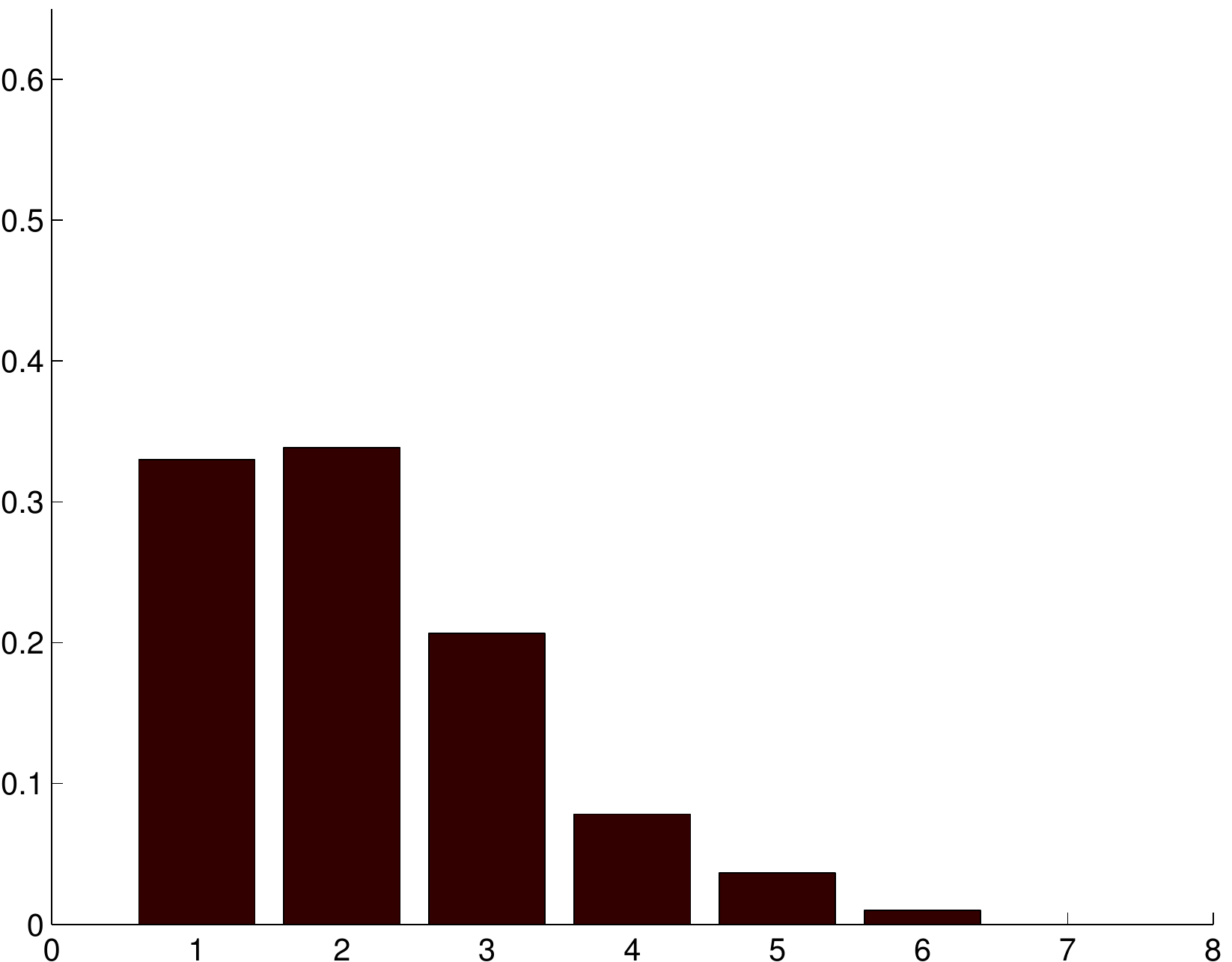}
\includegraphics[scale=0.2]{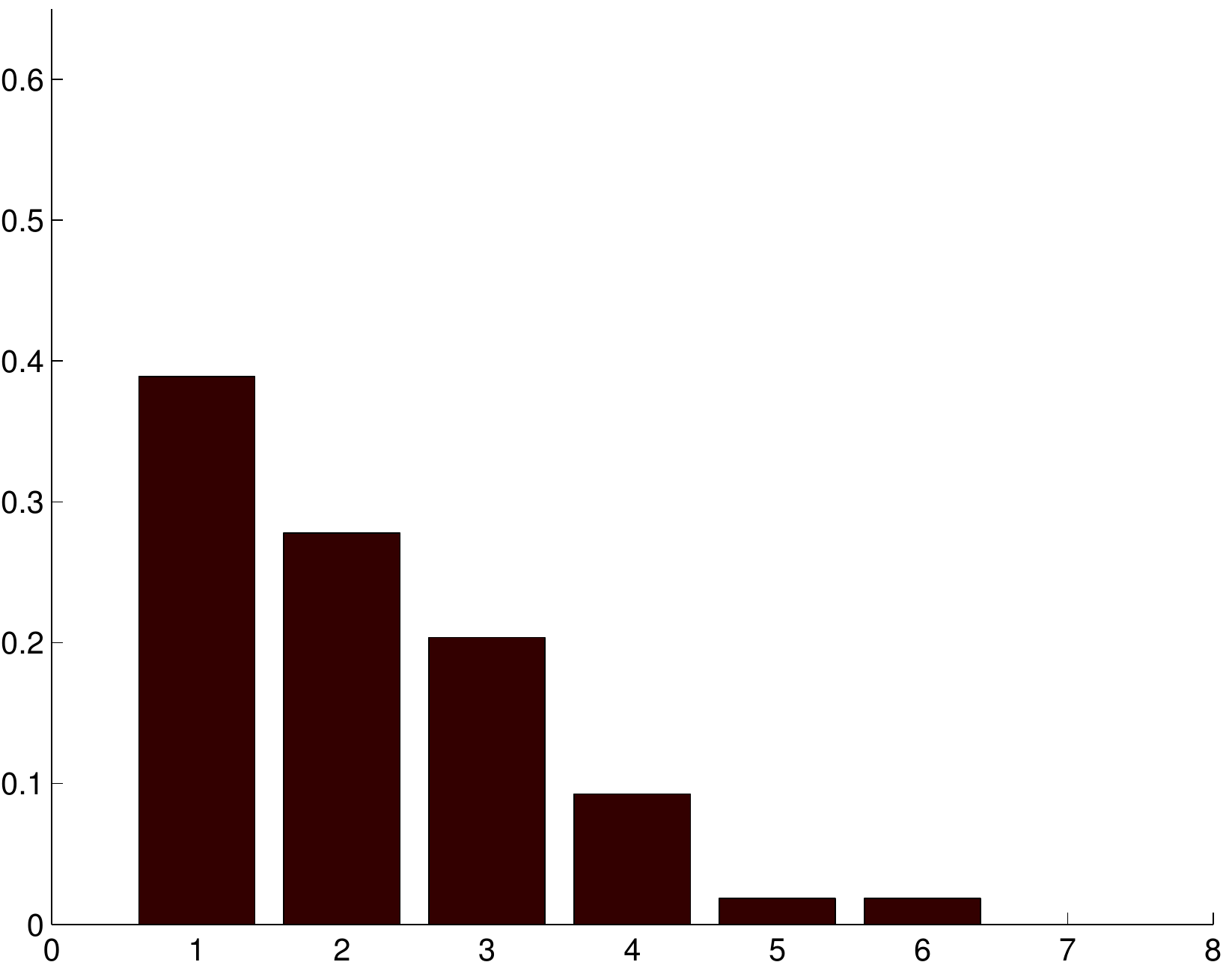}
\caption{Comparison between the numerical simulations (the results of the different orientation strategies are averaged and reported on a single plot) and the experimental data. In red a lineage from the central zone and in black a lineage from the periphery. Left images show numerical results (a), (c), (e), (g), right images report experimental data (b), (d), (f), (h). Figures (a), (b), (c), (d) summarize the analysis of sphericity of the lineage (x-axis $R_3$, y-axis frequency of the connected components with a given ratio area/perimeter). Figures (e), (f), (g), (h) summarize the analysis of the fragmentation of a lineage (x-axis $R_4$, y-axis frequency of the number of connected components per lineage).
\label{fig2bis}}
\begin{tikzpicture}[overlay]
    \node at (-3.6,13.6){
      {\Large {\bf a}}
    };
    \node at (-3.6,10.9){
      {\Large {\bf c}}
    };
    \node at (3.,13.6){
      {\Large {\bf b}}
    };
    \node at (3.,10.9){
      {\Large {\bf d}}
    };
    \node at (-3.6,8.2){
      {\Large {\bf e}}
    };
    \node at (3.,8.2){
      {\Large {\bf f}}
    };
    \node at (-3.6,5.5){
      {\Large {\bf g}}
    };
    \node at (3.,5.5){
      {\Large {\bf h}}
    };

  \end{tikzpicture}
\end{figure}

\subsection{Results of the experiments} 
\label{sec:3.2}
The same diagnostic analysis done for the mathematical model is performed on the experimental data. The results can be summarized as follows.\\ \textit{1) Circularity of the lineages is very low compared to that of the entire population} (the average value of $R_1\in[0,1]$ is $R_1=0.56$. \textit{2) Lineages are filamentous} (the average value of $R_3\in[0,1]$ is very small, around $0.12$). Central lineages are more filamentous than peripheral lineages: there is a larger number of lineages with $R_3=0$ in the central part compared to the periphery. \textit{3) There is a strong disparity (measured by $R_4$) between the central and the peripheral zones} which is expressed by a stronger fragmentation of the central lineages compared to the peripheral ones. In the central zone, the number of lineages divided in one, two or three pieces is almost the same while on the periphery the number of lineages divided in two and in three parts is lower than the number of lineages which did not divide. \textit{4) The central lineage fragments, measured by $R_5$, are much smaller than those of the periphery.} $2/3$ of the fragments are composed of one or two cells in the center while on the periphery they are around $40\%$. \textit{5) Half of the central lineages have a direction very close to the radial one,} while in the periphery no preferred orientation seems to arise (this value is measured by $R_6$). 

\subsection{Comparisons with the experimental data} 
\label{sec:3.3}
The comparisons between the simulation results and the experimental outcomes lead us to the following conclusions shown in Fig.~\ref{fig2bis}. In this Figure the results for all the different orientation strategies used in the numerical simulation are summarized, i.e. the results of the different simulations are averaged and a cumulative distribution is reported. \textit{1) Regarding the first two diagnostics, i.e. $R_1$ and $R_2$,} which are not reported in the Figure and which study the entire cell population, the experimental values are on average slightly lower than those obtained by the numerical simulations. In experiments cells organization is less regular. In particular boundaries are less smooth and the shapes are less round. This is likely due to the fact that real cells are less regular than the perfect disks chosen in the mathematical model. However, the results for these two diagnostics are qualitatively comparable and the choice of perfect disks does not seem to affect other results. \textit{2) Both simulations and experiments highlight a difference between center and boundaries of the growing cell population.} However, this disparity is expressed differently (indicator $R_6$): in the experiments, no preferred tangential orientation of the lineages appears at the periphery of the population (or more precisely only a slight preference for the tangential direction), while a clear preferred tangential orientation is obtained in the numerical simulations for all the different orientation strategies considered. On the other hand, a preferred radial orientation is visible for the central lineages in the experiments, which is very close to the results obtained when the constrained radial orientation strategy is used for the numerical simulations. \textit{3) In the mathematical model, the peripheral lineages are more filamentous} more fragmented into smaller fragments than the corresponding central lineages, while in the experiments the situation is reversed: peripheral lineages are less fragmented with larger size fragments than those in the center (from diagnostic $R_4$ and $R_5$). Globally we can conclude that the initial choices done for constructing the mathematical model permit to reproduce some of the observed features but they are not sufficient to correctly describe all the cells behaviors measured in the experiments. In particular, the difference observed at center of the aggregate between the data and simulations especially in diagnostic $R_4$ (lineages fragmentation) suggests that some additional mechanism is responsible for the displacement of the central cells to the periphery. This mechanism disrupts the organization of the cells and makes central lineages more fragmented and filamentous. In the next Section we propose such mechanism and show that it results in simulations being closer to experimental data.

\section{Improved model: bounded confinement force and cell-cell interchange}
\label{sec:4}

Here, we propose additional mechanisms to reconcile the simulation results with the experimental data. The Section is divided into three parts, in the first part we discuss two improvements to the model. In the second part we detail the modifications of the algorithm necessary to take into account these modifications. In the third part, we discuss comparisons between the new results and the experimental data. 
\subsection{ Model improvements}
\label{sec:4.1}
The first modification consists of the possibility for two adjacent cells to switch their positions. The second modification consists of modifying how the interaction potential depends on the distance from the center of the tumor. The first mechanism permits to switch the position between a new born cell and a neighboring cell. Indeed we hypothesize that compression by the other cell may induce deformations of the cell membrane and that the so deformed cell may be able to migrate into the extra-cellular medium. The random switch between two cells is a way to model this migration favored by cell deformation. We only allow newborn cells to shift their position with a neighboring cell because newborn cells have the smallest size and are more likely to find a migration path in the extra-cellular medium than mature cells. Finally, we only allow cells from the central region to perform this switch because cells from outer regions are subject to lower compression and weaker deformations which decreases their ability to migrate. We consider the possibility for a cell $C_i$ to switch if its position is inside a disk defined by $$\textmd{d}(0,X_i)\leq C_{int}\max_{0\leq j\leq N}(\textmd{d}(0,X_j))$$ with $ C_{int}\in [0,1]$ a modelling parameter. The permutation rule is then as follows: after a division of a progenitor which lies inside this disk, one of the two daughter cells has the possibility to switch its position with one of its neighbors. The decision to switch is modeled by a probability following a Bernoulli distribution of parameter $p$: with probability $p$ a newborn cell inside the sphere switches its position with a neighboring cell.\\
The second additional mechanism consists of applying a weaker attractive potential to the cells which are far from the center of the aggregate. With this aim, the quadratic potential is replaced by a linear potential when the tumor reaches a critical size and the linear potential applies only to cells which are farther than a critical distance. This reflects the hypothesis that cells are submitted to a lower mechanical compression at the periphery of the aggregate. However, for small-sized aggregates, we hypothesize that the adhesion forces between the cells are stronger (since a strong grouping enhances the survival chances of the cells) which motivates the use of a quadratic potential. The positioning rule is consequently modified as follows: as soon as the number of cells reaches a critical value $N_C$, a linear potential is used for the remote area while the same quadratic potential as defined in Section 2 is used for the other cells. The modified global adhesion potential is $$W_L(\xi(t))=\sum_{j,|X_j(t)|\leq R_c} |X_j(t)|^2+\sum_{j,|X_j(t)|\geq R_c} |X_j(t)|,$$ where $R_c=C_L \max_{0\leq j\leq N(t)}(\textmd{d}(0,X_j(t)))$ while the remote area is defined as the set containing the remote cells and consequently a cell at position $X(t)$ is said remote if $$\textmd{d}(0, X(t))\geq C_L \max_{0\leq j\leq N(t)}(\textmd{d}(0,X_j(t)))$$ with $C_L\in[0,1]$ a modelling parameter.

\subsection{Algorithm adaptation}
\label{sec:4.2}
The new structure of the numerical algorithm is the following
\begin{itemize}
 \item[a)] Initialization
 \item[b)] At each time step
 \begin{itemize}
 \item[i)] Growth step.
 \item[ii)] Test on size of the cell, cell by cell. If the threshold size is reached a division occurs.
 \item[iii)] For each mitosis up to the final division
 \begin{itemize}
  \item[1)] Partial division.
  \item[2)] Modified positioning step. 
  \item[3)] If necessary, depending on the orientation strategy chosen, orientation update.
 \end{itemize}
  \item[iv)]Permutation step.
 \item[v)] Modified positioning step.
 \end{itemize}
 \item[c)] Statistical quantifiers computation.
\end{itemize}
We discuss the modified positioning step and the permutation step. The modified positioning step consists in finding a saddle point of the new Lagrangian function $\mathcal L_L (\xi(t),\lambda(t)): (\R^2)^{N(t)}\times \R^M\rightarrow \R$ defined by 
\begin{equation*}
\begin{split}
&\mathcal L_L (\xi(t),\lambda(t))=W_L(\xi(t))\\
&+\sum_{1\leq i\leq j\leq N(t)}\lambda_{ij}(t)\phi_{ij}(\xi(t),\rho(t)), \ \forall (\xi(t),\lambda(t)),  
\end{split}
\end{equation*}
Thus, starting from an initial guess $(\xi(t)^{(0)},\lambda(t)^{(0)})$, the method reads, as in the previous case, as
\begin{equation*}
\begin{cases}
&\xi^{(p+1)} =X^{(p)}-\beta\nabla_x\mathcal L_L\left(\xi^{(p)},\lambda^{(p)}\right), \\
&\phi^{(p+1)}_{ij} =\phi_{ij}\left(\xi^{(p+1)}\right), \ \forall \ i,j\in[1,N], \ i<j, \\
&\lambda^{(p+1)}_{ij} = \max\left(0,\lambda^{(p)}_{ij}+\mu \phi^{(p)}_{ij}\right), \forall \ i,j\in[1,N], \ i<j,\\
\end{cases}
\end{equation*}  
where $\beta$ and $\mu$ are the same numerical parameters as those discussed in Section 3. After some computations, the first equation of the above system can be rewritten for all cells in the remote region, i.e. $|X_j|>R_c$, as $$X^{(p+1)}_k =\left(1-\frac{\beta}{|X_k^{(p)}|}\right)X_k^{(p)}+2\beta\sum_{j=1}^{N}\lambda^{(p)}_{kj}\left(X_k^{(p)}-X_j^{(p)}\right).$$ 
The same stopping criteria are used for this new positioning algorithm. During the research for a saddle point, it may happen that a cell close to the boundary between the central and the remote regions changes zone. Since, this displacement is typically very small, we choose not to change the potential energy to which this cell is submitted during the minimization procedure.\\ The permutation algorithm consists simply in choosing with probability $p$ if a newborn cell performs a switch. In order to do that, denoting by $C_{I}$ the cell which has decided to perform a switch, we first determine all the neighboring cells of $C_{I}$ and then we pick randomly one of them with uniform distribution and we perform the switch. The values of the new parameters added to the model are summarized in Table 3. The value of $C_L$ is chosen so that the central region coincides with the definition of a central lineage in the experiments.

\begin{table}
\centering
\begin{tabular}{lll}
\hline
 Parameter&  Value&  Meaning  \\
 \hline
 $p$ & 0.1,0.15,0.25 &Probability of permutation     \\
 \hline
 $C_{int}$&  0.5& Interior lineages delimitation  \\
\hline
$ N_s$ &  100, 200& Threshold for linear potential \\   
 \hline
  $C_L$&  0.5& External lineages delimitation    \\
\hline
\end{tabular}
\caption{Numerical parameters for the modified algorithm. 
\label{table3}}
 \end{table}

\subsection{Comparisons between the
experiments and the improved numerical model}
\label{sec:4.3}
\begin{figure}%[h!]
\vskip-.0cm
\hspace*{0.5cm}
\includegraphics[width=105mm,trim={0cm 0 0 0}]{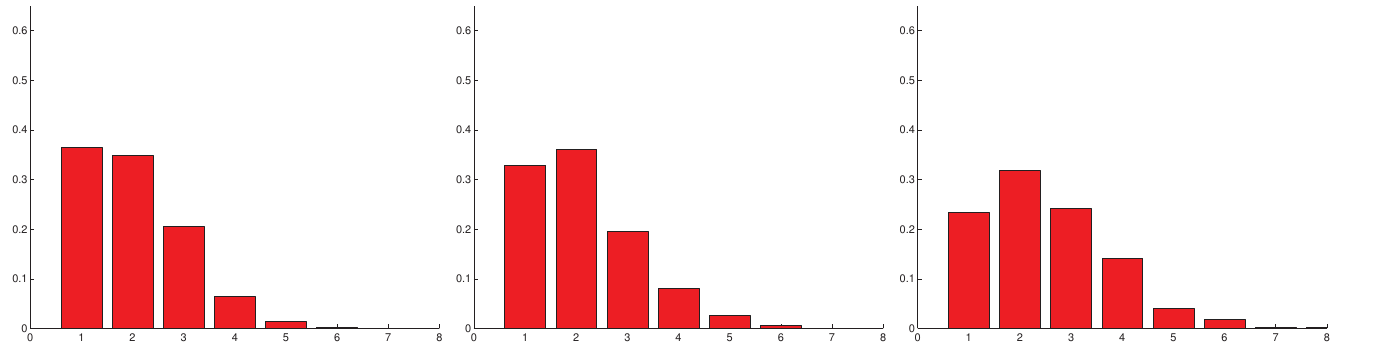}
%\vskip-3cm
\caption{Diagnostic $R_4$: fragmentation of a lineage in the central region with cell switch mechanism for different values of the switching probability $p$. Left: $p=0.1$. Middle: $p=0.15$. Right: $p=0.2$. The more $p$ increases, the more the distribution shifts to the right: the lineages are separated into a larger number of fragments.
\label{fig30}}
\end{figure}

\begin{figure}%[h!]
%\vskip-0.5cm
\hspace*{0.5cm}
\includegraphics[width=103mm,trim={0cm 0 0 0}]{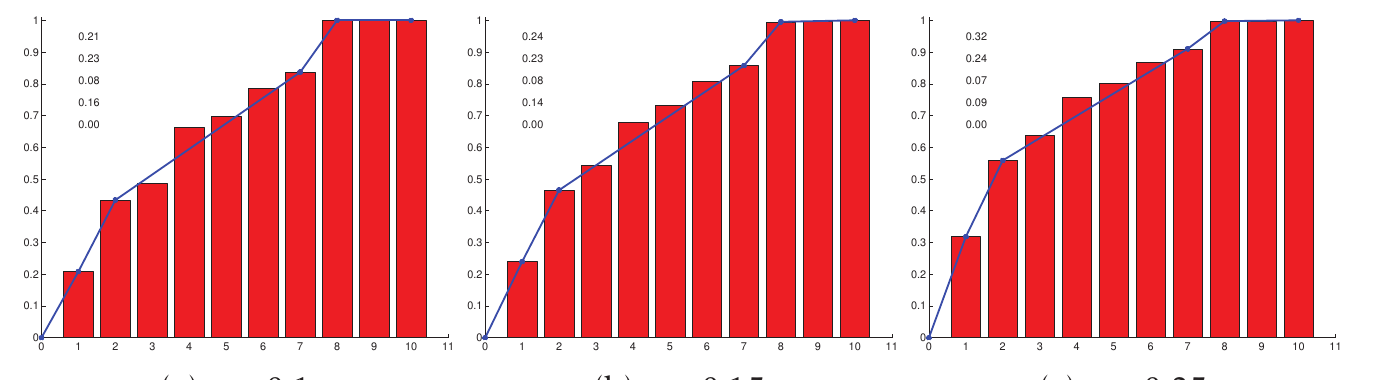}
%\vskip-3cm
\caption{Diagnostic $R_5$: number of cells per connected component of a given lineage in the central region with cell switching mechanism for different values of the
probability $p$. Left: $p=0.1$. Middle: $p=0.15$. Right: $p=0.2$. The more $p$ increases, the more the values are concentrated on the left of the distribution: the number of small connected components increases. A piecewise linear interpolation of the cumulative distribution is depicted in blue color. The values of this piecewise linear interpolation are indicated in the graph.
\label{fig31}}
\end{figure}

We first analyze separately the impacts of the two modifications introduced in the model. We start discussing the effect of the switch. The results for diagnostic $R_4$ (lineage fragmentation) and $R_5$ (size of the fragments of a lineage) are reported respectively in Fig. \ref{fig30} and \ref{fig31} for three different values of the permutation probability $p$: $p=0.1, \ p=0.15, \ p=0.2$. They show that the larger the switching probability, the more fragmented the lineages are and the smaller the fragments are. Cell switching even though it is performed only in the central region has also an influence on the peripheral region. However, the impact of these changes is low and we do not report it. The direction of lineages computed with diagnostic $R_6$ does not change noticeably with the switching. 

\begin{figure}%[h!]
%\vskip-.5cm
\hspace*{1cm}
\includegraphics[width=95mm,trim={0cm 0 0 0}]{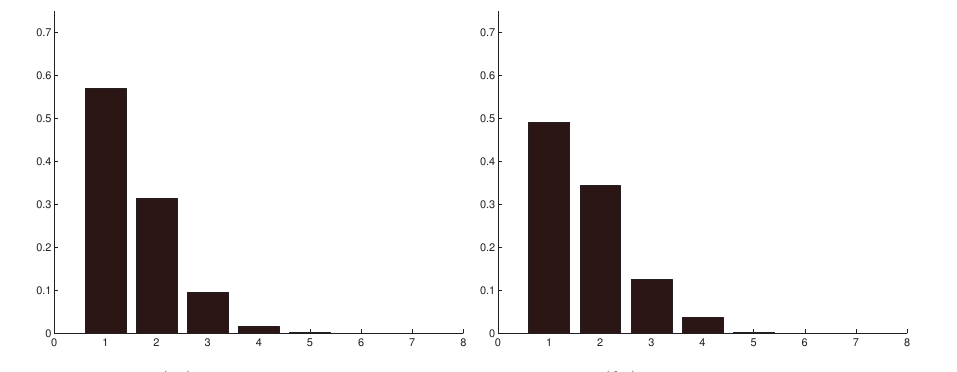}
%\vskip-3cm
\caption{Diagnostic $R_4$: fragmentation of a lineage with the modified potential at the periphery for different values of the threshold number $N_s$ which performs the switch from the quadratic potential to the modified one. Left: $N_s=100$. Right: $N_s=200$.
\label{fig33}}
\end{figure}

\begin{figure}%[h!]
%\vskip-.5cm
\hspace*{1cm}
\includegraphics[width=95mm,trim={0cm 0 0 0}]{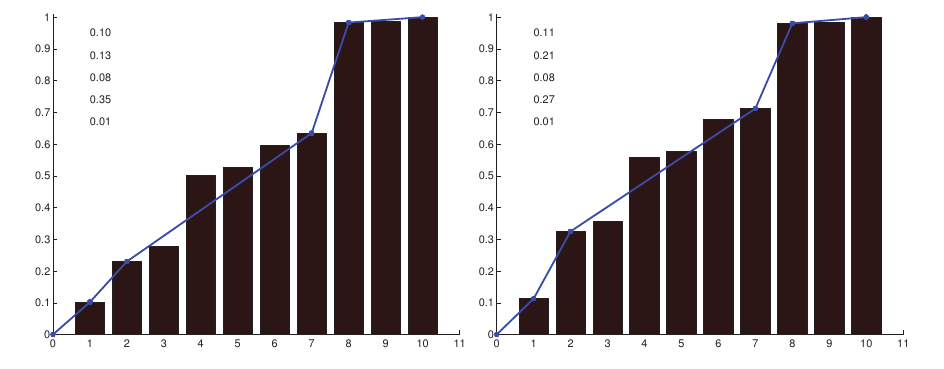}
%\vskip-3cm
\caption{Diagnostic $R_5$: number of cells per connected component of a given lineage with the modified potential law at the periphery for different values  of the threshold number $N_s$ which performs the switch from the quadratic potential to the modified one. Left: $N_s=100$. Right: $N_s=200$. A piecewise linear interpolation of the cumulative distribution is depicted in blue color. The values of this piecewise linear interpolation are indicated in the graph.
\label{fig34}}
\end{figure}

Now, we consider the influence of the change in the attractive potential. The results for diagnostics $R_4$ and $R_5$ for two different values of the parameter $N_s$, i.e. $N_s=100$ and $N_s=200$ are reported in Fig. \ref{fig33} and \ref{fig34}. They show the effect of this additional mechanism at the periphery of the tumor. In particular, we observe that the smaller $N_s$ is, the less segmented the lineages are and the bigger the fragments are at the periphery. The change in the potential energy has also an effect in the central zone, where the lineages are becoming less fragmented. Finally, regarding the direction of the lineages with diagnostic $R_6$, we do not observe noticeable compared to the quadratic potential. 

\begin{figure*}[ht]
\centering
\includegraphics[scale=0.17]{fig2_R4CS.pdf}
\includegraphics[scale=0.17]{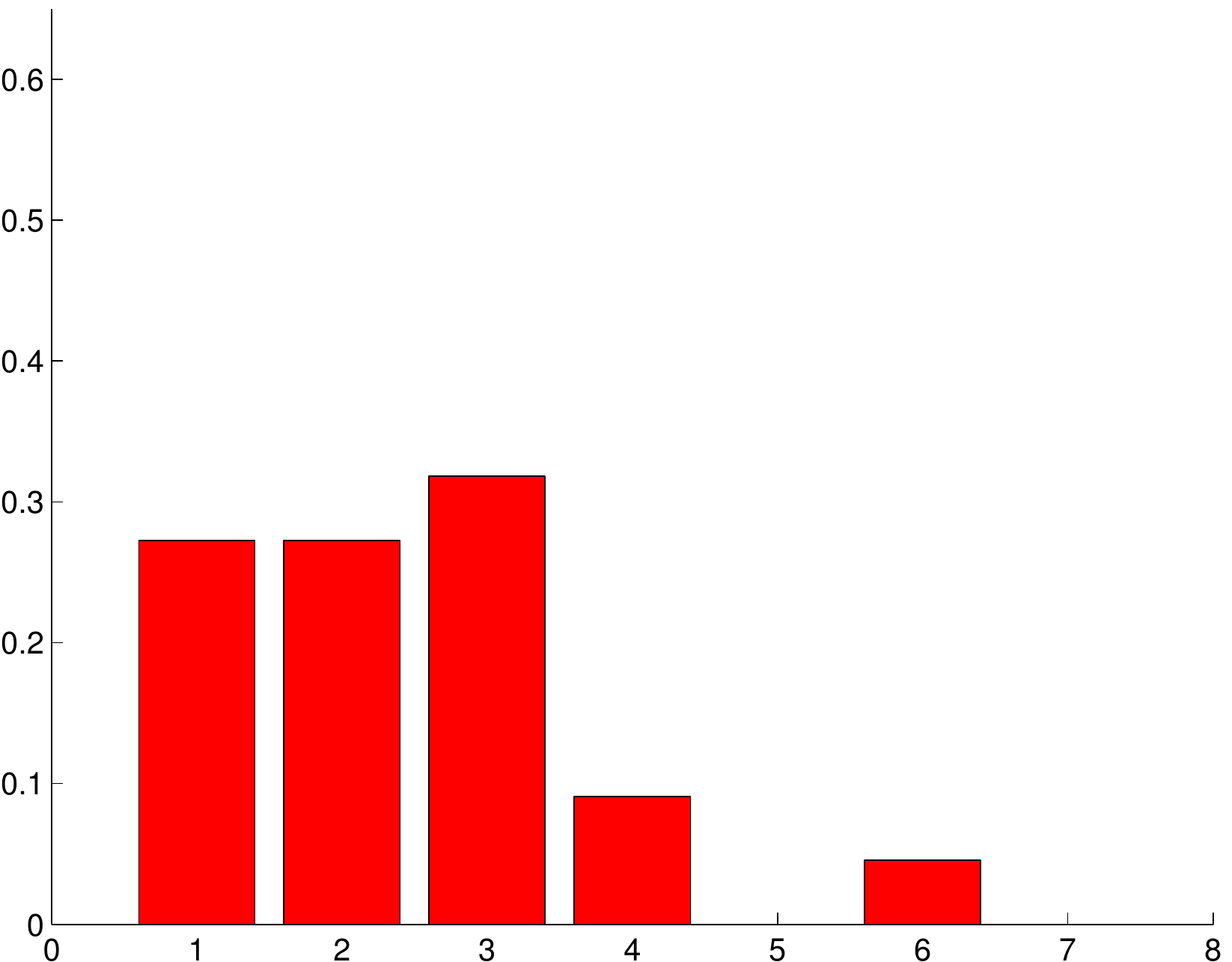}
\includegraphics[scale=0.17]{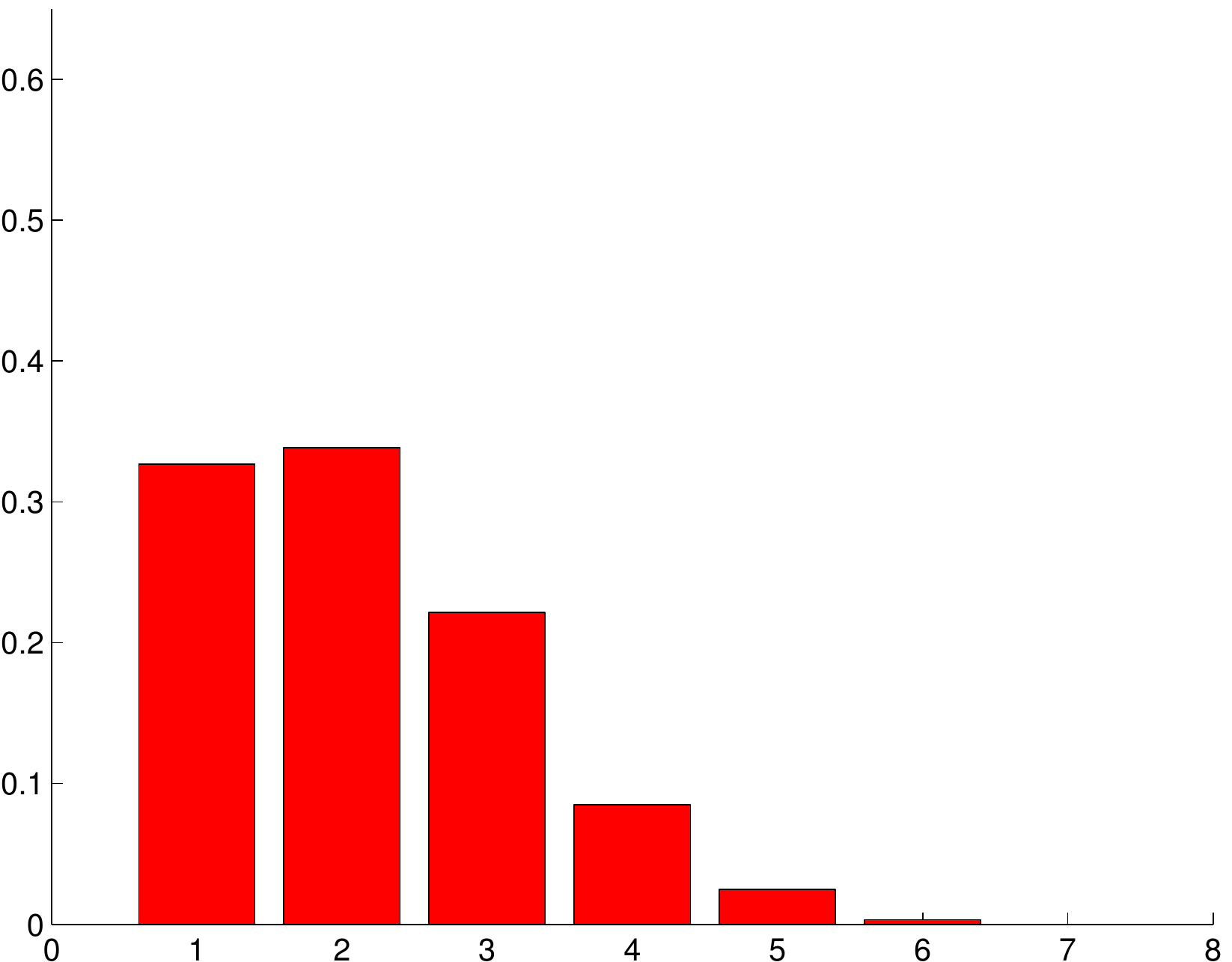}\\
\includegraphics[scale=0.17]{fig2_R4PS.pdf}
\includegraphics[scale=0.17]{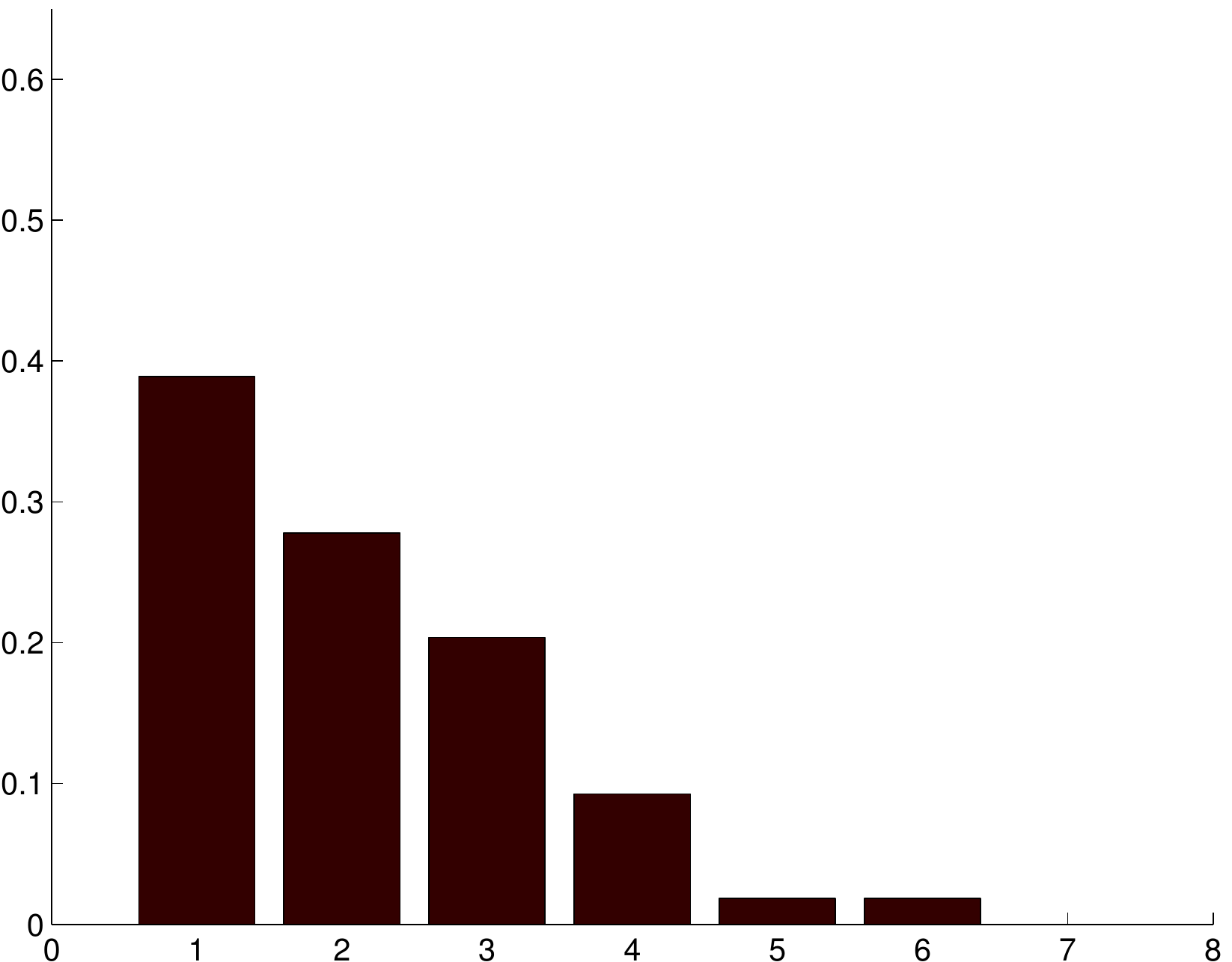}
\includegraphics[scale=0.17]{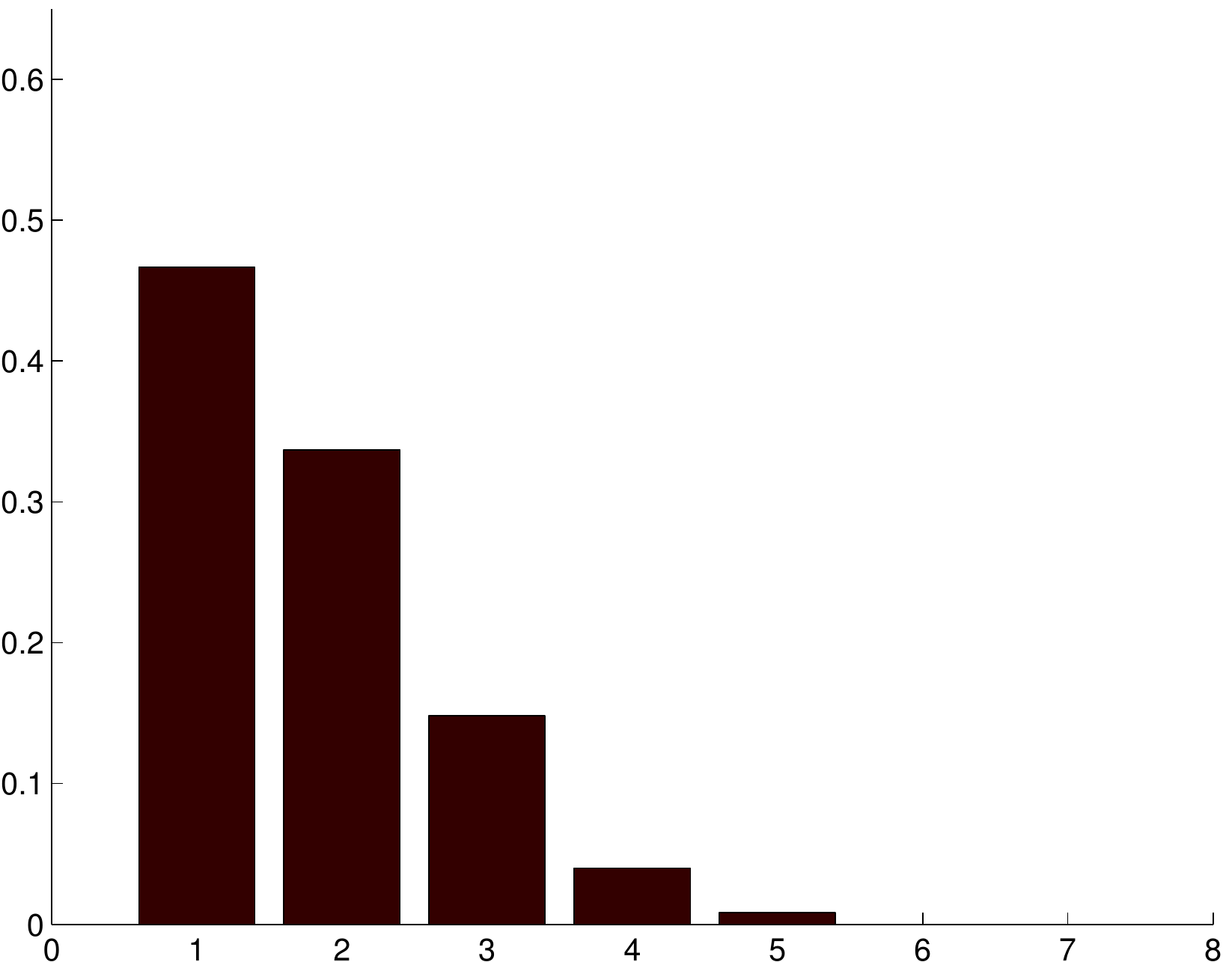}\\
\includegraphics[scale=0.17]{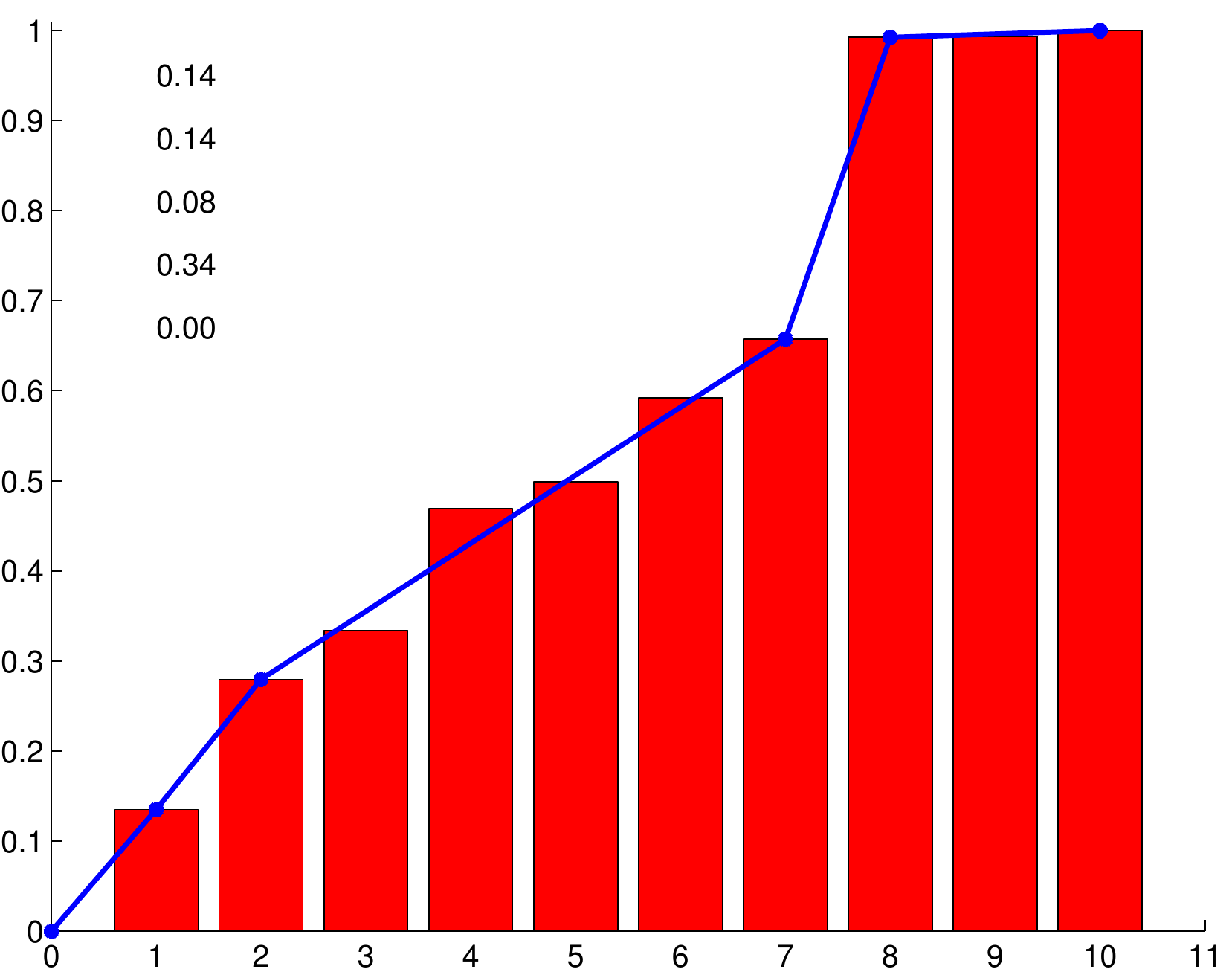}
\includegraphics[scale=0.17]{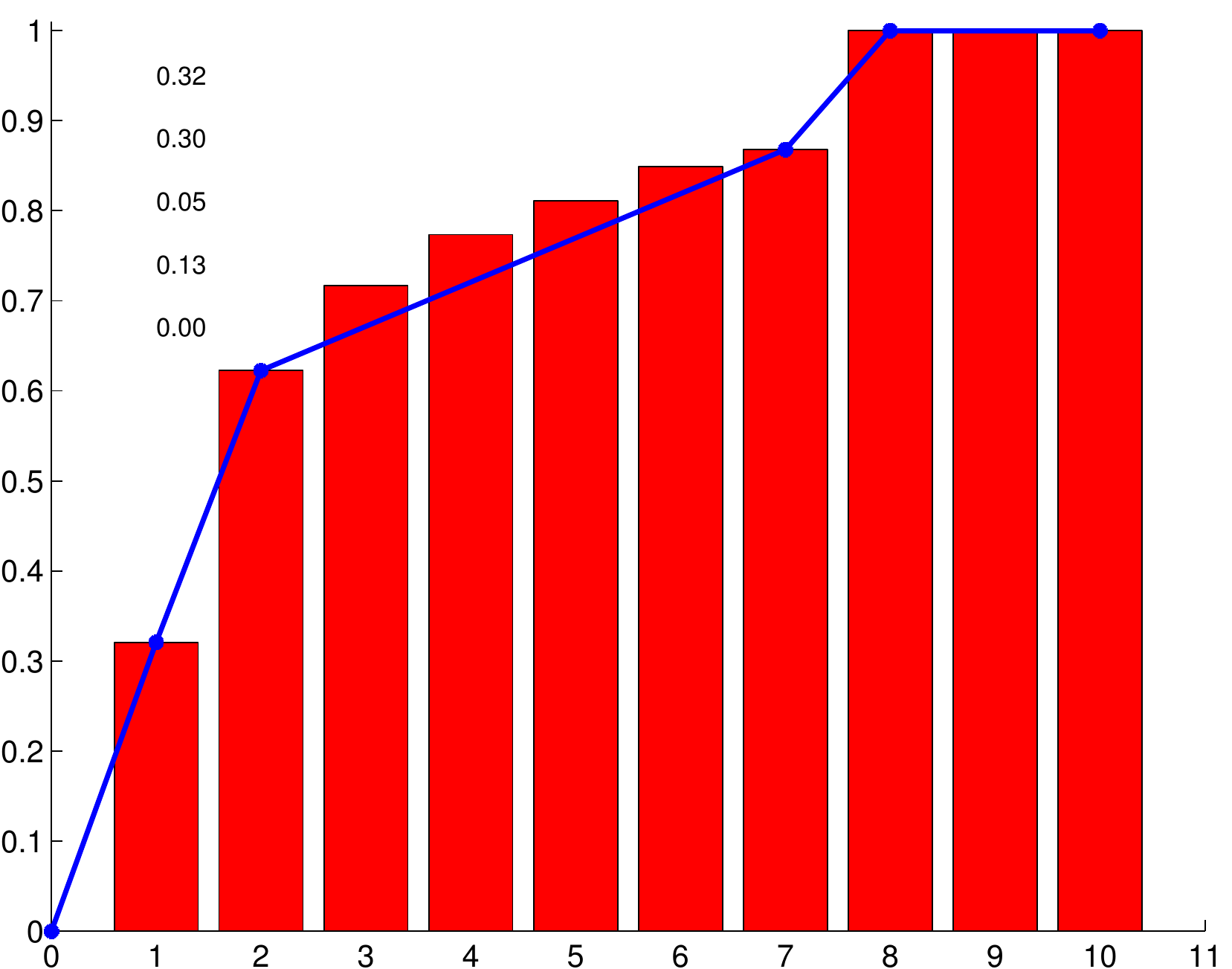}
\includegraphics[scale=0.17]{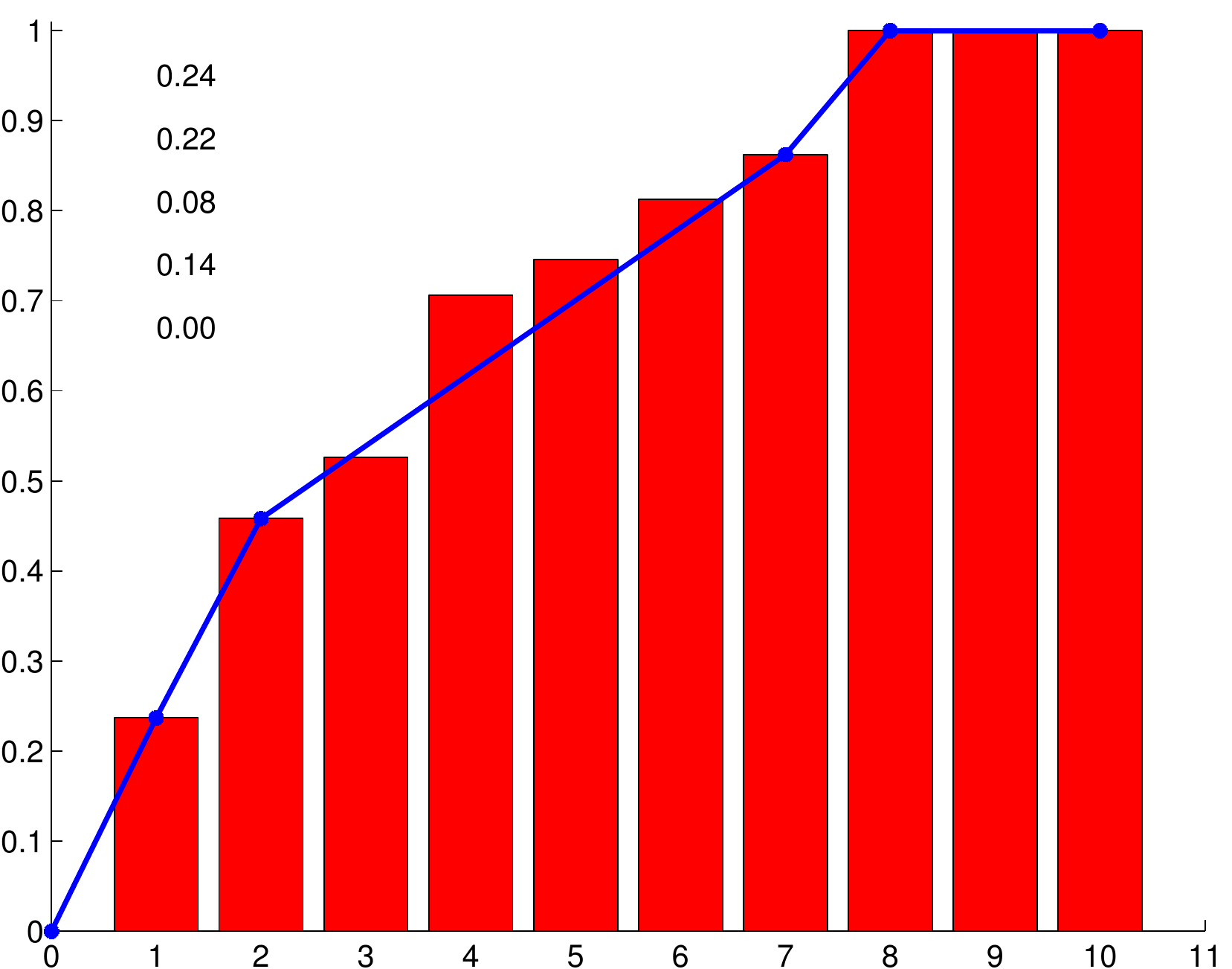}\\
\includegraphics[scale=0.17]{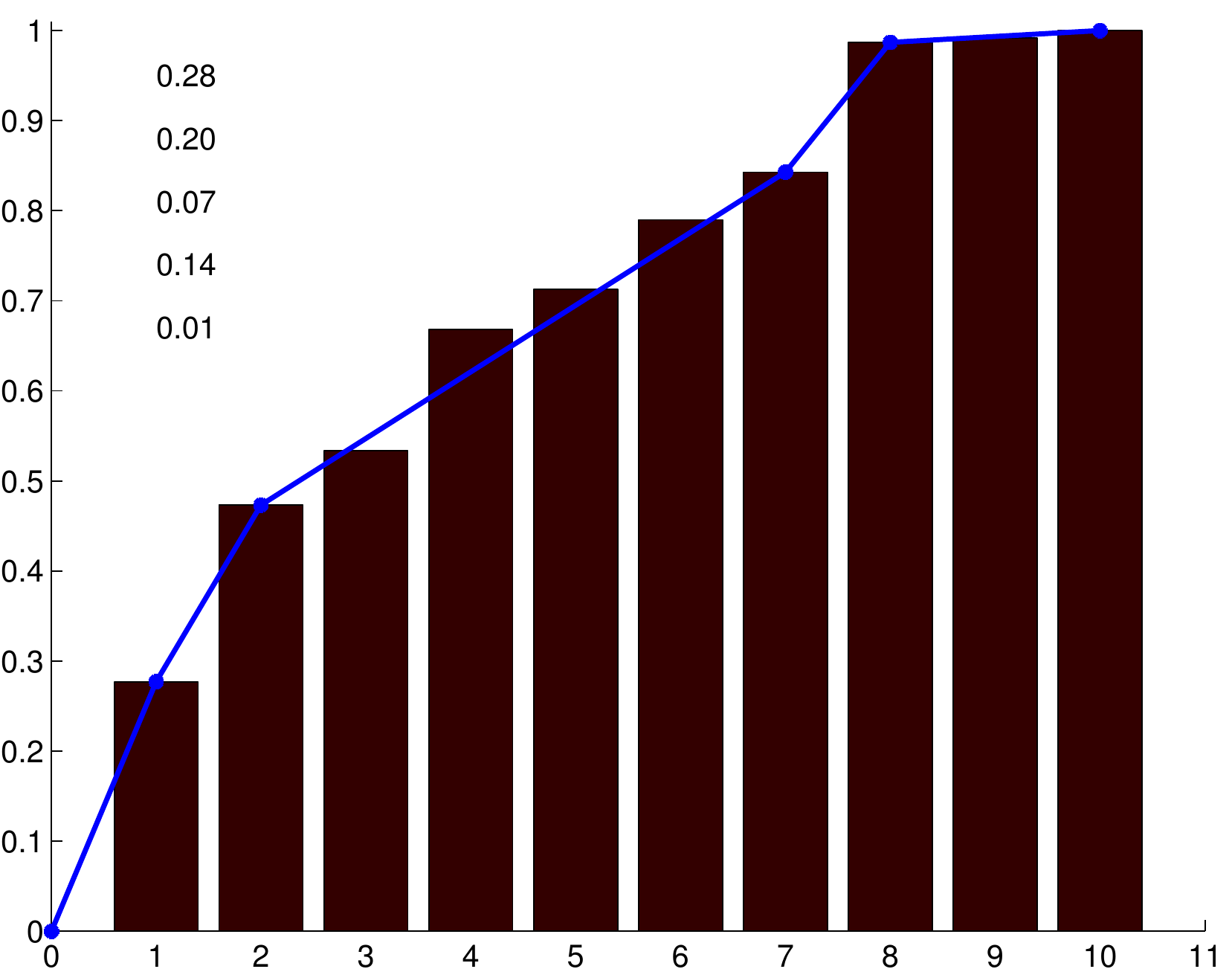}
\includegraphics[scale=0.17]{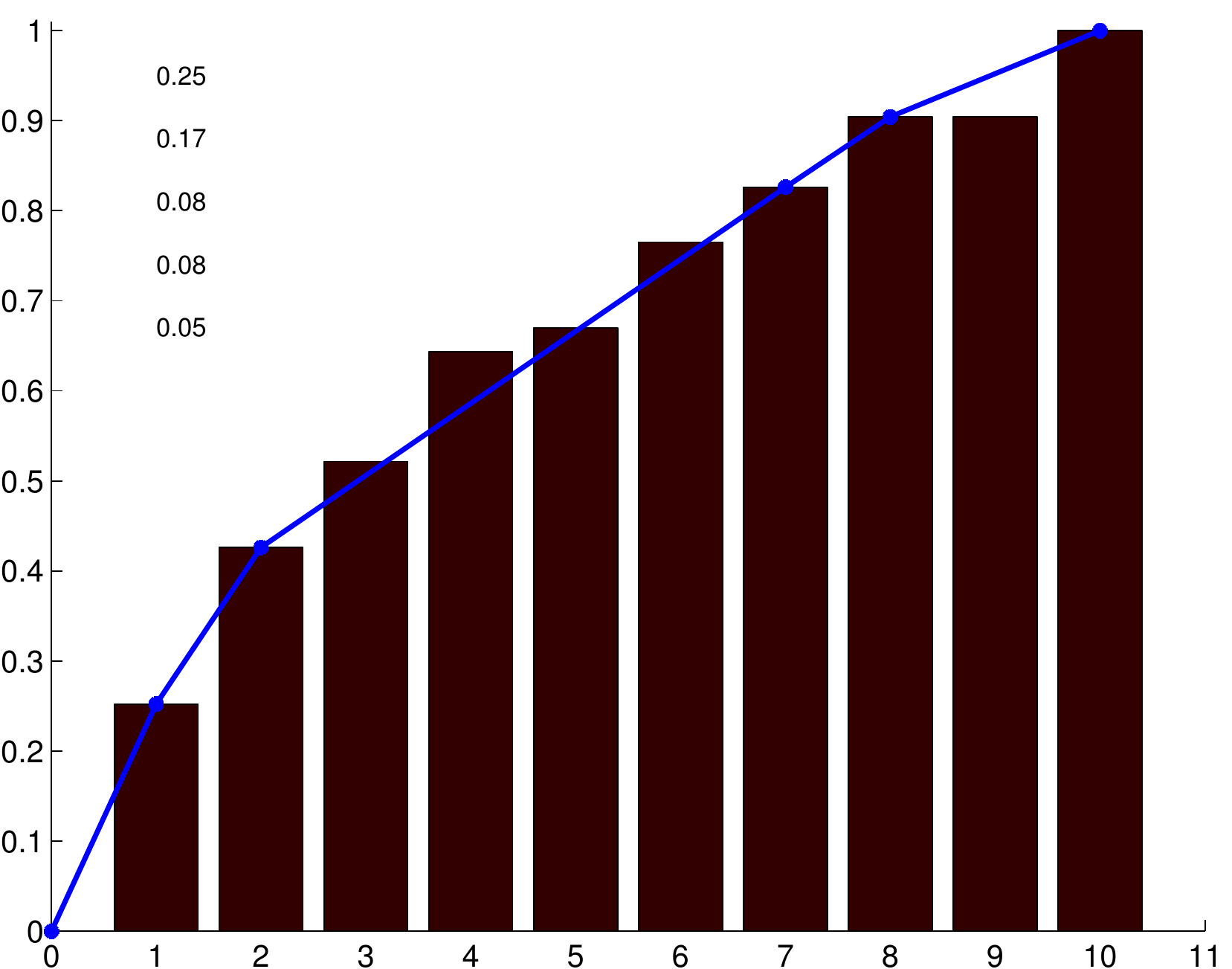}
\includegraphics[scale=0.17]{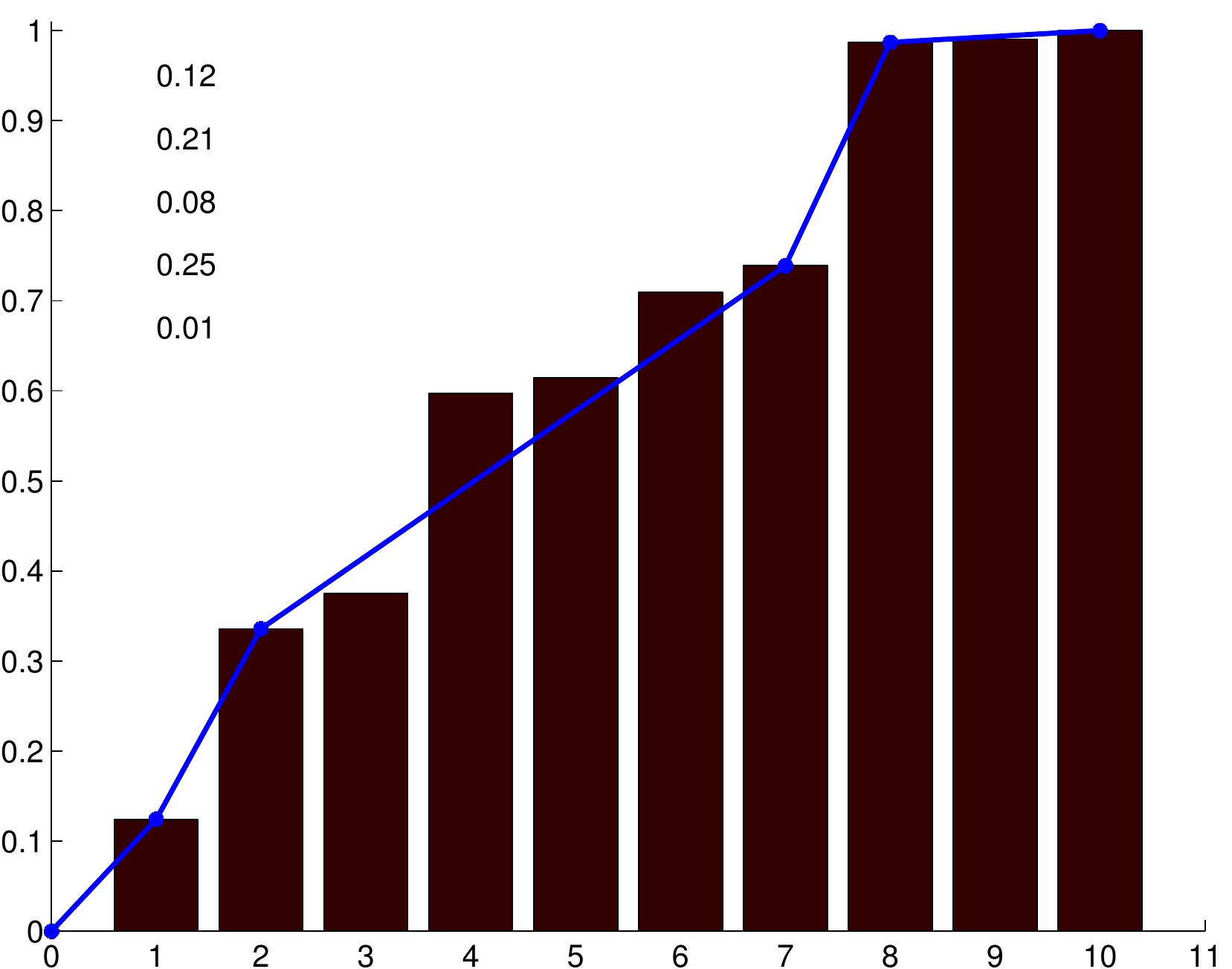}
\caption{Comparison between the numerical simulations and the experimental data. In red a lineage from the central zone and in black a lineage from the peripheral zone. Left images report original numerical results (a), (d), (g), (l) center images report experimental data (b), (e), (h), (m) right images report numerical results with the modified model (c), (f), (i), (n). Top and middle top, fragmentation of a lineage (x-axis $R_4$, y-axis frequency of the number of connected components per lineage). Middle bottom and bottom, number of cells per connected component (x-axis $R_5$, y-axis connected component cumulative frequency). 
\label{fig3bis}}
\begin{tikzpicture}[overlay]
    \node at (-4.6,13.1){
      {\Large {\bf a}}
    };
    \node at (0.,13.1){
      {\Large {\bf b}}
    };
    \node at (4.6,13.1){
      {\Large {\bf c}}
    };
    \node at (-4.6,10.6){
      {\Large {\bf d}}
    };
    \node at (0.,10.6){
      {\Large {\bf e}}
    };
    \node at (4.6,10.6){
      {\Large {\bf f}}
    };
    \node at (-4.6,8.5){
      {\Large {\bf g}}
    };
    \node at (0.,8.5){
      {\Large {\bf h}}
    };
    \node at (4.6,8.5){
      {\Large {\bf i}}
    };
    \node at (-4.6,6.2){
      {\Large {\bf l}}
    };
    \node at (0.0,6.2){
      {\Large {\bf m}}
    };
    \node at (4.6,6.2){
      {\Large {\bf n}}
    };
  \end{tikzpicture}

\end{figure*}

We finally  observe the results obtained while combining the two modifications.
As shown in Fig.~\ref{fig3bis}, the simulated results obtained with the two model improvements are more similar to experimental results, both in the central zone and at the periphery. In the Figure, the results of the Diagnostics $4$ and $5$ are reported. For these tests a switching probability for a daughter cell of $p=0.2$ and a threshold of $N=200$ to pass from a quadratic to a linear potential have been chosen. These parameters permit the best fit between the simulations and the data. We conclude that these modifications improve the match between the model and the experiments. The distributions of fragments number and size of lineages obtained with the modified model match those of the experimental data remarkably well.

\section{Conclusions}
\label{sec:5}
This work represents a step towards understanding the impact of division parameters on the growth of a cell population via comparison of mathematical models and experiments. The approach consists of a bottom-up strategy where the behavior of a growing population of cells and the structure of the associated lineages is modeled through simple interaction rules of mechanical type between cells. After observing that these simple rules could not explain the morphology of the cell population alone, we introduced additional phenomena, which, at first were considered negligible. This process of gradually increasing complexity was repeated until a good fit between experiments and simulations is obtained. This approach permits to assess which key mechanisms are more likely to be underlying the observed phenomena without introducing too many empirical parameters. From the mathematical point of view, the energy minimization considered here is motivated by the observation that physical principles are often expressed in variational forms. Our results suggest that this variational approach seems also at play here. Our numerical and experimental results show a wide disparity between central and peripheral lineages. Peripheral lineages are more fragmented and slightly tangentially oriented. With the simplest model, some differences between simulations and experimental data are found: experiments show that cells in the central region move over larger distances than in the model. We thus introduced the possibility for cells to switch positions inside the cell population. We also lowered the aggregation force at the periphery of the cellular aggregate to model weaker aggregation of peripheral cells. The corresponding numerical results match the experimental data very convincingly. Another interesting experimental observation is that in the central area, a preferred radial direction of the lineages emerges and that a similar feature can be found in the model if a radial division plane orientation is imposed. The analysis suggests that the disparity between the center and the periphery, found experimentally and verified numerically, could be explained by the simple hypotheses made in the model. More quantitative work is needed to understand the role of the division orientation on the observed emerging structures and on lineage shape and orientation. In the future, larger populations should be considered. On the other hand, considering very small populations permits to ignore the role played by nutrients, growth factors and oxygen, and to consider simple two dimensional settings for both the model and the experiments. This also made the image processing easier and more reliable avoiding problems related to the tracking of cells in a three dimensional structure and it allowed for the use of cpu-effective agent based model which would be too costly in three dimensions. However, these simplifying choices lead to a model with restricted validity and questionable applicability to three dimensional structures. To understand the role of the orientation of the division plane future experiments which track cells during mitosis would permit to further explore the biology of tumor aggregates and provide a better benchmark for the validation of three dimensional models. Another direction of research is to explore the impact of mechanical confinement on cell proliferation. Recent experiments \cite{Valerie} showed that proliferation gradients within mechanically confined spheroids are different from those in spheroids grown in suspension. This discovery strengthens the hypothesis of mechanical forces playing a central role in the morphologies of the lineages and requires further studies.

\appendix
\section{Detailed results of the simulations}
\begin{figure}[ht!]
%\vskip-.2cm
\hspace*{2cm}
\includegraphics[width=75mm,trim={-0cm 0 0 0}]{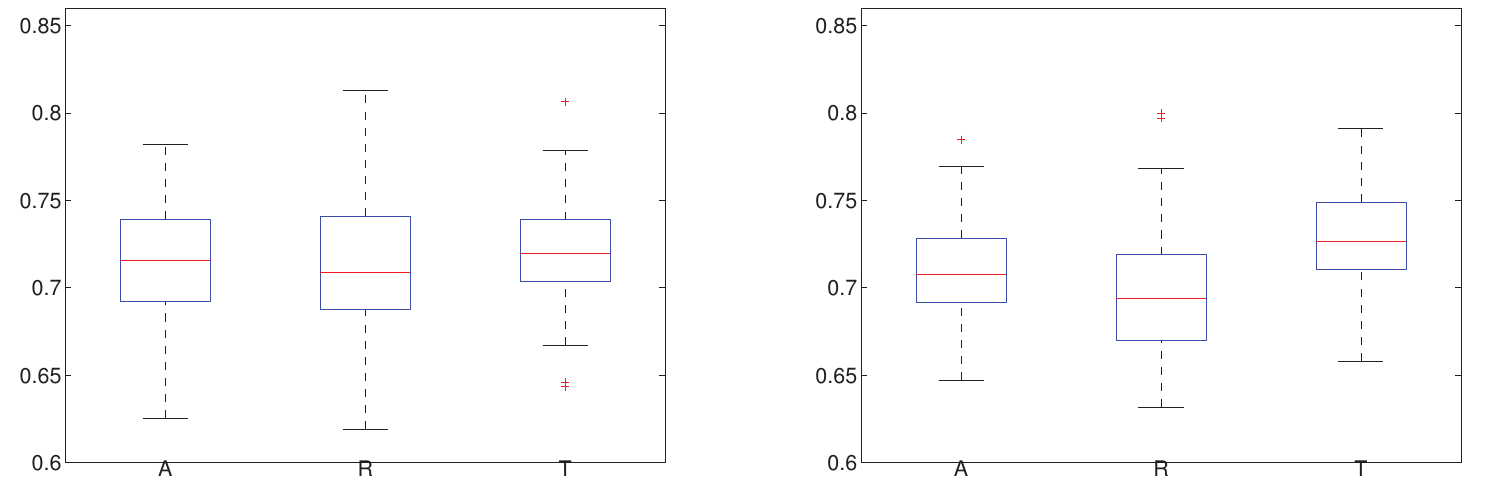}
%\vskip-3cm
\caption{Box diagram for the indicator $R_1$. Left picture: free orientation strategy. Right picture: constrained orientation strategy.
(A) random direction, (R) radial direction and (T) tangential direction of the division plane.
\label{fig8}}
\end{figure}

\begin{figure}[ht!]
%\vskip-.1cm
\hspace*{2cm}
\includegraphics[width=75mm,trim={0cm 0cm 0 0}]{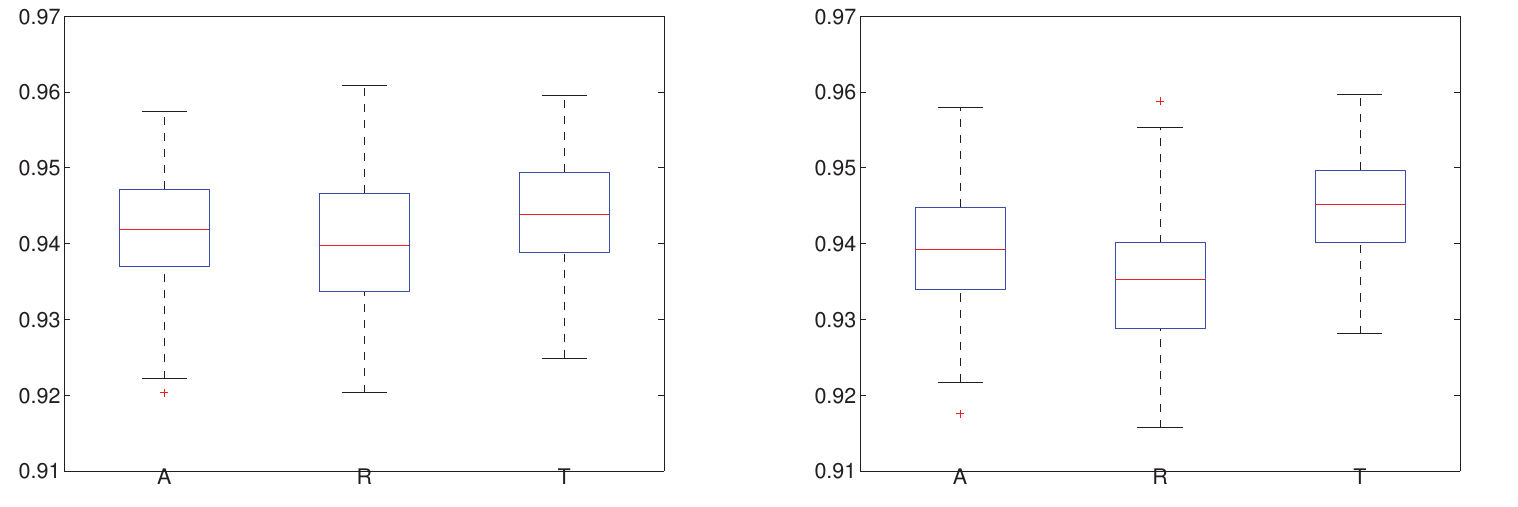}
%\vskip-3cm
\caption{Box diagram for the indicator $R_2$. Left picture: free orientation strategy. Right picture: constrained orientation strategy. (A) random direction, (R) radial direction and (T) tangential direction  of the division plane.
\label{fig9}}
\end{figure}

\begin{figure}[ht!]
%\vskip-0.6cm
\hspace*{1.6cm}
\includegraphics[width=75mm,trim={0cm 0 0 0}]{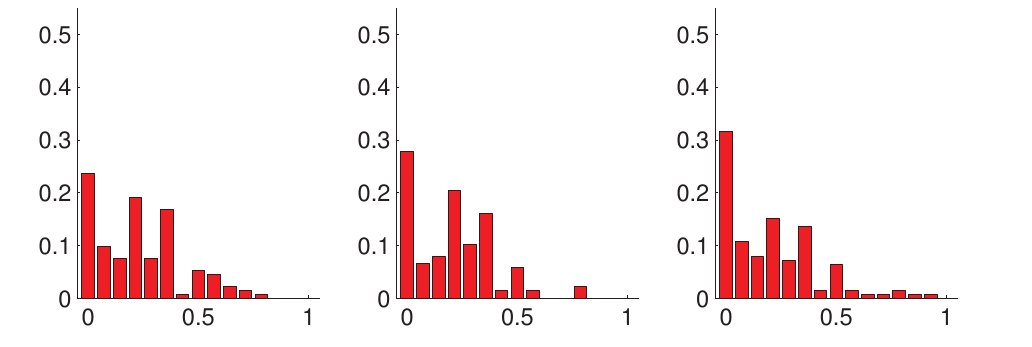}\\
\hspace*{1.6cm}
\includegraphics[width=75mm,trim={-0cm 0 0 0}]{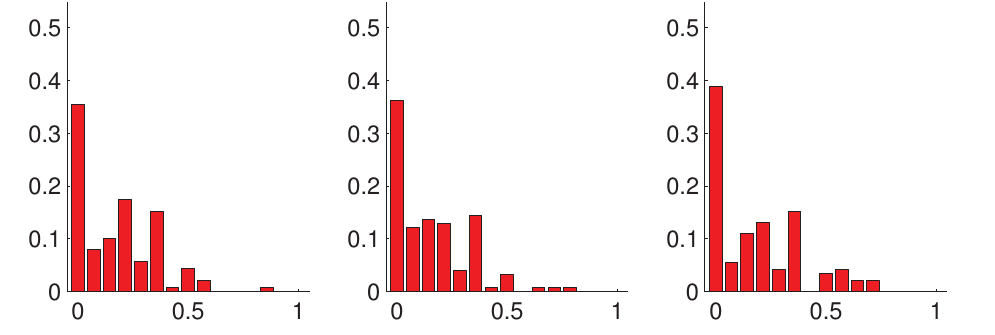}\\
\hspace*{1.6cm}
\includegraphics[width=77mm,trim={0.2cm 0 0 0}]{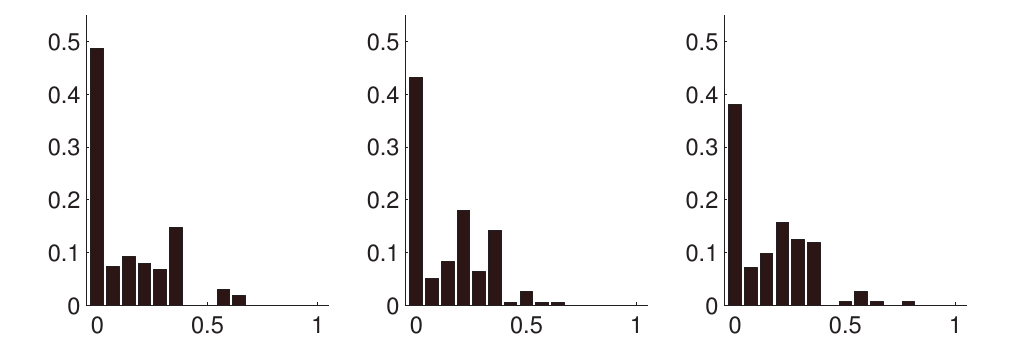}\\
\hspace*{1.6cm}
\includegraphics[width=75mm,trim={-0cm 0 0 0}]{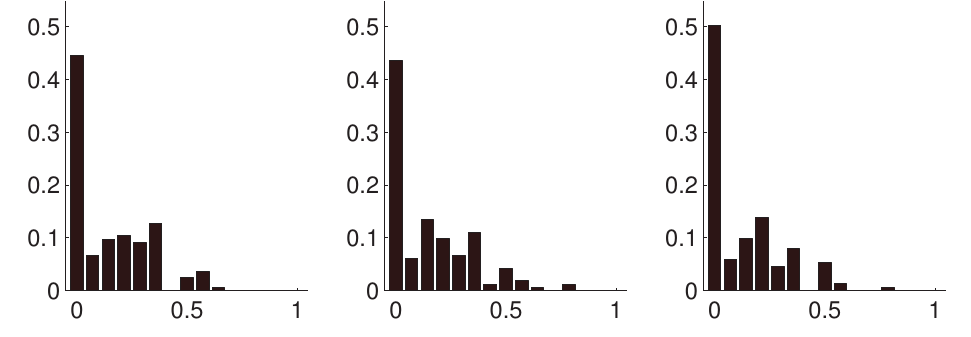}
%\vskip1cm
\caption{Area-perimeter squared ratio for connected component of a lineage. x-axis: $R_3$; y-axis: frequency of connected components. 
Left: random division orientation (a), (d), (g), (l). Middle: radial division orientation (b), (e), (h), (m). Right: tangential division orientation (c), (f), (i), (n). Figures (a), (b), (c): free orientation. Figures (d), (e), (f): constrained orientation. Figures: (g), (h), (i) free orientation strategy. Figures (l), (m), (n): constrained orientation. 
Figures (a)-(f): lineages which lie at the center of the tumor. Figures (g)-(n): lineages which lie at the boundary of the tumor. 
\label{fig11}}
\begin{tikzpicture}[overlay]
    \node at (2.8,13.8){
      {\small {\bf a}}
    };
    \node at (5.5,13.8){
      {\small {\bf b}}
    };
    \node at (8.,13.8){
      {\small {\bf c}}
    };
    \node at (2.8,11.4){
      {\small {\bf d}}
    };
    \node at (5.5,11.4){
      {\small {\bf e}}
    };
    \node at (8.,11.4){
      {\small {\bf f}}
    };
    \node at (2.8,8.6){
      {\small {\bf g}}
    };
    \node at (5.5,8.6){
      {\small {\bf h}}
    };
    \node at (8.,8.6){
      {\small {\bf i}}
    };
    \node at (2.8,5.9){
      {\small {\bf l}}
    };
    \node at (5.5,5.9){
      {\small {\bf m}}
    };
    \node at (8.,5.9){
      {\small {\bf n}}
    };
  \end{tikzpicture}
\end{figure}
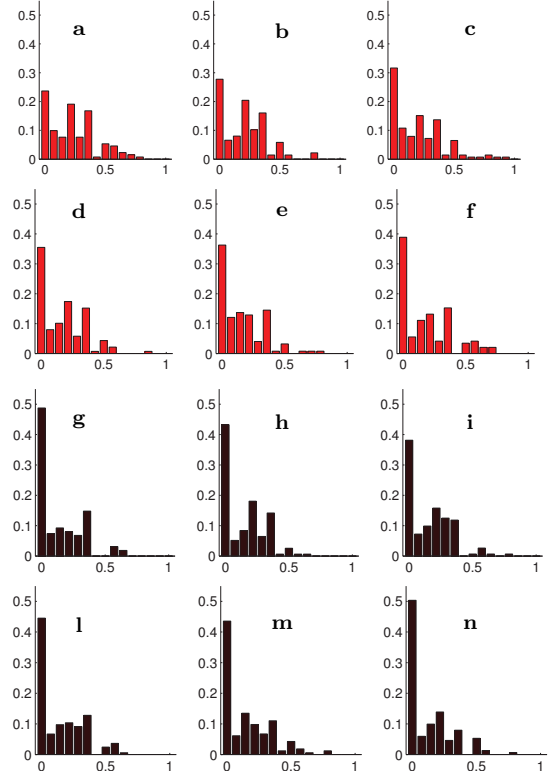

\begin{figure}%[h!]
\hspace*{1.9cm}
\includegraphics[width=75mm,trim={0.9cm 0 0 0}]{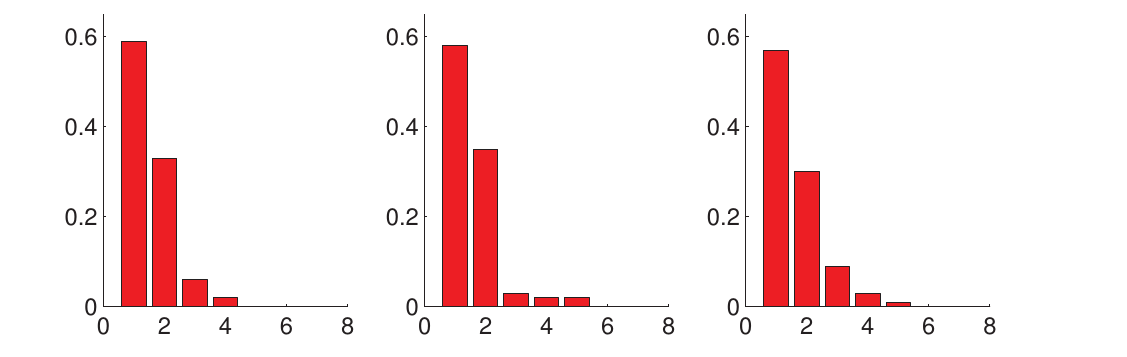}\\
\hspace*{1.5cm}
\includegraphics[width=75mm,trim={-0cm 0 0 0}]{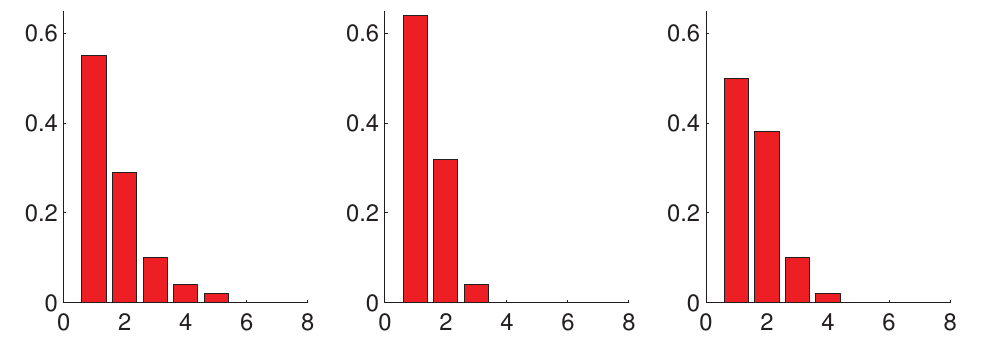}\\
\hspace*{1.5cm}
\includegraphics[width=75mm,trim={-0cm 0 0 0}]{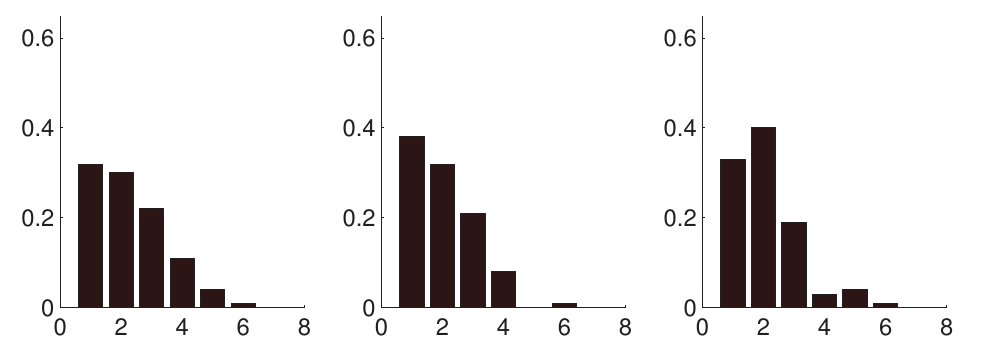}\\
\hspace*{1.5cm}
\includegraphics[width=75mm,trim={-0cm 0 0 0}]{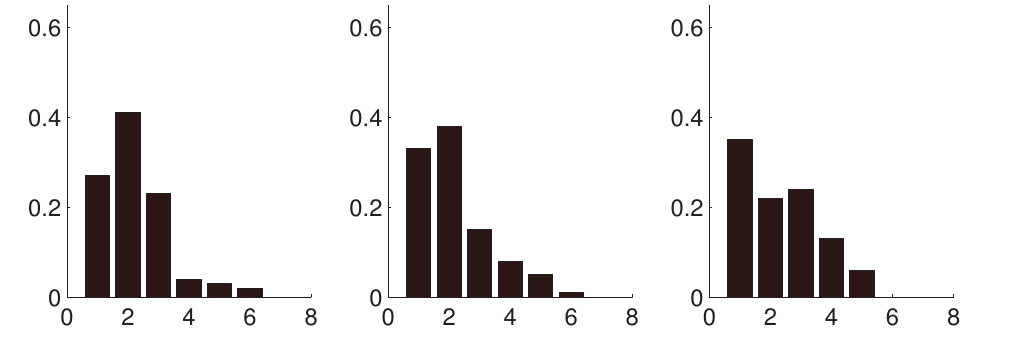}
%\vskip-3cm
\caption{Number of connected components of a lineage $R_4$. x-axis: $R_4$; y-axis: frequency of connected components. 
Left: random division orientation (a), (d), (g), (l). Middle: radial division orientation (b), (e), (h), (m). Right: tangential division orientation (c), (f), (i), (n). Figures (a), (b), (c): free orientation. Figures (d), (e), (f): constrained orientation. Figures: (g), (h), (i) free orientation strategy. Figures (l), (m), (n): constrained orientation. 
Figures (a)-(f): lineages which lie at the center of the tumor. Figures (g)-(n): lineages which lie at the boundary of the tumor.  
\label{fig13}}
\begin{tikzpicture}[overlay]
    \node at (2.8,13.8){
      {\small {\bf a}}
    };
    \node at (5.5,13.8){
      {\small {\bf b}}
    };
    \node at (8.,13.8){
      {\small {\bf c}}
    };
    \node at (2.8,11.4){
      {\small {\bf d}}
    };
    \node at (5.5,11.4){
      {\small {\bf e}}
    };
    \node at (8.,11.4){
      {\small {\bf f}}
    };
    \node at (2.8,8.6){
      {\small {\bf g}}
    };
    \node at (5.5,8.6){
      {\small {\bf h}}
    };
    \node at (8.,8.6){
      {\small {\bf i}}
    };
    \node at (2.8,5.9){
      {\small {\bf l}}
    };
    \node at (5.5,5.9){
      {\small {\bf m}}
    };
    \node at (8.,5.9){
      {\small {\bf n}}
    };
  \end{tikzpicture}
\end{figure}

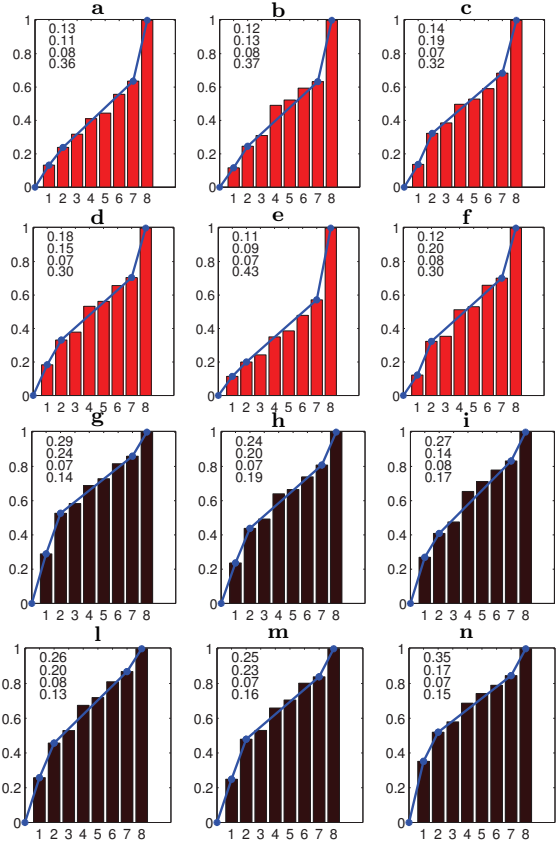
\begin{figure}%[h!]
\hspace*{2.cm}
\includegraphics[width=72mm,trim={-0.cm 0 0 0}]{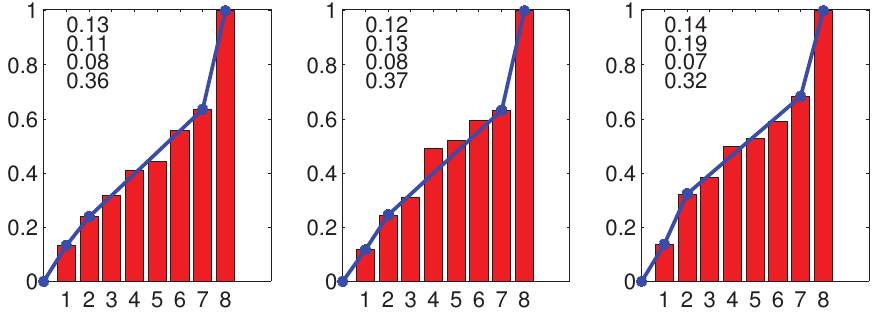}\vskip0.1cm
\hspace*{1.9cm}
\includegraphics[width=75mm,trim={-0.cm 0 0 0}]{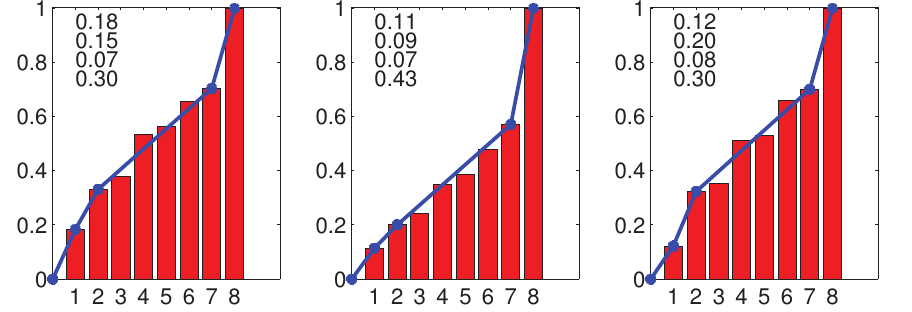}\vskip0.1cm
\hspace*{1.9cm}
\includegraphics[width=75mm,trim={-0.cm 0 0 0}]{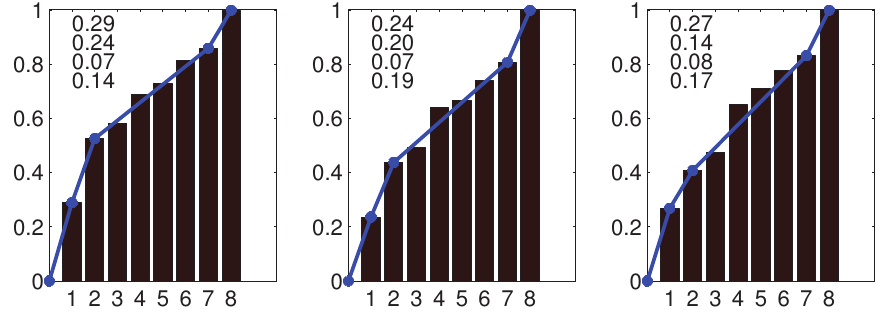}\vskip0.1cm
\hspace*{1.9cm}
\includegraphics[width=75mm,trim={-0.cm 0 0 0}]{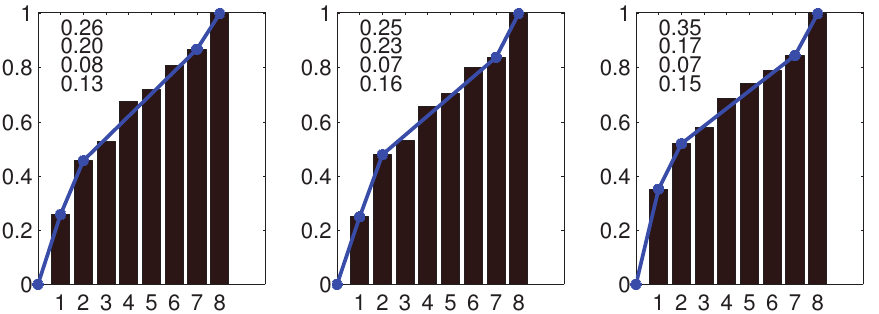}
%\vskip-3cm
\caption{Number of connected components of a lineage $R_5$. x-axis: $R_5$. y-axis: cumulative distribution of connected components. A piecewise linear interpolation of the cumulative distribution is depicted in blue color. The values of the slopes of this linear interpolation are reported in each subgraph. Left: random division orientation (a), (d), (g), (l). Middle: radial division orientation (b), (e), (h), (m). Right: tangential division orientation (c), (f), (i), (n). Figures (a), (b), (c): free orientation. Figures (d), (e), (f): constrained orientation. Figures: (g), (h), (i) free orientation strategy. Figures (l), (m), (n): constrained orientation. Figures (a)-(f): lineages which lie at the center of the tumor. Figures (g)-(n): lineages which lie at the boundary of the tumor.  
\label{fig15}}
\begin{tikzpicture}[overlay]
    \node at (3.3,15.65){
      {\small {\bf a}}
    };
    \node at (5.7,15.65){
      {\small {\bf b}}
    };
    \node at (8.2,15.65){
      {\small {\bf c}}
    };
    \node at (3.3,12.9){
      {\small {\bf d}}
    };
    \node at (5.7,12.9){
      {\small {\bf e}}
    };
    \node at (8.2,12.9){
      {\small {\bf f}}
    };
    \node at (3.3,10.2){
      {\small {\bf g}}
    };
    \node at (5.7,10.2){
      {\small {\bf h}}
    };
    \node at (8.2,10.2){
      {\small {\bf i}}
    };
    \node at (3.3,7.35){
      {\small {\bf l}}
    };
    \node at (5.7,7.35){
      {\small {\bf m}}
    };
    \node at (8.2,7.35){
      {\small {\bf n}}
    };
  \end{tikzpicture}
\end{figure}

\begin{figure}[ht!]
%\vskip-.5cm
\hspace*{2cm}
\includegraphics[width=75mm,trim={0cm 0 0 0}]{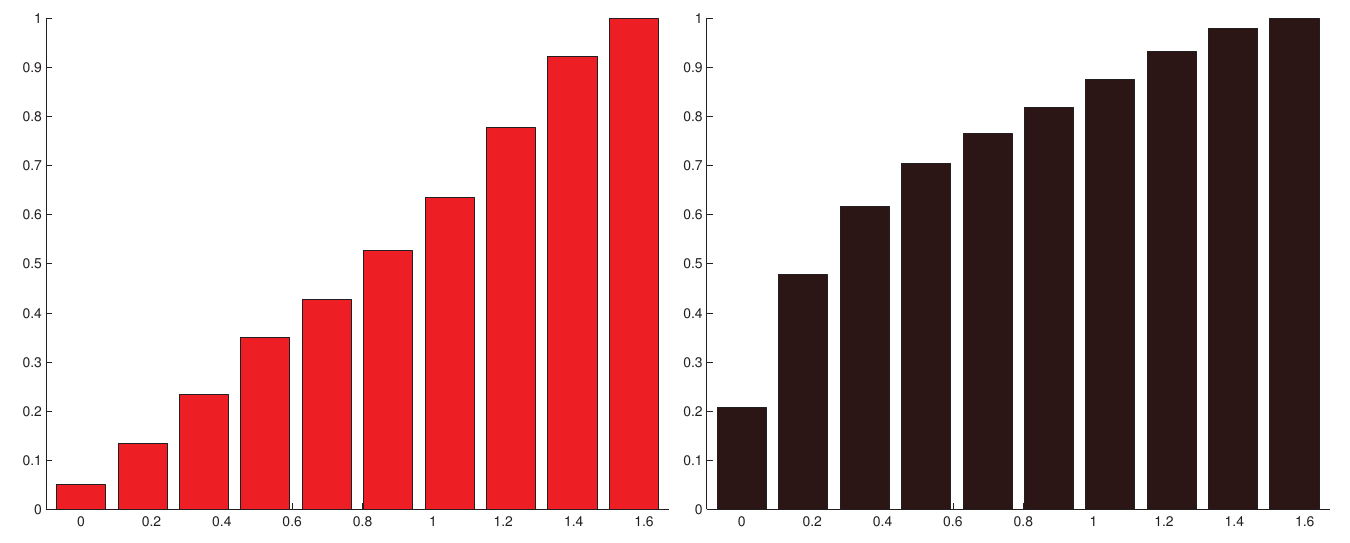}
%\vskip-3cm
\caption{Deviation of the main orientation of the lineages from the tangential direction $|R_6|$. x-axis: $|R_6|$. y-axis: cumulative distribution of connected components. Left: central lineages. Right: peripheral lineages. Results are averaged over all the different division strategies. At the periphery of the cellular aggregate $48\%$ of the lineages are within the two first columns which means a deviation from $0^\circ$ to $15^\circ$ degrees from the tangential direction. This proportion falls down to $14\%$ in the central region.
\label{fig16}}
\end{figure}

\begin{figure}[ht!]
%\vskip-1.1cm
\hspace*{1.9cm}
\includegraphics[width=75mm,trim={-0.cm 0 0 0}]{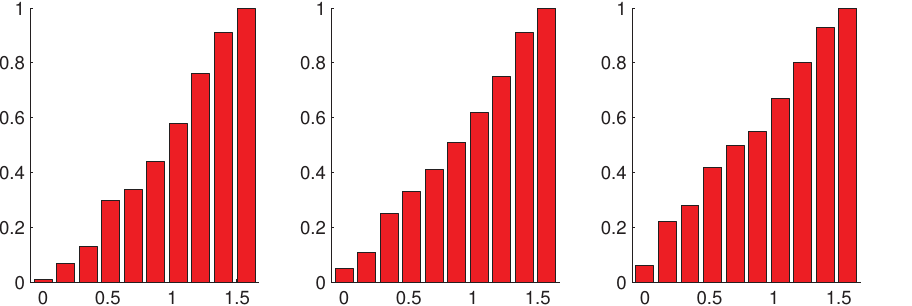}\vskip 0.19cm
\hspace*{1.9cm}
\includegraphics[width=75mm,trim={-0cm 0 0 0}]{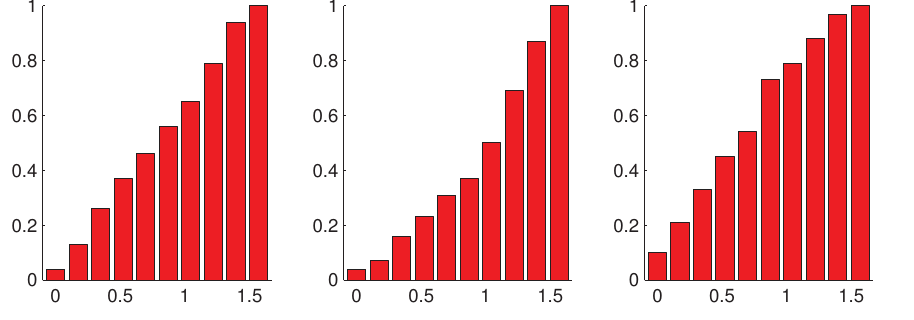}\vskip 0.1cm
\hspace*{1.9cm}
\includegraphics[width=74mm,trim={-0cm 0 0 0}]{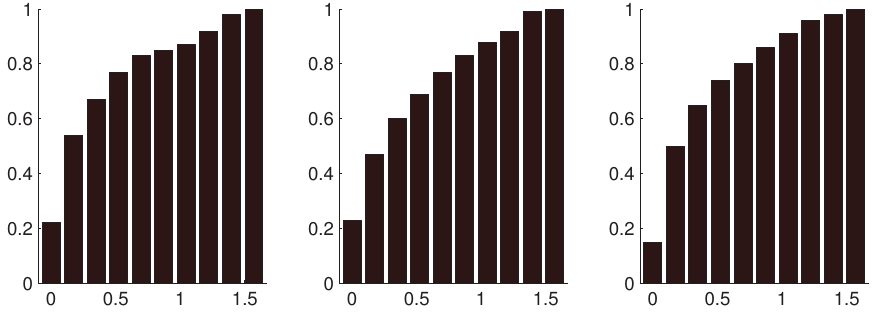}\vskip 0.1cm
\hspace*{1.9cm}
\includegraphics[width=75mm,trim={-0cm 0 0 0}]{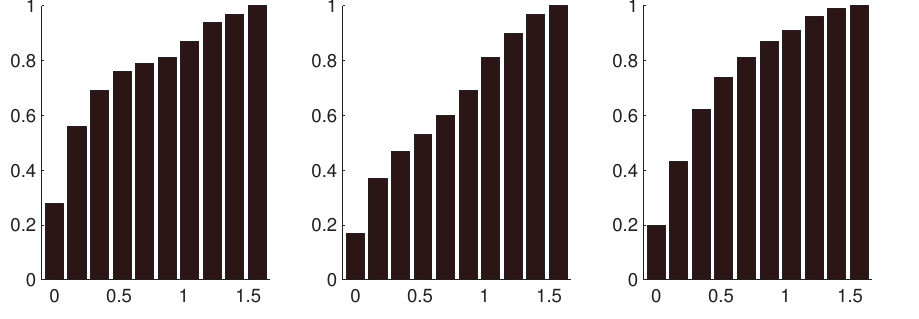}
%\vskip-3cm
\caption{Deviation of the main orientation of the lineages from the tangential direction $|R_6|$. x-axis: $|R_6|$. y-axis: cumulative distribution of connected components. Left: random division orientation (a), (d), (g), (l). Middle: radial division orientation (b), (e), (h), (m). Right: tangential division orientation (c), (f), (i), (n). Figures (a), (b), (c): free orientation. Figures (d), (e), (f): constrained orientation. Figures: (g), (h), (i) free orientation strategy. Figures (l), (m), (n): constrained orientation. Figures (a)-(f): lineages which lie at the center of the tumor. Figures (g)-(n): lineages which lie at the boundary of the tumor.  
\label{fig17}}
\begin{tikzpicture}[overlay]
    \node at (3.2,14.7){
      {\small {\bf a}}
    };
    \node at (5.9,14.7){
      {\small {\bf b}}
    };
    \node at (8.2,14.7){
      {\small {\bf c}}
    };
    \node at (3.2,12.){
      {\small {\bf d}}
    };
    \node at (5.9,12.){
      {\small {\bf e}}
    };
    \node at (8.2,12.){
      {\small {\bf f}}
    };
    \node at (3.2,9.2){
      {\small {\bf g}}
    };
    \node at (5.9,9.2){
      {\small {\bf h}}
    };
    \node at (8.2,9.2){
      {\small {\bf i}}
    };
    \node at (3.2,6.4){
      {\small {\bf l}}
    };
    \node at (5.9,6.4){
      {\small {\bf m}}
    };
    \node at (8.2,6.4){
      {\small {\bf n}}
    };
  \end{tikzpicture}
\end{figure}

We analyze here the detailed results of the simulations.\\
\textbf{Diagnostic 1. Sphericity of the population $R_1$:}
for all different cases tested, the distribution of $R_1$ shows a Gaussian profile with mean value approximately around $0.7$. These distributions are illustrated in Fig. \ref{fig8}. This result suggests that the tested orientation divisions have a very limited impact on the overall shape of the population, in other words the positioning rule has much stronger impact on the overall shape of the population that the division law. However, some differences clearly emerge from the results. In particular we can observe that the differences in sphericity between the three orientation directions are much stronger when the constrained strategy is employed. This is expected since in the free strategy case the cells have more freedom to adapt to reduce the energy of the system. In addition, when the tangential direction strategy is used larger sphericity values are obtained than with the radial orientation strategy. This can be explained by the fact that a cell at the boundary which divides radially generates more irregularity compared to the tangential division. This irregularity increases the perimeter of the population keeping the area almost fixed.\\
\textbf{Diagnostic 2. Convexity of the population $R_2$:}
results are consistent with the ones of the previous diagnostic. Whatever the strategy and division direction chosen, the values of $R_2$ are very similar, around a mean value of 0.94, as reported in Fig. \ref{fig9}. Giving a closer look at the results, we see that the differences between the three division orientations are more pronounced when the constrained strategy is employed. In particular we can state that, if the free strategy is adopted no difference between the three orientations of the division plane is observed. In the constrained case, results show larger $R_2$ values for the tangential direction, which can be interpreted by saying that the shape of the boundaries is more regular when the division plane has tangential direction.\\
\textbf{Diagnostic 3. Sphericity of the lineage $R_3$:} in Fig. \ref{fig2bis} $(a)$ and $(c)$ we plot the distribution of connected components of a given lineage as a function of $R_3$. On the left picture the results for a lineage situated close to the center of the tumor are shown while on the right picture those for a lineage at the periphery of the tumor are displayed. All the different adopted strategies are averaged together. The results clearly evidence differences between the center and the peripheral zones on the morphology of the lineages. The sphericity of the lineages is much lower at the periphery than in the central region. In particular, almost half of connected components have $R_3 = 0$ at the boundary of the aggregate while only one third of the components have $R_3 = 0$ in the central region. We can conclude that the lineages are more likely to remain grouped in the central region. This is true for all the tested division orientations and strategies. For each of these, the results are reported separately in Fig. \ref{fig11}.\\
\textbf{Diagnostic 4. Lineage fragmentation $R_4$:}
the number of connected component for central and peripheral lineages $R_4$ is reported in Fig. \ref{fig2bis} $(d)$ and $(e)$ for all the different strategies averaged together. The main result we can draw is that a strong disparity occurs between the central and peripheral lineages. In particular, almost all lineages are divided from one up to four connected components in the central region and from one to five connected components when at the periphery. In the central region around $57\%$ of lineages remain connected while $33\%$ of the lineages are separated into two connected components. On the periphery of the tumor, the proportions are $33\%$ and $34\%$ respectively. We can conclude that lineages are more fragmented at the periphery. This is probably due to the fact that divisions inside the aggregate push cells farther from the central region which results in their intercalation between cells of the lineages of the periphery. The complete set of data for all the studied different strategies is reported in Fig. \ref{fig13}. We can notice that two cases give clearly remarkable results: in the central region, the radially constrained division orientation presents lineages that are divided almost always in one or two components. On the other hand, at the periphery, the constrained tangential orientation of the division plane presents many lineages divided in four or five components.\\
\textbf{Diagnostic 5. Fragment sizes of a lineage $R_5$:}
the number of cells for the connected components of the central and peripheral lineages is reported in Fig. \ref{fig3bis} $(g)$ and $(l)$. The Figure shows the cumulative distribution of connected components as a function of the number $R_5$ of cells of the connected component. The Figure reports also the profile of this distribution between five zones: from zero to one cell, from one to two, from two to seven, from seven to eight cells and from eight to larger connected components. This profile is obtained by computing the linear interpolation lines of the cumulative distribution between the ends of these five zones. On the periphery of the cellular aggregate, there are $50\%$ of small connected components and $16\%$ of large connected components.
By contrast, in the central region, there is a higher percentage (almost $30\%$) of small connected components and a similar percentage ($34\%$) of large connected components. In the central and peripheral regions, the percentage of lineages of medium size (three to seven cells) is almost identical (around $7.5\%$). There is a sort of phase transition from small to large connected components. It can be concluded that, at the periphery, the connected components are more numerous and composed of a fewer number of cells. In other words, more cells are isolated from their lineage.
Full data, strategy per strategy, are shown in Fig. \ref{fig15}. From this Figure, we can observe that the free division strategy for the three considered division directions give very similar results, whereas the fixed strategy enhances the differences between the different possible orientations of the division. Two situations are remarkable. The first one is the radially constrained strategy which causes the formation of many large connected components in the central region. The second remarkable situation is obtained for the constrained tangential orientation strategy. For this situation, on the boundary of the aggregate we obtain many isolated cells separated from the rest of their lineage.\\ 
% \begin{figure}%[h!]
% \vskip-.5cm
% \hspace*{1.cm}
% \includegraphics[width=105mm,trim={0cm 0 0 0}]{diagnostic5.pdf}
% %\vskip-3cm
% \caption{Number of cells per connected component of lineage, $R_5$. x-axis: $R_5$. y-axis: cumulative distribution of connected components. Left: central lineages. Right: peripheral lineages. Results are averaged over all the different division strategies. A piecewise linear interpolation of the cumulative distribution is depicted in blue color. The values of this piecewise linear interpolation are indicated in the graph. The connected components with eight cells make $34\%$ of the total number of components in the central region and only $16\%$ on the periphery.
% \label{fig14}}
% \end{figure}
\textbf{Diagnostic 6. Lineage orientation $R_6$:}
in Fig. \ref{fig16} we report the cumulative distribution of lineages as a function of $|R_6|$. A concentration of the frequencies near $0$ indicates that the main direction of the lineages is tangential while a concentration close to $\pi/2$ indicates a radially oriented lineage. At the periphery there is a high percentage (almost $50\%$) of lineages the orientation of which is in the tangential direction. In the central region, no direction seems privileged. The tangential direction is strongly favored by the lineages at the periphery probably because of the influence of the positioning rule. The complete data are shown in Fig. \ref{fig17}. From this Figure, we can observe that for the free strategy the three orientations of the division plane give very similar results whereas for the constrained strategy differences become more visible. Moreover, for the radially constrained division strategy the results depart significantly from those of the other strategies. In the central region, this orientation gives $50\%$ of lineages whose direction angle is less than $25^\circ$. At the periphery, for tangential or random division orientations, the three-quarters of the lineages exhibit a difference with respect to the tangential direction of less than $35^\circ$, while a similar result is only reached for the radial division strategy  when considering all the lineages until the value of $60^\circ$.\\

\textbf{acknowledgements}
The support of the TRI-Genotoul and ITAV imaging facility is gratefully acknowledged. The work in the laboratory of BD and VL is supported by the University of Toulouse, the CNRS and la Ligue contre le cancer. PD acknowledges support by the Engineering and Physical Sciences Research Council (EPSRC) under grant ref. EP/M006883/1 and EP/N014529/1, by the Royal Society and the Wolfson Foundation through a Royal Society Wolfson Research Merit Award ref. WM130048 and by the National Science Foundation (NSF) under grant RNMS11-07444 (KI-Net). PD is on leave from CNRS, Institut de Math\'ematiques
de Toulouse, France.
%\end{acknowledgements}

\medskip
\noindent
{\bf Data accessibility.} Data supporting this work are available on : TBC

% Non-BibTeX users please use

\end{document}